\pdfoutput=1
\documentclass{elsarticle}
\newcommand*{\rom}[1]{\uppercase\expandafter{\romannumeral #1\relax}}

\newcommand{\bx}{\boldsymbol{\textbf{x}}}

\newcommand{\bR}{\boldsymbol{\textbf{R}}}

\newcommand{\Rthree}{ {\mathbb{R}^{3}} }

\providecommand{\e}[1]{\ensuremath{\times 10^{#1}}}  

\usepackage{graphicx}
\usepackage{multirow}
\usepackage{fancyhdr, ifpdf}
\usepackage{amsxtra}
\usepackage{stmaryrd}
\usepackage{mathrsfs}
\usepackage{psfrag}
\usepackage{epsfig}
\usepackage{rotating}
\usepackage{setspace}
\usepackage{float}
\usepackage{amsfonts}
\usepackage{amsmath}
\usepackage{amsbsy}
\usepackage{amscd}
\usepackage{amsthm}
\usepackage{color}
\usepackage{tabularx}
\usepackage{caption}
\usepackage{sidecap}
\usepackage{dcolumn}
\usepackage[ruled]{algorithm2e}
\usepackage{algorithmic}
\usepackage{subfigure}
\usepackage{tablefootnote}
\usepackage[bottom]{footmisc}
\usepackage{scrextend}
\usepackage{hyperref}

\definecolor{hellgruen}{rgb}{0.2,0.7,0.2}

\usepackage{lineno,hyperref}
\modulolinenumbers[5]

\usepackage[margin=0.75in]{geometry}

\journal{Journal of the Mechanics and Physics of Solids}

\biboptions{authoryear}

\begin{document}



\begin{frontmatter}

\title{Electronic structure study of screw dislocation core energetics in Aluminum and core energetics informed forces in a dislocation aggregate}

\author[umichAddress1]{Sambit Das}
\author[umichAddress2]{Vikram Gavini\corref{mycorrespondingauthor}}
\cortext[mycorrespondingauthor]{Corresponding author. \\ \textit{E-mail address}: vikramg@umich.edu (V. Gavini).}
\address[umichAddress1]{Department of Mechanical Engineering, University of Michigan, Ann Arbor, MI 48109, USA}
\address[umichAddress2]{Department of Mechanical Engineering, Department of Materials Science and Engineering, University of Michigan, Ann Arbor, MI 48109, USA}

\begin{abstract}
We use a real-space formulation of orbital-free DFT to study the core energetics and core structure of an isolated screw dislocation in Aluminum. Using a direct energetics based approach, we estimate the core size of a perfect screw dislocation to be $\approx$ 7 $|{\bf b}|$, which is considerably larger than previous estimates of $1-3~|{\bf b}|$ based on displacement fields. The perfect screw upon structural relaxation dissociates into two Shockley partials with partial separation distances of 8.2 \AA~and 6.6 \AA~ measured from the screw and edge component differential displacement plots, respectively. Similar to a previous electronic structure study on edge dislocation, we find that the core energy of the relaxed screw dislocation is not a constant, but strongly dependent on macroscopic deformations. Next, we use the edge and screw dislocation core energetics data with physically reasonable assumptions to develop a continuum energetics model for an aggregate of dislocations that accounts for the core energy dependence on macroscopic deformations. Further, we use this energetics model in a discrete dislocation network, and from the variations of the core energy with respect to the nodal positions of the network, we obtain the nodal core force which can directly be incorporated into discrete dislocation dynamics frameworks. We analyze and classify the nodal core force into three different contributions based on their decay behavior. Two of these contributions to the core force, both arising from the core energy dependence on macroscopic deformations, are not accounted for in currently used discrete dislocation dynamics models which assume the core energy to be a constant excepting for its dependence on the dislocation line orientation. Using case studies involving simple dislocation structures, we demonstrate that the contribution to the core force from the core energy dependence on macroscopic deformations can be significant in comparison to the elastic Peach-Koehler force even up to distances of $10-15$ nm between dislocation structures. Thus, these core effects, whose origins are in the electronic structure of the dislocation core, can play an important role in influencing dislocation-dislocation interactions to much larger distances than considered heretofore. 
\end{abstract}


\end{frontmatter}

\section{Introduction}\label{sec:intro}
Crystalline plasticity is governed by the collective behavior of dislocations, which includes their nucleation, kinetics, evolution, and interaction with other defects such as point-defects (vacancies, interstitials), solute atoms, precipitates, grain boundaries, surfaces and interfaces (cf.~\citet{argon2008,Hirth1982,Reed1973}). This dislocation mediated deformation response at macroscale is governed by a vast span of interacting length-scales (cf. ~\cite{LeSar2014}) from electronic-structure effects at the dislocation core to continuum elastic effects, with mesoscale physics of the dislocation microstructure (cf. ~\cite{Kubin2013}) playing an important role. Various mesoscopic modeling approaches like the discrete dislocation dynamics (DDD)~\citep{Arsenlis2002,Arsenlis2007,Aubry2016,Bulatov2004,Kubin1992,Martinez2008,Nicola2003,Po2014,Schwarz1999,Zbib1998}, phase field methods~\citep{Beyerlein2016,Hunter2014,Koslowski2002,Lee2011,mianroodi2015,Wang2001}, and continuum theories based on continuously distributed dislocations~\citep{Acharya2001,Head1993,kroner1981,Mura1963,Needleman2001,Wang2016} have been developed to connect dislocation microstructure evolution to macroscale crystal plasticity quantities like yield stress, flow stress and strain hardening rate. While these mesoscopic models correctly account for the elastic interactions outside the dislocation core using linear elastic theories of dislocations~\citep{Gutkin1999,Hirth1982,Lazar2013,Lazar2014} or non-linear elastic theories of dislocations~\citep{Wang2016,Willis1967}, the physics inside the dislocation core has been supplemented by atomistic calculations of dislocation core structure, dislocation core energetics, dislocation mobilities, and solute strengthening (cf. e.g.~\cite{Curtin2006,Marian2006,Martinez2008,Morris2000,Olmsted2005,Olmsted2006,Srinivasan2005,Wu2015}). Atomistic models have also been coupled with DDD (cf.~\cite{Cho2015,Curtin2005,Curtin2002}). However, the major drawback of these atomistic calculations is that they are based on interatomic potentials, which may not adequately describe the significant electronic structure effects inside the dislocation core. In fact, recent investigations of dislocation core structure in Al and Mg~\citep{Ghazisaeidi2014,Shin2009,Shin2012,Shin2013,Woodward2008} using explicit electronic structure calculations based on plane-wave implementations of density functional theory (DFT) have demonstrated that atomistic predictions of the dislocation core structure properties are widely sensitive to the choice of the interatomic potential, and also show discrepancies in comparison to the more transferable and accurate electronic structure calculations. For instance, ~\cite{Shin2013}, using orbital-free DFT, have found a metastable undissociated core structure of screw dislocation in Aluminum, which has not been reported in prior atomistic simulations. Furthermore, dislocation core structure and energetics significantly influence dislocation mobilities and dislocation-solute interactions, which have been recently investigated using electronic structure calculations (cf.~\cite{Shin2013,Dallas2005,Tsuru2015,Ventelon2013,Yasi2010}). Thus, with the increasing computational resources and the recent development of efficient numerical algorithms, there is a rapidly growing interest in electronic structure studies of fundamental dislocation core properties in a wide range of crystalline materials. We refer to a recent overview article by ~\cite{Rodney2016}, which discusses the unique contributions of electronic structure calculations towards the prediction of dislocation core properties in metals and semiconductors, and the challenges that still remain. 

One significant challenge that still remains with commonly used plane-wave based DFT implementations (cf. ~\cite{Gonze2002,Hung2010}) is that the displacement fields of an isolated dislocation are not compatible with the periodic boundary conditions inherent in plane-wave based methods. Such studies are thus limited to either using dipole and quadrapole configurations of dislocations~\citep{Weinberger2013,Yadav2014}, or introducing a vacuum region around the isolated dislocation ~\citep{Ghazisaeidi2014,Shin2009,Woodward2008}. While these approaches have been useful to predict the dislocation core structures~\citep{Ghazisaeidi2014,Shin2009,Woodward2008,Yadav2014}, and to compute Peierls stress~\citep{Shin2013,Weinberger2013,Yadav2014} and interaction energetics such as dislocation-solute binding energies~\citep{Dallas2005,Yasi2010}, a direct calculation of dislocation core energy using plane-wave based DFT implementations has been beyond reach. We note that some studies (cf. ~\cite{Clouet2009,Dezerald2014,Li2004}) have used an indirect approach to extract the core energy of the isolated dislocation from the energy of the dipole or quadrapole configuration by subtracting out the total elastic interaction energy between the dislocations in the simulation domain and its periodic images. However, these studies have employed simulation cells containing a few hundred atoms, where the spacing between dislocations is much smaller than the dislocation core-sizes estimated from direct calculations on isolated dislocations as reported in a recent study for the case of an edge dislocation in Aluminum~\citep{Iyer2015} and as will be demonstrated in the present study for the case of a screw dislocation in Aluminum. In order to address the aforementioned limitation of the plane-wave basis, recent efforts have focussed on real space formulations of DFT and their numerical implementations using finite-element discretization~\citep{Das2015, Gavini2007c, Motamarri2012,Motamarri2013, Motamarri2014, Suryanarayana2010} and finite-difference discretization~\citep{Ghosh2016,Ghosh2016b,Suryanarayana2014}.        
 
In the present work, we employ a local real-space formulation of orbital-free density functional theory~\citep{Das2015,Gavini2007c,Radhakrishnan2010} implemented using finite-element discretization to study a screw dislocation in Aluminum. In orbital-free DFT, the kinetic energy of non-interacting electrons is modeled as an explicit functional of the electron density. The widely used WGC kinetic energy functional~\citep{Wang1999} is in good agreement with Kohn-Sham DFT over a wide range of bulk and defect properties in Aluminum as demonstrated in~\citet{Carling2003,Ho2007,Shin2009,Wang1999}, and is employed in this work. As the energy is an explicit functional of the electron density in orbital-free DFT, the computational complexity scales linearly with respect to the system size in orbital-free DFT, as opposed to the cubic-scaling complexity of Kohn-Sham DFT. This enables consideration of larger computational domain sizes using orbtial-free DFT, which are required to accurately study the energetics of isolated defects as demonstrated in prior studies~\citep{Das2015,Iyer2015,Radhakrishnan2010} and is also underscored in the present study. Furthermore, the finite-element basis used to discretize the orbital-free DFT formalism is a natural choice as it can handle the complex non-periodic geometries of isolated dislocations. The finite element-basis also allows for consideration of arbitrary boundary conditions (Dirichlet, periodic, semi-periodic), which in conjunction with the local real-space formulation enables the direct calculation of the dislocation core energy. Such direct calculations of dislocation core energetics were conducted recently on an edge dislocation in Aluminum using orbital free DFT~\citep{Das2016,Iyer2015}, where the core size of the isolated edge dislocation was computed directly from the energetics, and the dislocation core energy was shown to strongly depend on the macroscopic deformation. Our present study on a screw dislocation follows along similar lines, but the quantum-mechanical effects at the dislocation core can be significantly influenced by the dislocation character, which is evidenced by comparing the results in the present study with the corresponding results from the edge dislocation study. Equally important, in this work, we develop a framework to use the screw and edge dislocation core energetics data computed from electronic structure calculations to inform the forces on dislocations in a dislocation aggregate, thus providing a phenomenological approach to incorporate the electronic structure effects into continuum DDD simulations.  

In the first part of this work, we start by studying the dislocation core size of a perfect screw dislocation in Aluminum. For this study, the atomic positions of the perfect screw dislocation are determined from the classical Volterra solution of screw dislocation~\citep{Hirth1982} using isotropic elasticity. We employ mixed boundary conditions on the electronic fields that represents an isolated line dislocation embedded in bulk. While holding the atomic positions frozen, we find the distance from the dislocation line up to which the contribution from electronic structure perturbations to the dislocation energy is non-trivial, and thus directly estimate, from an energetic viewpoint, the core-size of the dislocation. The core-size of the perfect screw dislocation in Aluminum from this study is found to be $\approx$ 7 $|{\bf b}|$, which is much larger than conventional displacement field based estimates of $1-3~|{\bf b}|$~\citep{Banerjee2007,Hirth1982,Peierls1940}. A similarly large core size of $\approx$ 10 $|{\bf b}|$ was also observed in a recent study ~\citep{Iyer2015} of a perfect edge dislocation in Aluminum. Subsequently, we relax the internal core structure, resulting in the dissociation of the perfect screw dislocation into two Shockley partials. The computed partial separation distances of 6.6 \AA~corresponding to the edge component in the differential displacement plots and 8.2 \AA~corresponding to the screw component are found to be in close agreement with previous electronic structure studies~\citep{Shin2009,Woodward2008}. As a next step, we apply macroscopic deformations to the perfect screw dislocation, relax the internal core structure, and compute the core energy of the relaxed screw dislocation as a function of the macroscopic deformation. In particular, we consider equi-triaxial volumetric strains, uniaxial strains, and shear strains over a considerable range of values. Interestingly, we find that the core energy of relaxed screw dislocation is strongly dependent on the macroscopic deformation. Further, the core energy dependence on macroscopic deformations has a non-zero slope at zero deformation, which has important implications even at continuum scales. Similar dependence of the core energy of relaxed edge dislocation on macroscopic deformations was reported in~\citet{Das2016,Iyer2015}. However, we find that the dependence of the screw core energy on macroscopic deformations is in general monotonic and exhibits a close to linear dependence, whereas the dependence of the edge core energy on macroscopic deformations was, in general, found to be non-linear and non-monotonic. The strong dependence of the dislocation core energy on macroscopic deformations leads to an additional configurational force in an inhomogeneous strain field. This has important consequences towards the force experienced by a dislocation in a dislocation aggregate, which is the focus in the second part of this work. 

In the second part of this work, we develop a model that takes into account the core energy dependence on macroscopic deformation and derive the expressions for the force experienced by a dislocation in an arbitrary dislocation aggregate present inside an infinite isotropic elastic continua. To this end,  we first build an energetics model, where the total energy of the dislocation aggregate is partitioned into the elastic energy obtained using the non-singular linear elastic model formulated by ~\cite{Cai2006} and the contribution to the dislocation core energy beyond the non-singular core approximation that includes the non-elastic contributions governed by quantum mechanical interactions. This partition is obtained from the core energetics of isolated edge and screw dislocations computed using electronic structure calculations. We then apply this energetics model to a nodal discrete dislocation network, and obtain the nodal force as a generalized configurational force conjugate to the nodal degrees of freedom. The key aspect of this model is that it accounts for the aforementioned dislocation core energy dependence on macroscopic deformation, which has not been considered in prior formulations that are the basis for state-of-the-art DDD frameworks (cf. ~\cite{Arsenlis2007,Martinez2008}), where the common assumption is that the core energy is a constant. We next analyze the nodal force, and find that the contributions coming from the core energy dependence on the macroscopic deformation contain terms that are proportional to the external strain field and its spatial gradient at the dislocation core. These additional contributions, whose origins are atomistic and quantum mechanical in nature and which we refer to as the \emph{core force}, can be significant in regions of inhomogeneous deformations. In order to elucidate the relative importance of these contributions to the total force on a dislocation, we consider case studies on interaction forces between simple dislocation structures such as an infinite straight edge dislocation, dislocation glide loop, and low angle tilt grain boundary. We find that, even for the relatively simple dislocation structures we considered in our case studies, the core force can be significant in comparison to the Peach-Koehler force and the Peierls-Nabarro force even up to distances ranging from $10-15$ nm between these dislocation structures, thus underscoring the importance of these contributions in governing dislocation behavior.

The remainder of this paper is organized as follows. Section~\ref{sec:OFDFT} presents an overview of the real-space orbital-free DFT methodology employed in this work. Section~\ref{sec:screw} presents our electronic-structure study of the screw dislocation in Aluminum along with a discussion of the findings, a comparison with corresponding findings from the previous edge dislocation study, and the implications towards governing dislocation behavior in inhomogeneous strain fields. Next, in Section~\ref{sec:coreForceModel}, we use the electronic structure data of dislocation core energetics to inform the forces on dislocations in an arbitrary dislocation aggregate, in particular, focusing on a nodal discrete dislocation network. Section~\ref{sec:caseStudies} applies the developed force model to various case studies and numerically investigates the significance of the additional core force contribution manifesting from the dislocation core energy dependence on macroscopic deformation. We finally conclude in Section~\ref{sec:Conclusions} with a summary and an outlook.

\section{Overview of orbital-free DFT}\label{sec:OFDFT}
In order to keep the discussion self-contained, we now provide a brief overview of the local real-space formulation of orbtial-free DFT, the configurational forces associated with structural relaxations, and the finite-element discretization of the formulation employed in this work. We refer to ~\cite{Das2015,Motamarri2012} for a detailed discussion of the methodology.

\subsection{Local real-space formulation of orbital-free DFT}
The ground-state energy of a charge neutral materials system containing $M$ nuclei and $N$ valence electrons in density functional theory \citep{Martin,ParrYang} is given by
\begin{equation}\label{eq:DFT}
E(\rho,\bR)=T_s(\rho)+E_{xc}(\rho)+J(\rho,\bR)\,,
\end{equation}
where $\rho$ denotes the electron-density and $\bR=\{\bR_{1},\bR_{2},\ldots,\bR_{M}\}$ denotes the vector containing the positions of $M$ nuclei (ions). In the above, $T_s$ denotes the kinetic energy of non-interacting electrons, $E_{xc}$ is the exchange-correlation energy, and $J$ denotes the classical electrostatic interaction energy involving the nuclei and the electrons.


The first term in equation~\eqref{eq:DFT}, $T_s$, is computed exactly in the Kohn-Sham formalism using the single electron wavefunctions computed from the Kohn-Sham eigenvalue problem, which conventionally scales as $\mathcal{O}(N^3)$. While there has been progress in developing reduced order scaling algorithms for the Kohn-Sham approach~\citep{bow2012,goedecker99}, the accessible system sizes have still been limited to a few thousand atoms for metallic systems~\citep{Motamarri2014}. On the other hand, in orbital-free DFT, $T_s$ is modeled as an explicit functional of the electron density~\citep{ParrYang}. In the present work, we use the Wang-Goving-Carter (WGC) density-dependent orbital-free kinetic energy functional~\citep{Wang1999}, which is a widely used kinetic energy functional for ground-state calculations on materials systems with an electronic structure close to a free electron gas. In particular, it has been shown to be transferable for the Al, Mg and Al-Mg materials systems~\citep{Carling2003,Das2015,Ho2007}. The functional form of the WGC orbital-free kinetic energy functional is given by
\begin{equation}\label{eq:KE}
T_s(\rho)=C_F\int \rho^{5/3}(\bx)\,{\rm d}\bx + \frac{1}{2}\int |\nabla \sqrt{\rho(\bx)}|^2\,{\rm d}\bx + T_{K}(\rho)\,,
\end{equation} 
where
\begin{eqnarray}
T_{K}(\rho)=C_F\int\int \rho^{\alpha}(\bx)\,K(\xi_{\gamma}(\bx,\bx'),|\bx-\bx'|)\,\rho^{\beta}(\bx')\,{\rm d}\bx\,{\rm d}\bx'\,,\notag\\
\xi_{\gamma}(\bx,\bx')=\Big(\frac{k_F^{\gamma}(\bx)+k_F^{\gamma}(\bx')}{2}\Big)^{1/\gamma}, \quad k_F(\bx)=\big(3\pi^2\rho(\bx)\big)^{1/3}\,.\notag
\end{eqnarray}
In equation~\eqref{eq:KE}, the first term denotes the Thomas-Fermi energy with $C_F=\frac{3}{10}(3\pi^2)^{2/3}$, and the second term denotes the von-Weizs$\ddot{a}$cker correction~\citep{ParrYang}. The last term denotes the density dependent kernel energy, $T_{K}$, where the kernel $K$ is chosen such that the linear response of a uniform electron gas is given by the Lindhard response~\citep{Finnis}. In the WGC functional, the parameters in the density dependent kernel are chosen to be $\{\alpha,\beta\}=\{5/6+\sqrt{5}/6,5/6-\sqrt{5}/6\}$ and $\gamma=2.7$. Further, the density dependent kernel is expanded as a Taylor series about a reference electron density ($\rho_0$), considered to be the average electron density of the bulk crystal. Numerical investigations have suggested that the Taylor expansion to second order provides a good approximation of the density dependent kernel for materials systems with electronic structure close to a free electron gas~\citep{Choly2002,Wang1999}. The use of an explicit kinetic energy density functional makes orbital-free DFT inherently linear-scaling with system size and enables the consideration of large system sizes, which is necessary for an accurate study of the energetics of dislocations.

The second term in equation~\eqref{eq:DFT}, the exchange-correlation energy, denoted by $E_{xc}$, incorporates all the quantum-mechanical interactions in the ground-state energy of a materials system. While the existence of a universal exchange-correlation energy as a functional of electron-density has been established by~\cite{kohn64,kohn65}, its exact functional form has been elusive to date, and various models have been proposed over the past decades. For solid-state calculations, the local density approximation (LDA)~\citep{Ceperley1980,Perdew1981} and the generalized gradient approximation (GGA)~\citep{Langerth1983,Perdew1992} have been widely adopted across a range of materials systems. The last term in equation~\eqref{eq:DFT}, the classical electrostatic interaction energy $J$ is composed of the electron-electron, electron-ion and ion-ion interactions. 

We note that among the terms in the energy functional~\eqref{eq:DFT}, the electrostatic interaction energy and the kernel energy are non-local, or, in other words, they comprise of extended interactions in real-space. Local reformulations of these extended interactions have been developed recently~\citep{Das2015,Gavini2007c,Radhakrishnan2010} as they are crucial for an efficient and scalable numerical implementation of orbital-free DFT in real space, as well as, enabling the consideration of general boundary conditions which is exploited in the present work. The main idea behind these local reformulations is to take recourse to the solution of a partial differential equation whose Green's function corresponds to the kernel of the extended interaction, and further recast these extended interactions using a variational form of the corresponding partial differential equations. First, considering the extended interactions in electrostatics, we note that they are governed by the $\frac{1}{|\bx-\bx'|}$ kernel, which is the Green's function of the Laplace operator. Thus, the electrostatic interaction energy can be reformulated as the following local variational problem:
\begin{equation}\label{eq:elReformulation}
J
= -\min_{\phi \in \mathcal{Y}} \left\{\frac{1}{8\pi}\int |\nabla \phi(\bx)|^2\, {\rm d}\bx - \int \big(\rho(\bx) + \sum_{I=1}^{M}b_{I}(\bx,\bR_I)\big)\phi(\bx) \,{\rm d}\bx\right\}-E_{self}\,,
\end{equation}
where $b_I(\bx,\bR_I)$ denotes the charge distribution corresponding to the ionic pseudopotential of the $I^{th}$ nucleus, $\phi$ denotes the electrostatic potential corresponding to the total charge distribution comprising of the electrons and the nuclei, and $\mathcal{Y}$ denotes a suitable function space incorporating appropriate boundary conditions for the problem being solved. In the above, $E_{self}$ denotes the self energy of the nuclear charge distributions which can also be evaluated by taking recourse to the Poisson equation (cf.~\cite{Das2015}).

Next, we briefly outline the local reformulation of the extended interactions in the kernel kinetic energy functional. The local reformulation follows the ideas put forth by~\cite{Choly2002} that the series of density independent kernels obtained from the Taylor series expansion of the WGC density dependent kernel can each be approximated in the Fourier-space using a sum of partial fractions. Using this approximation, a real-space reformulation for these extended interactions is obtained by recasting it as a variational form of complex Helmholtz equations. In particular, if $K_0(|\bx-\bx'|)$ denotes the zeroth order density independent kernel in the Taylor expansion of the density dependent kernel, and is approximated in the Fourier-space using the approximation $\hat{K}_0(|\mathbf{q}|)\approx \sum_{j=1}^{m}\frac{A_j |\mathbf{q}|^2}{|\mathbf{q}|^2+B_j}$, the local reformulation of this component of the kernel energy is given by the following saddle point problem~\citep{Das2015,Radhakrishnan2010}:
\begin{subequations}\label{eq:kernel_variational}
\begin{align}
T_{K_0}(\rho)= C_F\int\int \rho^{\alpha}(\bx) K_{0}(|\bx-\bx'|)\rho^{\beta}(\bx') \,{\rm d}\bx \,{\rm d}\bx' = \min_{\omega^0_{\alpha_j}\in \mathcal{Y}}\max_{\omega^0_{\beta_j}\in \mathcal{Y}}\, \mathcal{L}_{K_0} (\omega^0_{\alpha}, \omega^0_{\beta}, \rho)\,,
\end{align}
where
\begin{align}
\mathcal{L}_{K_0} (\omega^0_{\alpha}, \omega^0_{\beta}, \rho) =  \sum_{j=1}^{m}C_F\Big\{ \int\big[ & \frac{1}{A_jB_j}\nabla\omega^0_{\alpha_j}(\bx) \cdot\nabla\omega^0_{\beta_j}(\bx) + \frac{1}{A_j}\omega^0_{\alpha_j}(\bx)\omega^0_{\beta_j}(\bx) \notag\\ & + \omega^0_{\beta_j}(\bx)\rho^{\alpha}(\bx)+\omega^0_{\alpha_j}(\bx)\rho^{\beta}(\bx)+A_j\rho^{(\alpha+\beta)}(\bx)\big]\,{\rm d}\bx\Big\}\,.
\end{align}
\end{subequations}
In the above, ${\omega}^0_{\alpha}=\{\omega^0_{\alpha_1},\omega^0_{\alpha_2},\ldots,\omega^0_{\alpha_m}\}$ and ${\omega}^0_{\beta}=\{\omega^0_{\beta_1},\omega^0_{\beta_2},\ldots,\omega^0_{\beta_m}\}$,  where $\omega^0_{\alpha_j}$ and $\omega^0_{\beta_j}$ for $j=1\ldots m$ are the auxiliary potential fields---referred to as the \emph{kernel potentials}---introduced in the local reformulation of $T_{K_0}$. Following a similar procedure, the  local variational reformulations are constructed for the higher order kernel terms in the Taylor series expansion of the density dependent kernel. We generically denote the auxiliary potential fields resulting from each density independent kernel as $\omega^s_{\alpha_j}$ and $\omega^s_{\beta_j}$, where the index $s$ denotes the density independent kernel, and denote the vectors containing these kernel potentials by $\omega_{\alpha}$ and $\omega_{\beta}$. 

Finally, using the local variational reformulations of the extended electrostatic and kernel energies, the problem of computing the ground-state energy for a given positions of atoms is given by the following  local variational problem in the \emph{electronic fields} comprising of electron-density, electrostatic potentials, and kernel potentials (cf. \cite{Das2015,Radhakrishnan2010}):
\begin{equation}\label{eq:locVar}
E_0(\mathbf{R}) = \min_{\rho \in \mathcal{Y}} \max_{\phi \in \mathcal{Y}} \min_{\omega^s_{\alpha_j}\in \mathcal{Y}}\max_{\omega^s_{\beta_j}\in \mathcal{Y}}\,\, \mathcal{L} (\rho,\phi,\omega_{\alpha},\omega_{\beta};\mathbf{R}) \qquad \mbox{subject to}: \int \rho(\bx) \,{\rm d} \bx = N,
\end{equation} 
where $\mathcal{L}$ denotes the resulting local real space functional. Such a local reformulation in real-space is key to employing the non-homogeneous Dirichlet boundary conditions for studying the energetics of isolated defects in bulk, which is exploited in this study.

\subsection{Configurational forces}
The configurational forces derived in ~\cite{Das2015} are used for the geometry optimization in the present work. These configurational forces are computed from inner variations of the orbital-free DFT local variational problem in equation~\eqref{eq:locVar}, which are generalized forces corresponding to perturbations of the underlying space. In particular, infinitesimal perturbations of the underlying space $\psi_{\epsilon}: \Rthree \to \Rthree$ are considered corresponding to a generator $\Gamma(\bx)$ given by $\Gamma=\frac{d\psi_{\epsilon}(\bx)}{d\epsilon}|_{\epsilon=0}$ such that $\psi_{0}=I$, along with  the additional constraint of only allowing rigid body deformations of the regularized nuclear charge distribution. The configurational force is thus given by the G\^{a}teaux derivative of $E_{0}(\psi_{\epsilon})$: $\frac{d E_{0}(\psi_{\epsilon})}{d\epsilon}\Big{|}_{\epsilon=0}$. Using the configurational force expression, the force on any atom is computed by restricting the compact support of $\Gamma$ to only include the atom of interest. On the other hand, the stress tensor associated with cell relaxation are computed by restricting $\Gamma$ to affine deformations.  We refer to ~\cite{Das2015} for details of the derivation of these configurational forces, and benchmarking studies of atomic forces and cell stress tensors in Aluminum and Magnesium with respect to Kohn-Sham DFT.  

\subsection{Finite-element discretization} 
As noted previously in the introduction, plane-wave discretization based electronic structure methods are not amenable to direct studies on the energetics of isolated dislocations in bulk as the geometry of an isolated dislocation is not compatible with the restrictive periodic geometries and boundary conditions inherent in plane-wave discretizations. In order to overcome these limitations, real-space methods for the orbital-free DFT problem have been developed recently that are based on finite-element discretization~\citep{Das2015,Gavini2007c,Motamarri2012,Radhakrishnan2010} and finite difference methods~\citep{Garcia2007,Ghosh2016,Mi2016}. In the present study, we use the finite-element basis~\citep{Brenner2002} as it has many features desirable for electronic structure calculations. Notably, the finite-element basis can naturally handle complex geometries and arbitrary boundary conditions, which is crucial for studying isolated dislocations as discussed in the next section. Further, ~\citet{Motamarri2012} have demonstrated that the use of higher-order finite-element discretizations for orbital-free DFT calculations can provide significant computational savings, which has bridged the tremendous efficiency gap that previously existed with respect to the plane-wave basis. Moreover, the good scalability of finite-element discretization on massively parallel computing platforms makes it ideal for large scale electronic structure studies of dislocations. We refer to~\cite{Das2015,Motamarri2012} for a comprehensive discussion on the convergence properties of the discrete real-space orbital-free DFT problem and the numerical aspects of its solution procedure.

\section{Isolated Screw Dislocation Energetics using orbital-free DFT}\label{sec:screw}
In this section, we present our study on an isolated screw dislocation in Aluminum using the local real-space orbital-free DFT framework (RS-OFDFT) discussed in section~\ref{sec:OFDFT}. We adopt the approach proposed in~\cite{Iyer2015}, where bulk Dirichlet boundary conditions have been applied on the electronic fields to simulate an isolated edge dislocation in the bulk in Aluminum. Following along similar lines, we compute the core size of the isolated screw dislocation in Aluminum directly from energetics, by identifying the region up to which the contribution from electronic structure perturbations (beyond those that can be accounted for in a nonlinear continuum theory) is significant to the energetics. The dislocation energy corresponding to this core size, computed from the proposed electronic structure calculations, is identified as the core energy. Further, we study the influence of external macroscopic deformations on the dislocation core energy and core structure for a wide range of macroscopic deformations. The RS-OFDFT calculations are conducted using the following choices: Wang-Govind-Carter (WGC) model for the kinetic energy functional~\citep{Wang1999} (second order Taylor expansion of the density dependent kernel, cf. ~\cite{Wang1999}), a local density approximation (LDA) for the exchange-correlation energy~\citep{Perdew1981}, and the Goodwin-Needs-Heine pseudopotential~\citep{Goodwin1990}. For the finite-element discretization, we use quadratic hexahedral elements, where the basis functions are constructed as a tensor product of basis functions in one dimension. Numerical parameters like the finite-element mesh size, quadrature rules and stopping tolerances for iterative solvers are chosen such that the error in the computed dislocation energies per unit length of the dislocation line is less than 0.001 eV/\AA. Atomic relaxations are performed till the force components in all directions on the atoms are less than 2.5\e{-3} eV/\AA.

\subsection{Dislocation core size and core energy}\label{sec:coreEnergy}
\begin{figure}[htbp]
\centering
\includegraphics[width=0.5\textwidth]{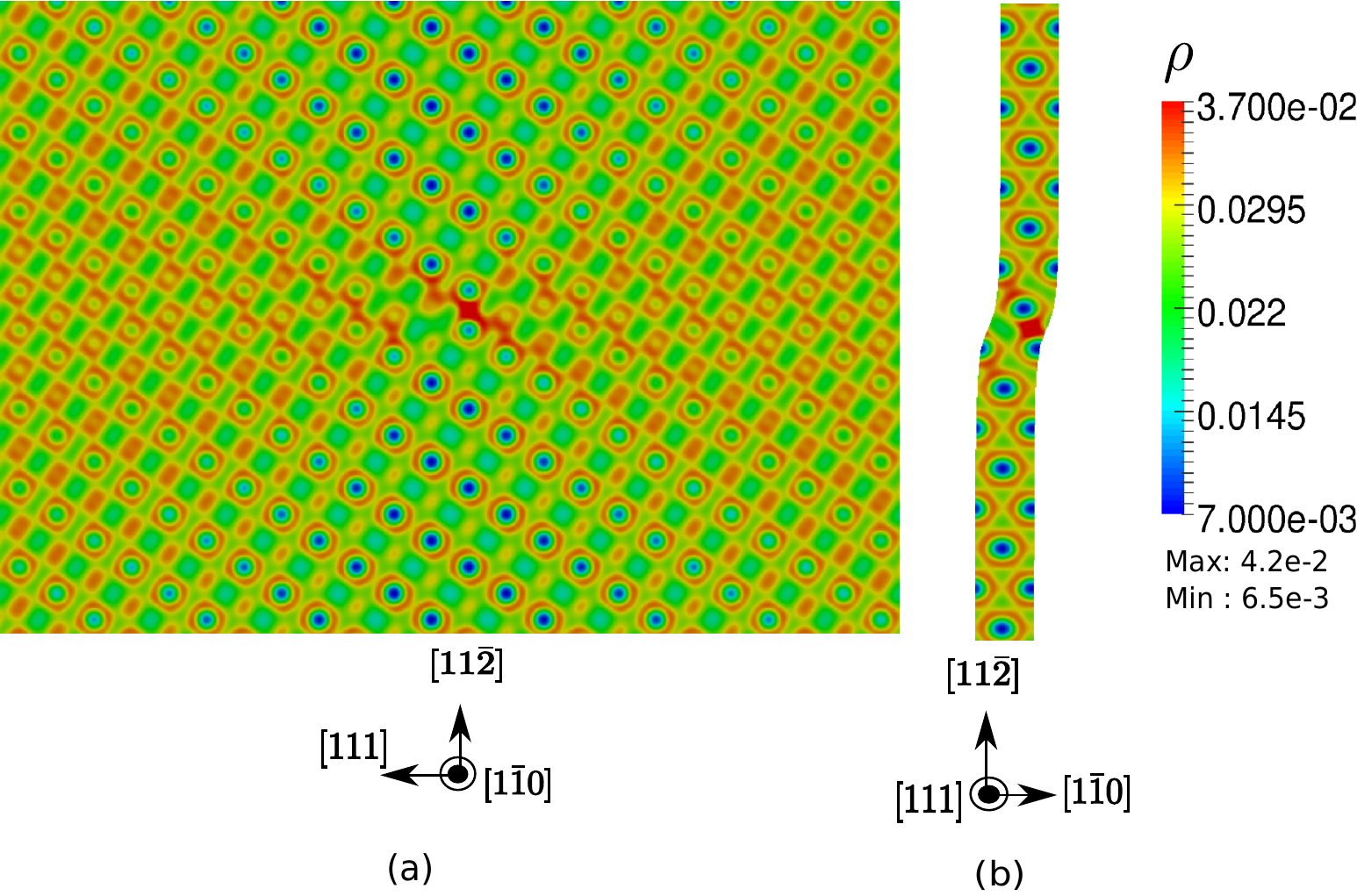}
\caption{\label{fig:contour}\small{Electron density contours on a) a ($1$$\bar{1}$0) plane, and b) a (111) plane of a perfect screw dislocation in Aluminum. The (111) plane passes through the dislocation center.}}
\end{figure}

\begin{figure}[htbp]
\centering
\includegraphics[width=0.5\textwidth]{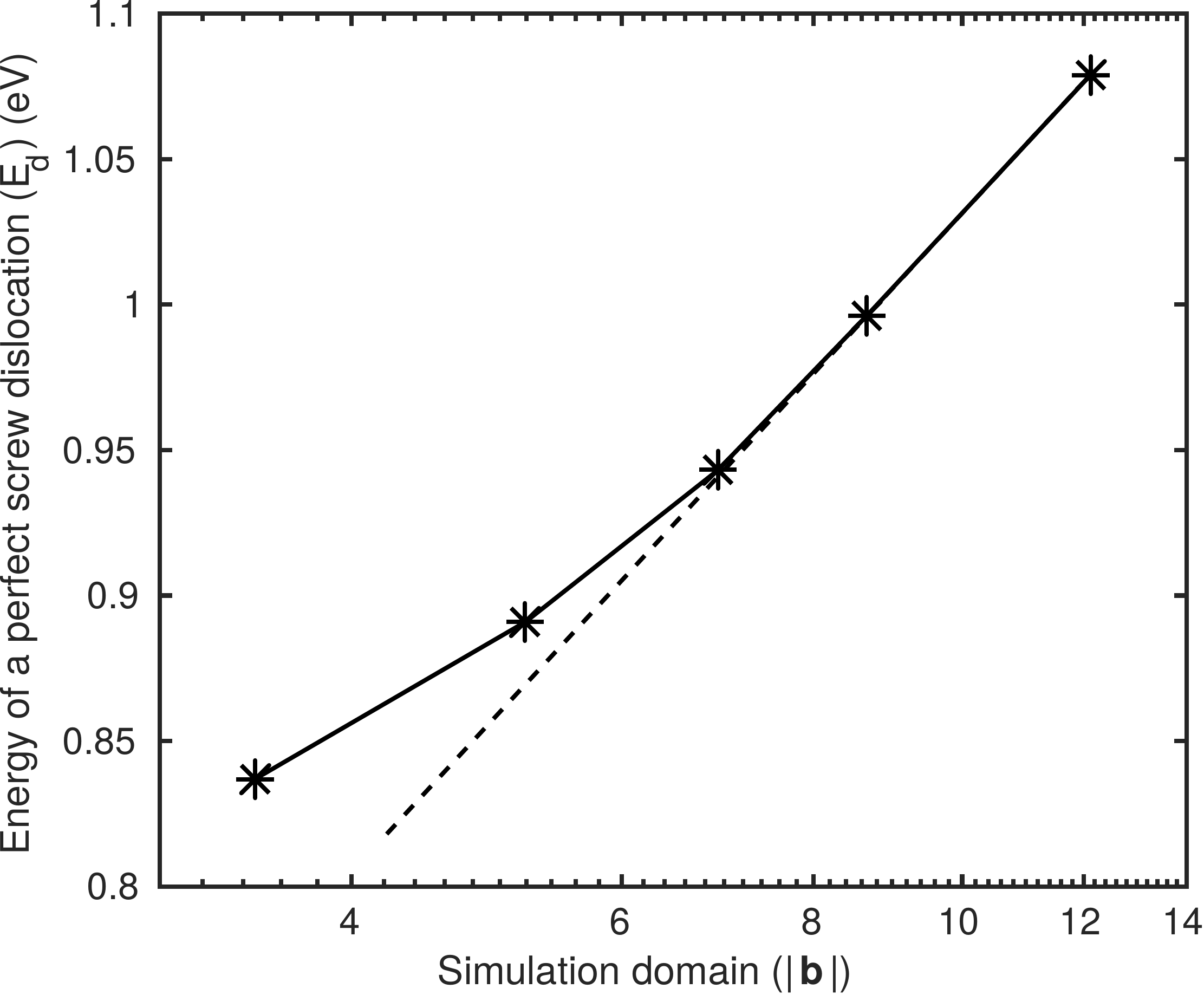}
\caption{\label{fig:domain}\small{Semi-log plot of dislocation formation energy of the screw dislocation as a function of simulation domain size. The dashed line demonstrates the asymptotic logarithmic divergence of the computed dislocation energies, as expected from continuum estimates, beyond simulation domains of $\approx$ 7$|{\bf b}|$.}}
\end{figure}

We begin by estimating the core size of a perfect screw dislocation in face-centered-cubic (fcc) Aluminum explicitly from the energetics, and subsequently calculate the core energy for the perfect screw dislocation as well as the core energy after atomic relaxation. The coordinate system, X --- Y --- Z axes (or equivalently 1---2---3), is aligned along [$1$ $1$ $\bar{2}$]---[$1$ $1$ $1$]---[$1$ $\bar{1}$  0] crystallographic directions. With this coordinate system, we create a perfect fcc crystal of size $R \sqrt{6}a_0 \times 2R\sqrt{3}a_0 \times \frac{a_0}{\sqrt{2}}$, where $a_0$ denotes the lattice parameter and $R$ is an integer-valued scaling factor which sets the simulation domain size. Then we introduce a perfect screw dislocation with Burgers vector ${\bf b} = \frac{a_0}{2} [ 1 \bar{1}  0]$ and line direction along [$1$ $\bar{1}$  0] at the center of the simulation domain by applying isotropic Volterra displacement fields~\citep{Hirth1982} of a screw dislocation to the positions of atoms. We employ the bulk Dirichlet boundary conditions on electronic fields, proposed in ~\cite{Iyer2015}, to simulate the isolated screw dislocation in bulk. More specifically, in the X and Y directions, we employ Dirichlet boundary conditions on the electronic fields comprising of electron density, electrostatic potential and kernel potentials, and, in the Z direction, we use periodic boundary conditions on these fields. The Dirichlet boundary values for the electronic fields are determined using the Cauchy-Born approximation, wherein the values are obtained by projection of orbital-free DFT computed electronic fields on periodic unit cells which are deformed using the elastic field of the screw dislocation. These boundary conditions on the electronic fields correspond to an isolated dislocation in bulk with the electronic structure on the boundary of (and outside of) the simulation domain given by the Cauchy-Born hypothesis. The local real space formulation along with the finite-element discretization are crucial to realizing these bulk Dirichlet boundary conditions. Using these boundary conditions, we compute the electronic-structure and ground-state energy of the perfect screw dislocation for varying simulation domains with $R=$ 2, 3, 4, 5, 7 while keeping the atomic positions fixed. This enables us to unambiguously delineate the contribution from electronic-structure perturbations to the dislocation energy, the origins of which are quantum mechanical in nature and are beyond the scope of any homogenized non-linear continuum theory, and also identify the region where this contribution is significant. Figure~\ref{fig:contour} shows the contours of the electron density for $R=4$. For each of these simulation domains, the dislocation energy ($E_{\rm d}$) is computed as
\begin{equation}
E_{\rm d}(N,V)=E_{\rm disloc}(N,V)-E_0(N,V),\label{eq:formationEng}
\end{equation}
where $E_{\rm disloc}(N,V)$ denotes the energy of a dislocation system containing N atoms with a volume V, and $E_0(N,V)$ denotes energy of a perfect crystal of the same volume and containing the same number of atoms. From a thermodynamic standpoint, $E_{\rm d}$ is the dislocation formation energy at constant volume. We note that there is no volume change due to the perfect screw dislocation Volterra field. Table~\ref{tab:table1} shows the computed dislocation energies for the various domains.  We note that the domain-size is measured to be $\sqrt{3}R|{\bf b}|$, which is the distance from dislocation line to the boundary along [$1$ $1$ $\bar{2}$]. In order to identify the region in which the electronic-structure perturbations arising from the dislocation are significant, we consider the dislocation energy change for every successive increase in the domain size, and denote this change by $\Delta E_{\rm d}$. This has two contributions: (i) the elastic energy of the region between the two domains, which we denote by $\Delta E_{\rm d}^{\rm elas}$ ; (ii) contribution from electronic-structure changes due to the change in the location of the bulk Dirichlet boundary conditions on electronic fields, which we denote by $\Delta E_{\rm d}^{\rm elec}$. $\Delta E_{\rm d}^{\rm elas}$ is computed using the Cauchy-Born approximation, wherein the elastic energy density at each point is computed from RS-OFDFT calculations on periodic unit-cells deformed by screw dislocation elastic fields. We note that $\Delta E_{\rm d}^{\rm elas}$, thus computed, corresponds to the elastic contributions from the non-linear continuum elastic theory derived from OFDFT. We subsequently infer $\Delta E_{\rm d}^{\rm elec}$, by subtracting $\Delta E_{\rm d}^{\rm elas}$ from $\Delta E_{\rm d}$. The computed $\Delta E_{\rm d}^{\rm elas}$ and $\Delta E_{\rm d}^{\rm elec}$ are reported in Table~\ref{tab:table1}, where the computed values are accurate up to 0.001 eV. We find that $\Delta E_{\rm d}^{\rm elec}$ remains significant in comparison to $\Delta E_{\rm d}^{\rm elas}$, i.e. more than 10 \% of $\Delta E_{\rm d}^{\rm elas}$, until a domain-size of $\approx$ 7$|{\bf b}|$. This suggests that the electronic-structure perturbations due to the dislocation are significant up to $\approx$ 7$|{\bf b}|$ from the dislocation line, which represents the core size of the perfect screw dislocation. Graphically, this is also evident from figure~\ref{fig:domain}, with the energy of the dislocation deviating from the logarithmic dependence for domain sizes below 7$|{\bf b}|$. Importantly, we note that this domain size is much larger than previous core size estimates based on continuum displacement or strain fields of ˆ$\sim1-3|{\bf b}|$~\citep{Banerjee2007,Hirth1982,Peierls1940}, underscoring the longer ranged nature of electronic structure perturbations from defects than previously believed, and its potential significance in governing the energetics of dislocations. The present finding is consistent with a recent electronic structure study on edge dislocation in Aluminum by ~\cite{Iyer2015}, wherein the core size was computed to be 10$|{\bf b}|$. We note that real-space studies on point defects have also shown strong cell-size effects owing to the electronic structure perturbations from the defect~\citep{Das2015, QC2007, GBO2007, Radhakrishnan2010, Radhakrishnan2016}. 


%
%

\begin{table}
\centering
\caption{\label{tab:table1}\small{Computed dislocation energy of perfect screw dislocation in Aluminum for varying domain-sizes, where N denotes the number of atoms in the simulation domain. $\Delta E_{\rm d}$ denotes the change in the dislocation energy from the previous domain-size. $\Delta E_{\rm d}^{\rm elas}$ } and $\Delta E_{\rm d}^{\rm elec}$ denote the elastic and electronic contributions to $\Delta E_{\rm d}$.}
\begin{tabular}{|c|c|c|c|c|c|c|}
\hline
Domain & N & $E_{\rm d}$  & $\Delta E_{\rm d}$ & $\Delta E_{\rm d}^{\rm elas}$
 &  $\Delta E_{\rm d}^{\rm elec}$  \\ 
size ($R \sqrt{3} |{\bf b}|$) & (atoms) & (eV) &  (eV) & (eV)
 &  (eV) \\
\hline
3.5$|{\bf b}|$ & 96 & 0.837  & -  & - & - \\

5.2$|{\bf b}|$ & 216 & 0.891 & 0.054 & 0.100 & -0.046 \\

6.9$|{\bf b}|$ & 384 &  0.943 & 0.052 & 0.069 & -0.017 \\

8.7$|{\bf b}|$ & 600 & 0.996 & 0.053 & 0.051 & 0.002 \\

12.1$|{\bf b}|$ & 1176 &  1.079 & 0.083 & 0.080 & 0.003 \\
\hline
\end{tabular}
\end{table}

\begin{figure}[htbp]
\centering
\subfigure[]{\includegraphics[width=0.5\textwidth]{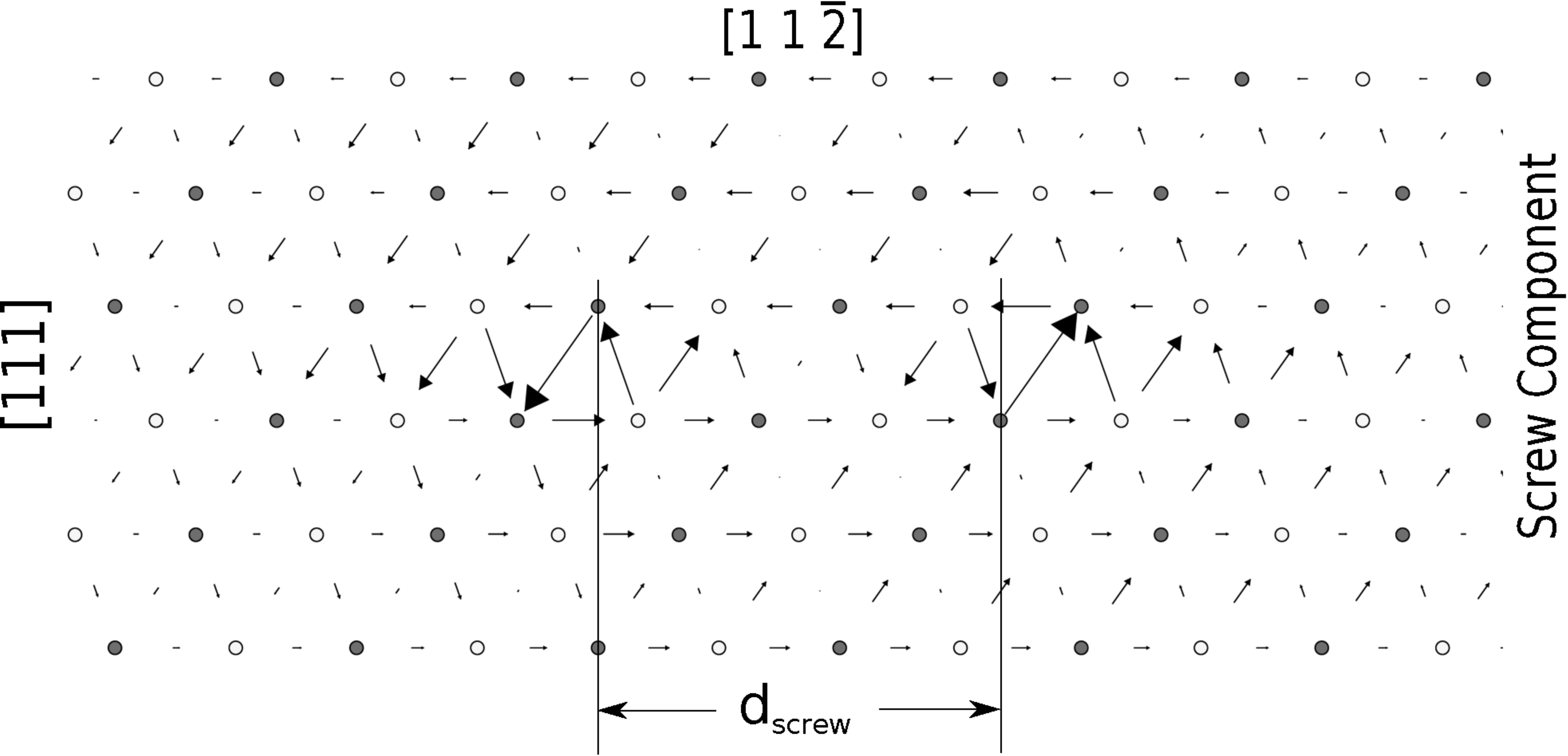}}\\
\subfigure[]{\includegraphics[width=0.5\textwidth]{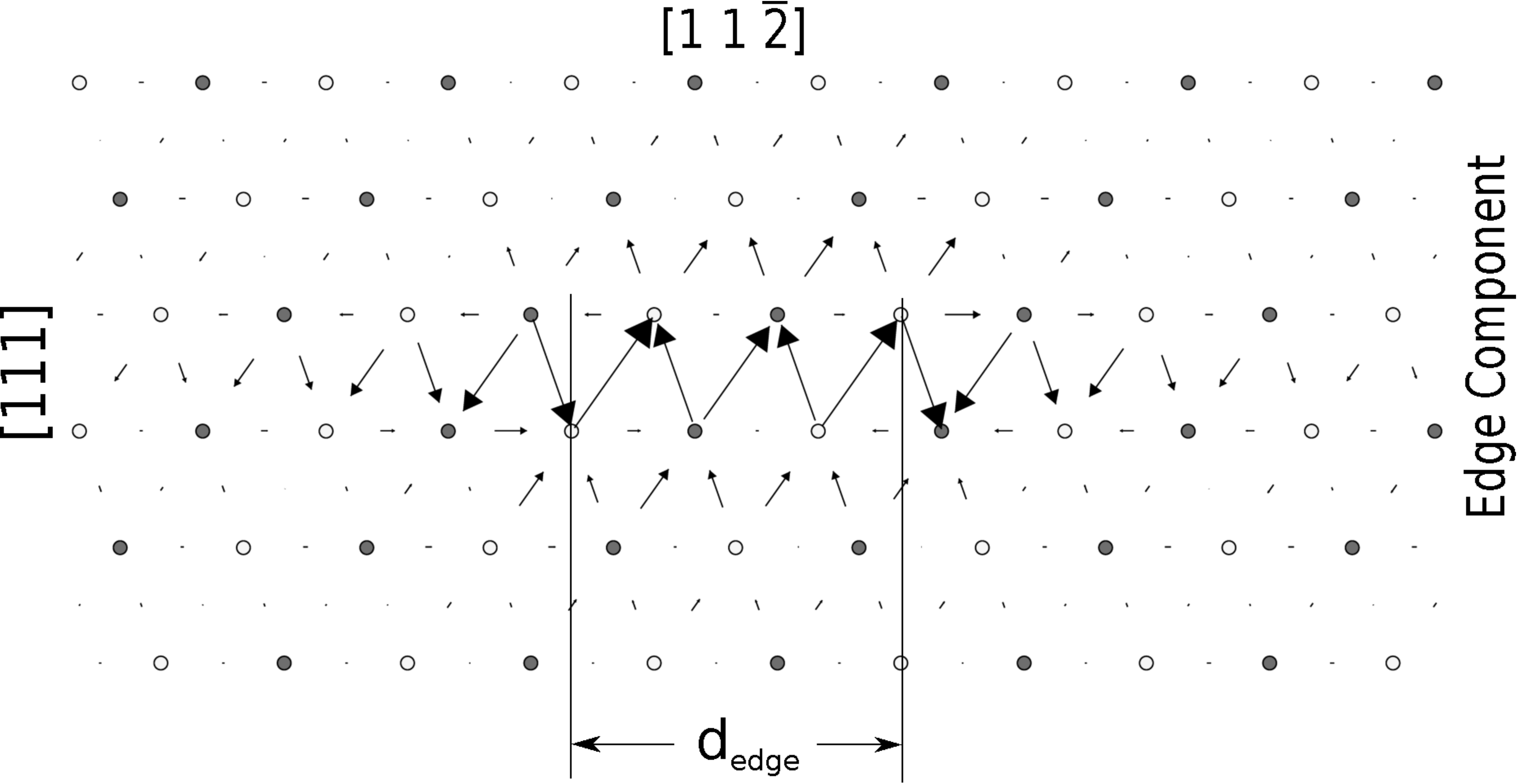}}
\caption{\label{fig:dd1}\small{Differential displacement plots of the a) screw and b) edge components of Shockley partials of a screw dislocation.}}
\end{figure}

Next we investigate the relaxed core structure of the screw dislocation by relaxing the positions of atoms interior to the simulation domain, while holding fixed the positions of atoms on the Dirichlet boundary. We find that the relaxed structure and the corresponding reduction in the energy from the perfect screw core energy (denoted by $E_{\rm d}^{\rm relax}$) are sensitive to the simulation domain size up to the domain size of 8.7$|{\bf b}|$. Beyond 8.7$|{\bf b}|$, the change in the core structure is negligible and the change in $E_{\rm d}^{\rm relax}$ is within the tolerance 0.001 eV/\AA. This suggests that electronic structure perturbations are not significant beyond 8.7$|{\bf b}|$ for the relaxed screw dislocation representing Shockley partials. Thus, we consider 8.7$|{\bf b}|$ to be the core size of the relaxed screw dislocation, and the dislocation energy corresponding to this core size as the dislocation core energy of the Shockley partials. The core energy of Shockley partials is computed to be 0.811 eV, or, equivalently, the core energy per unit length of dislocation line is 0.284 eV/\AA, and $E_{\rm d}^{\rm relax}$ is 0.065 eV/\AA. Figure~\ref{fig:dd1} shows the edge and screw components of the differential displacements~\citep{Vitek1970} and indicates the approximate location of the Shockley partials. The partial separation distance computed from the edge-component differential displacement (DD) plot is 6.59 \AA~(2.3 $|{\bf b}|$), and that computed from the screw-component differential displacement plot is 8.24 \AA~(2.9 $|{\bf b}|$). An uncertainty magnitude equal to twice the spacing between atomic planes in the X direction, which is $\frac{a_0}{\sqrt{6}}=1.65$ \AA~ (or 0.58 $|{\bf b}|$) is expected in computing the partial separation using this procedure. Comparing the partial separation distance with other DFT studies on a screw dislocation in Aluminum~\citep{Shin2009,Woodward2008}, we are in good agreement with these studies that have reported partial separation distances between 5.0--7.5 \AA. 

\subsection{Effect of macroscopic deformations}\label{sec:macroDeform}

This study is motivated from recent investigations which suggest that macroscopic deformations can play a significant role in governing the electronic structure and subsequently the energetics of defects~\citep{Gavini2008,Gavini2009,Iyer2014,Pizzagalli2009}. In the recent study on an edge dislocation in Aluminum~\citep{Das2016,Iyer2015}, it was observed that the core energy of the edge dislocation strongly depends on the external macroscopic deformation. In the present work, we perform a similar investigation on a screw dislocation in Aluminum to understand the effect of macroscopic deformation on its core energy and core structure. We begin with a perfect screw dislocation in a 8.7$|{\bf b}|$ simulation domain, which corresponds to the core size of the Shockley partials determined in Section~\ref{sec:coreEnergy}. On that simulation domain, we apply an affine deformation corresponding to a macroscopic strain ${\boldsymbol \epsilon}$, and, while holding the positions of the atoms fixed, we compute the electronic structure and the relaxed positions of atoms. The dislocation core energy, following equation~\eqref{eq:formationEng}, is computed as a function of macroscopic strain 
\begin{equation}
E_{\rm c}({\boldsymbol \epsilon})=E_{\rm disloc}({\boldsymbol \epsilon})-E_0({\boldsymbol \epsilon}), \label{eq:formationEngAffine}
\end{equation}
where $E_{\rm disloc}({\boldsymbol \epsilon})$ denotes the ground-state energy of the 8.7$|{\bf b}|$ simulation domain containing the dislocation under an affine deformation corresponding to macroscopic strain $\boldsymbol \epsilon$, and $E_0({\boldsymbol \epsilon})$ denotes the energy of a perfect crystal under the same affine deformation, containing the same number of atoms and occupying the same volume.

\begin{figure}[htbp]
\centering
\includegraphics[width=0.45\textwidth]{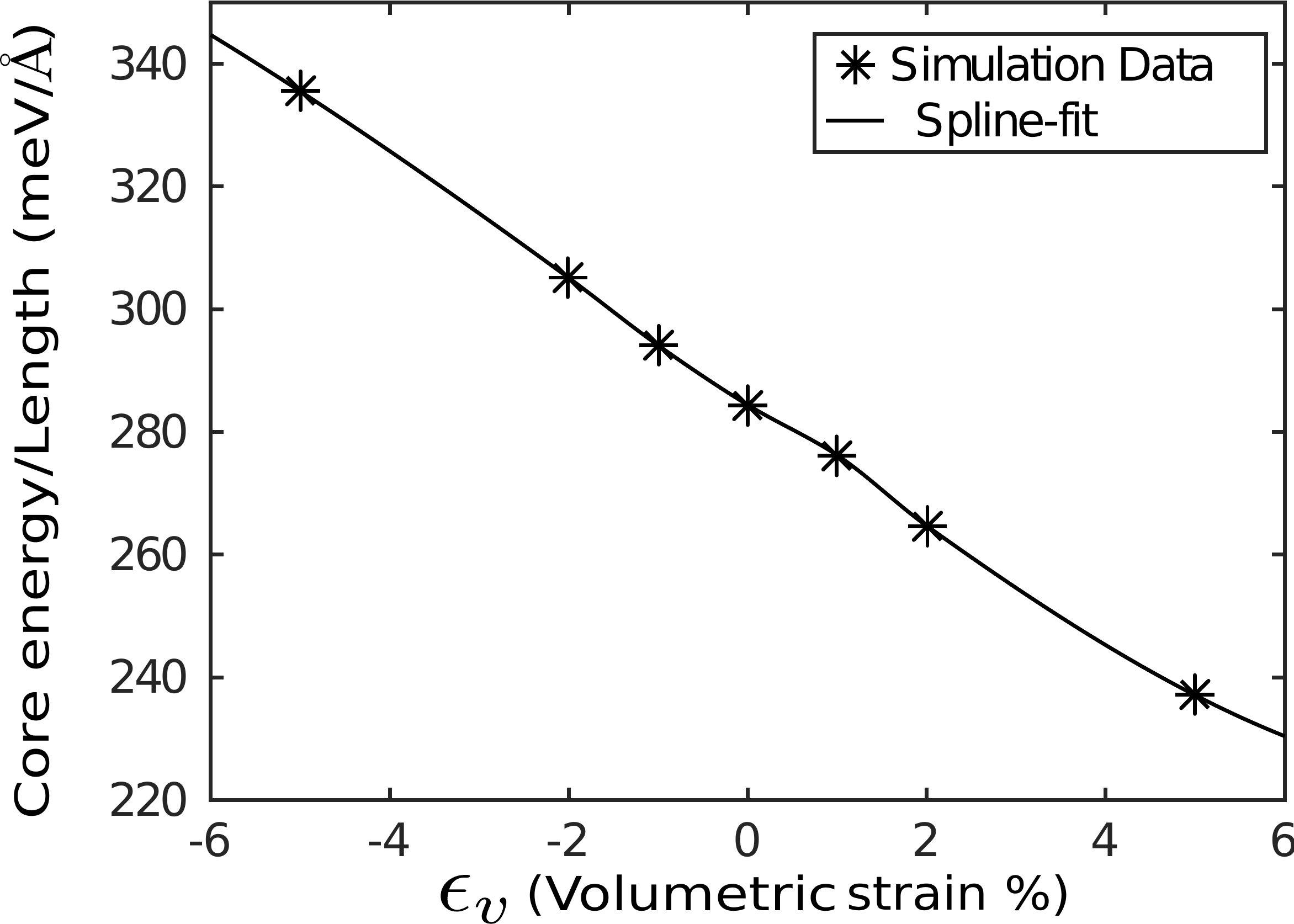}
\caption{\label{fig:coreEngV}\small{Core-energy per unit length of dislocation line of relaxed Shockley partials as a function of volumetric strain.}}
\end{figure}

We begin by studying the effect of macroscopic volumetric strain $\epsilon_{v}$, corresponding to equi-triaxial strain, on the core energy and the core structure of screw dislocation Shockley partials. In this study, we consider volumetric strains of -5\%, -2\%, -1\%, 1\%, 2\% and 5\%. Figure~\ref{fig:coreEngV} shows the core energy (per unit length of dislocation line) for the different volumetric strains. The core energy changed monotonically and almost linearly from 0.34 eV/\AA~at -5\% volumetric strain to 0.24 eV/\AA~at 5\% volumetric strain, showing a strong dependence on volumetric strain, which is in sharp contrast to widely used continuum based dislocation models where the dislocation core energy is assumed to be independent of macroscopic deformation. However, for the range of volumetric strains considered, the core structure only changed marginally. In particular, the partial separation distance in the edge-component DD plots is found to be 2.3--2.6 $|{\bf b}|$, and the partial separation distance in the screw-component DD plot remained unchanged at 2.9 $|{\bf b}|$. We note that $|{\bf b}|$ used here to quantify the partial separation distance is computed with respect to the fcc lattice under the applied macroscopic strain.

\begin{figure}[htbp]
\centering
\subfigure[]{\includegraphics[width=0.45\textwidth]{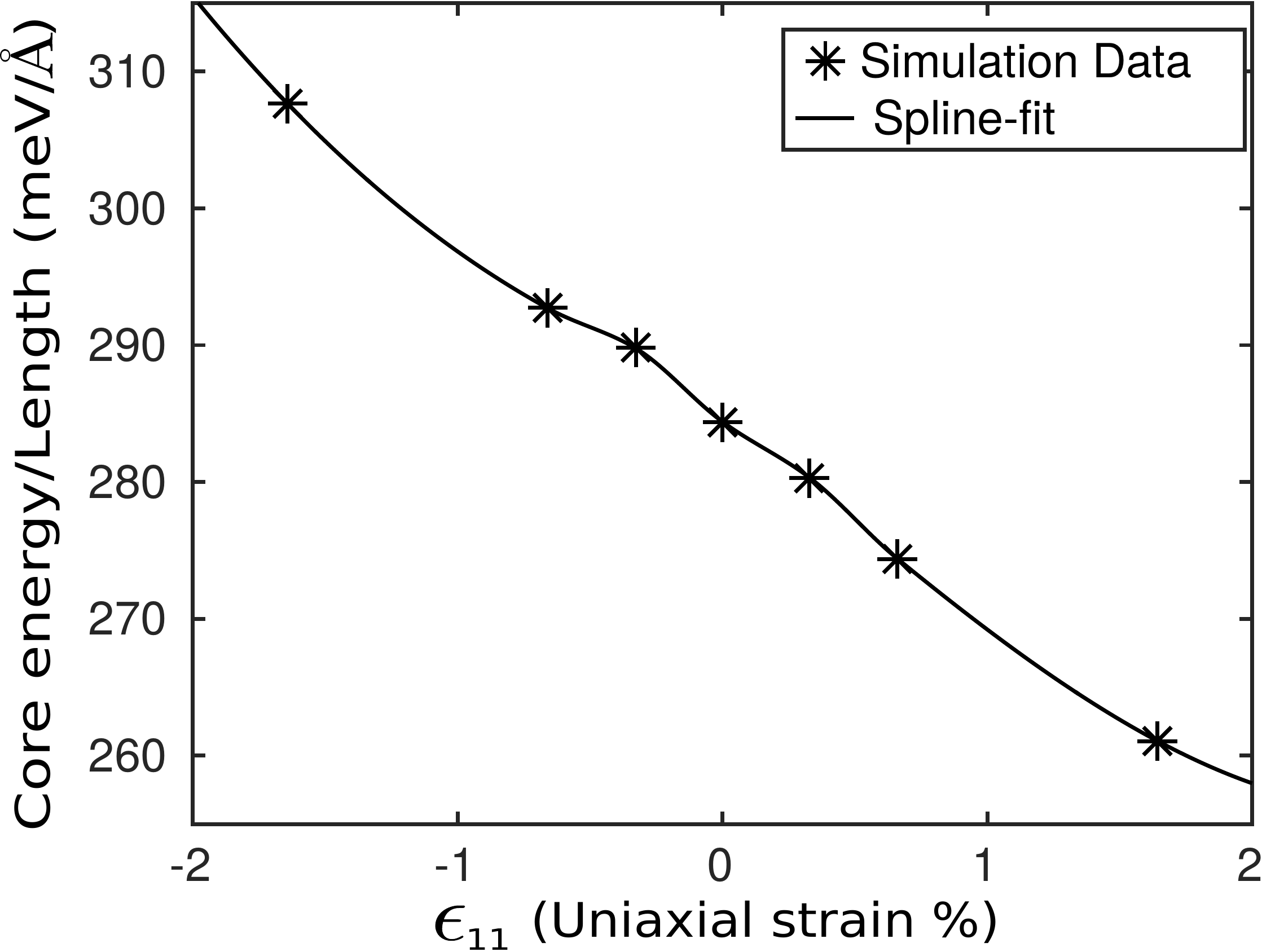}\label{fig:coreEnergy11}}
\subfigure[]{\includegraphics[width=0.45\textwidth]{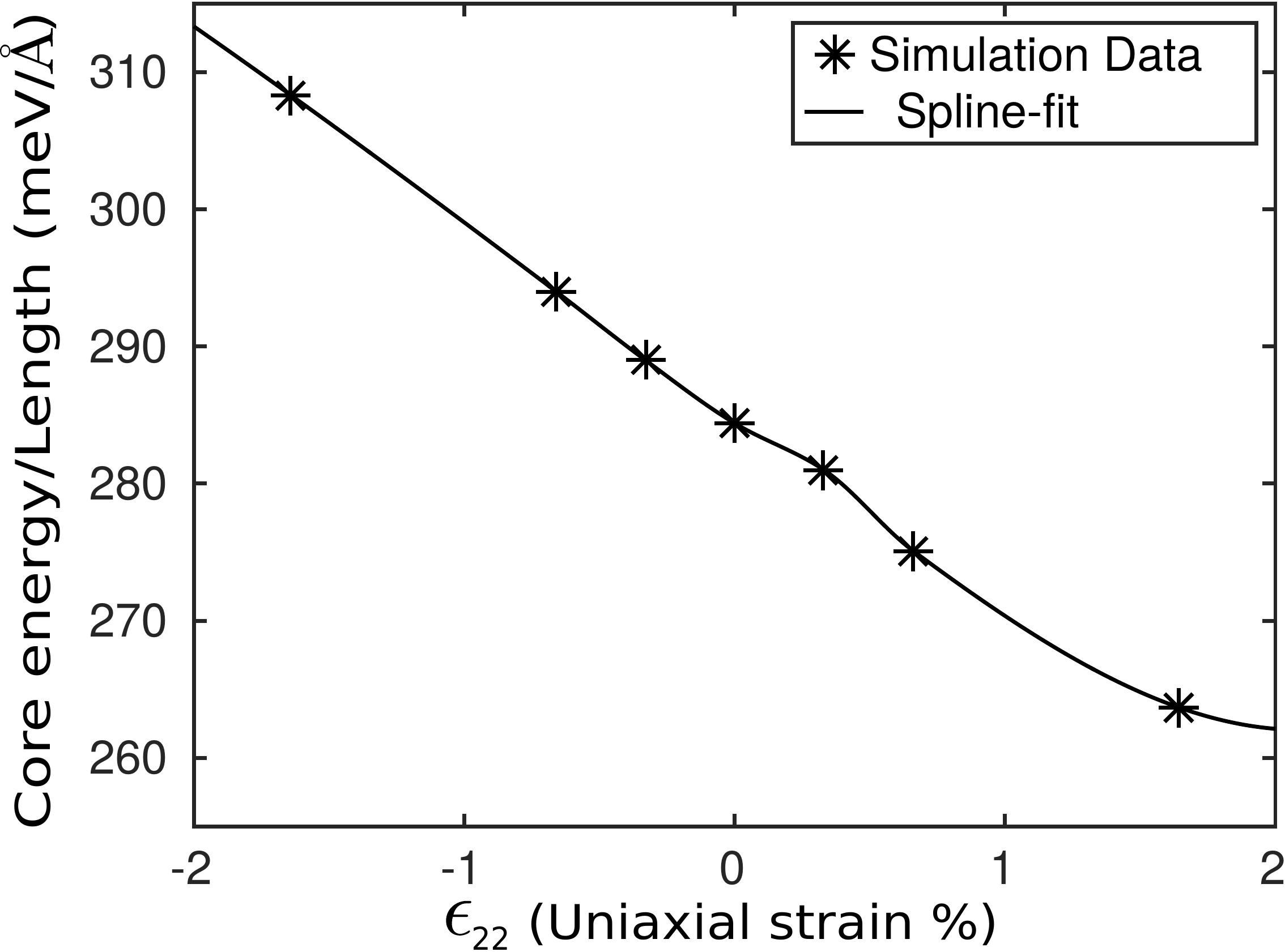}\label{fig:coreEnergy22}}\\
\subfigure[]{\includegraphics[width=0.45\textwidth]{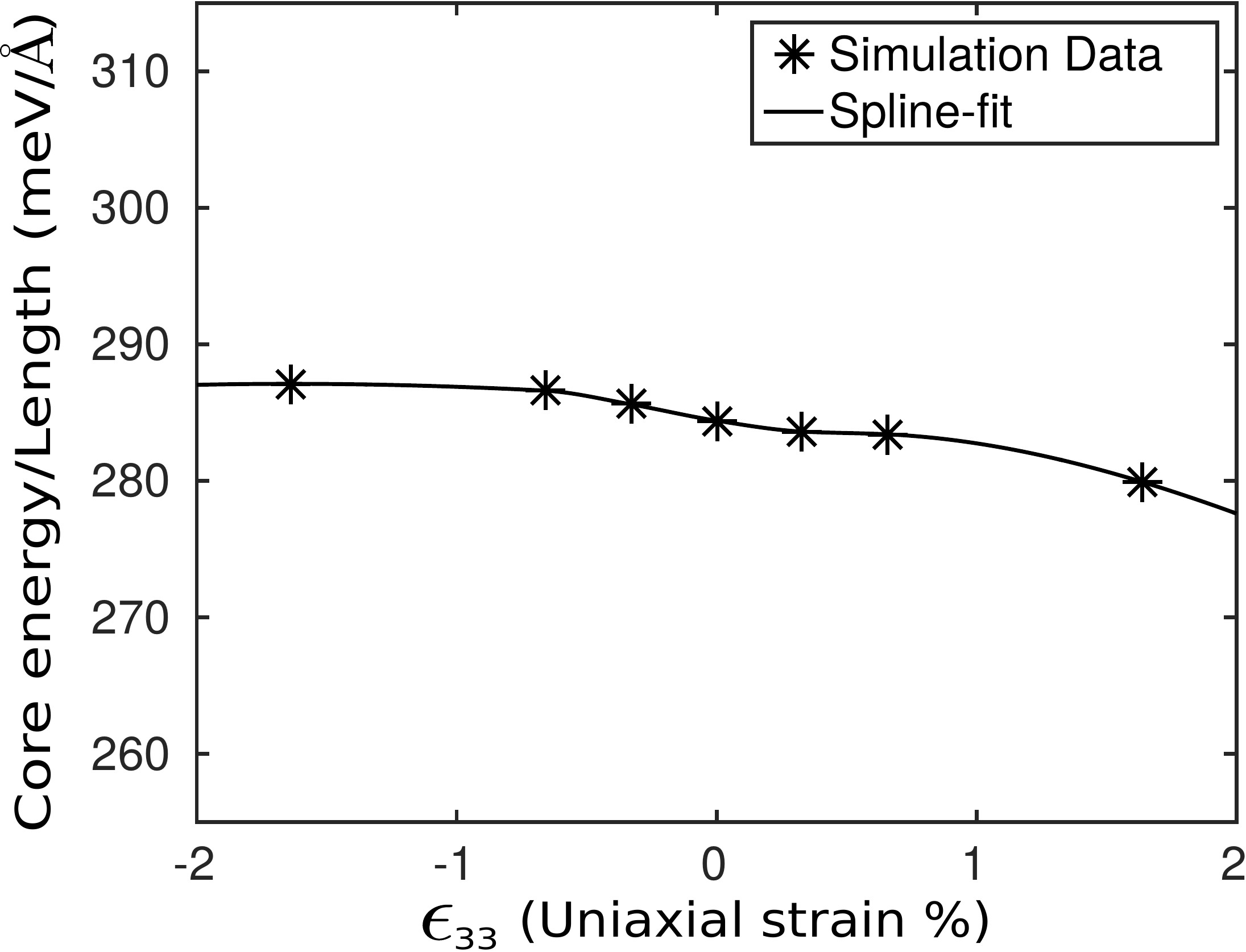}\label{fig:coreEnergy33}}
\caption{\label{fig:coreEnergyUniax}\small{Core-energy per unit length of dislocation line of relaxed Shockley partials as a function of uniaxial strains: (a) $\epsilon_{11}$; (b) $\epsilon_{22}$; (c) $\epsilon_{33}$.}}
\end{figure}

We next study the influence of macroscopic uniaxial strains along the coordinate directions, [$1$ $1$ $\bar{2}$]---[$1$ $1$ $1$]---[$1$ $\bar{1}$  0]. For each of the coordinate directions we consider uniaxial strain values of -1.64\%,-0.66\%,-0.33\%, 0.33\%, 0.66\% and 1.64\%. Figures~\ref{fig:coreEnergy11},~\ref{fig:coreEnergy22}, and~\ref{fig:coreEnergy33} show the core energy dependence on $\epsilon_{11}$ (uniaxial strain along [$1$ $1$ $\bar{2}$]), $\epsilon_{22}$ (uniaxial strain along [$1$ $1$ $1$]) and $\epsilon_{33}$ (uniaxial strain along [$1$ $\bar{1}$  0]), respectively.  Similar to volumetric strain, the core energy dependence on uniaxial strains, for the range of strains considered in this study, is found to be monotonically decreasing from compressive to tensile strains. However, the core energy dependence on $\epsilon_{33}$ is significantly weaker compared to the other two uniaxial strains. Interestingly, the monotonic and almost linear dependence of the screw dislocation core energy on uniaxial strains is in contrast to the edge dislocation results~\citep{Das2016,Iyer2015}, where the core energy dependency on uniaxial strains was found to be non-monotonic and non-linear. This suggests that the dislocation character can play an important role in influencing the dependence of core energies on macroscopic deformation. In contrast to the effect of volumetric strains, we find that $\epsilon_{22}$ and $\epsilon_{33}$ uniaxial strains have a more significant influence on the core structure of a screw dislocation, while the $\epsilon_{11}$ uniaxial strain has a smaller influence compared to $\epsilon_{22}$ and $\epsilon_{33}$ strains. For the range of uniaxial strains considered in this study, the partial separation distance in the edge-component DD plots varies monotonically in going from compressive to tensile strains. The ranges of the variation for $\epsilon_{11}$, $\epsilon_{22}$ and  $\epsilon_{33}$ strains are 2.3--2.1$|{\bf b}|$, 2.6--1.7$|{\bf b}|$, and  2.1--2.6 $|{\bf b}|$ respectively. The screw component partial separation distance remained unchanged for all uniaxial strains. These changes in the core structure cannot be rationalized using the linear elastic theory as these uniaxial strains  do not result in any glide forces on the Shockley partials, thus underscoring the role of electronic structure in governing the core structure and energetics.

\begin{figure}[htbp]
\centering
\subfigure[]{\includegraphics[width=0.47\textwidth]{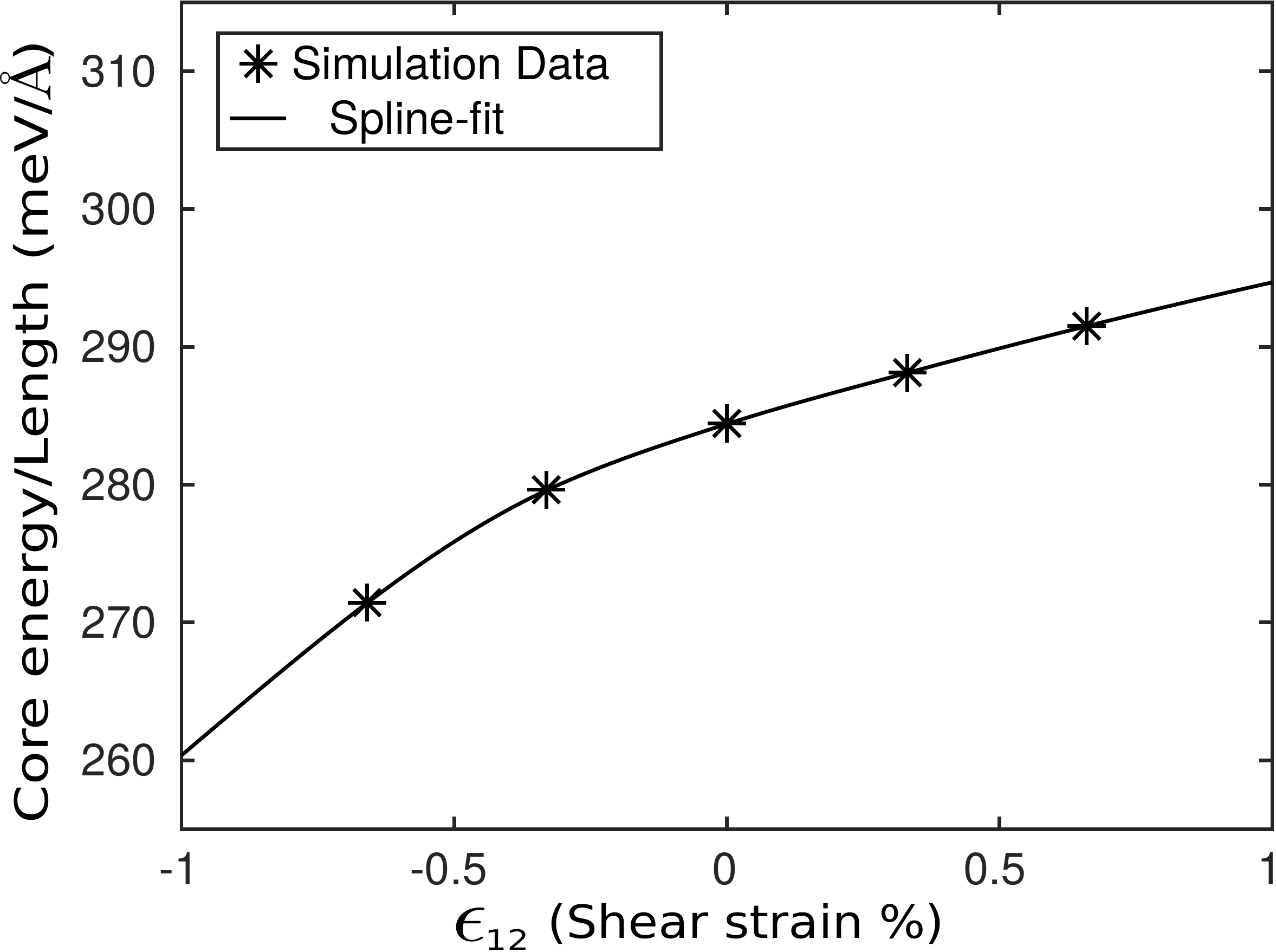}\label{fig:coreEnergy12}}
\subfigure[]{\includegraphics[width=0.45\textwidth]{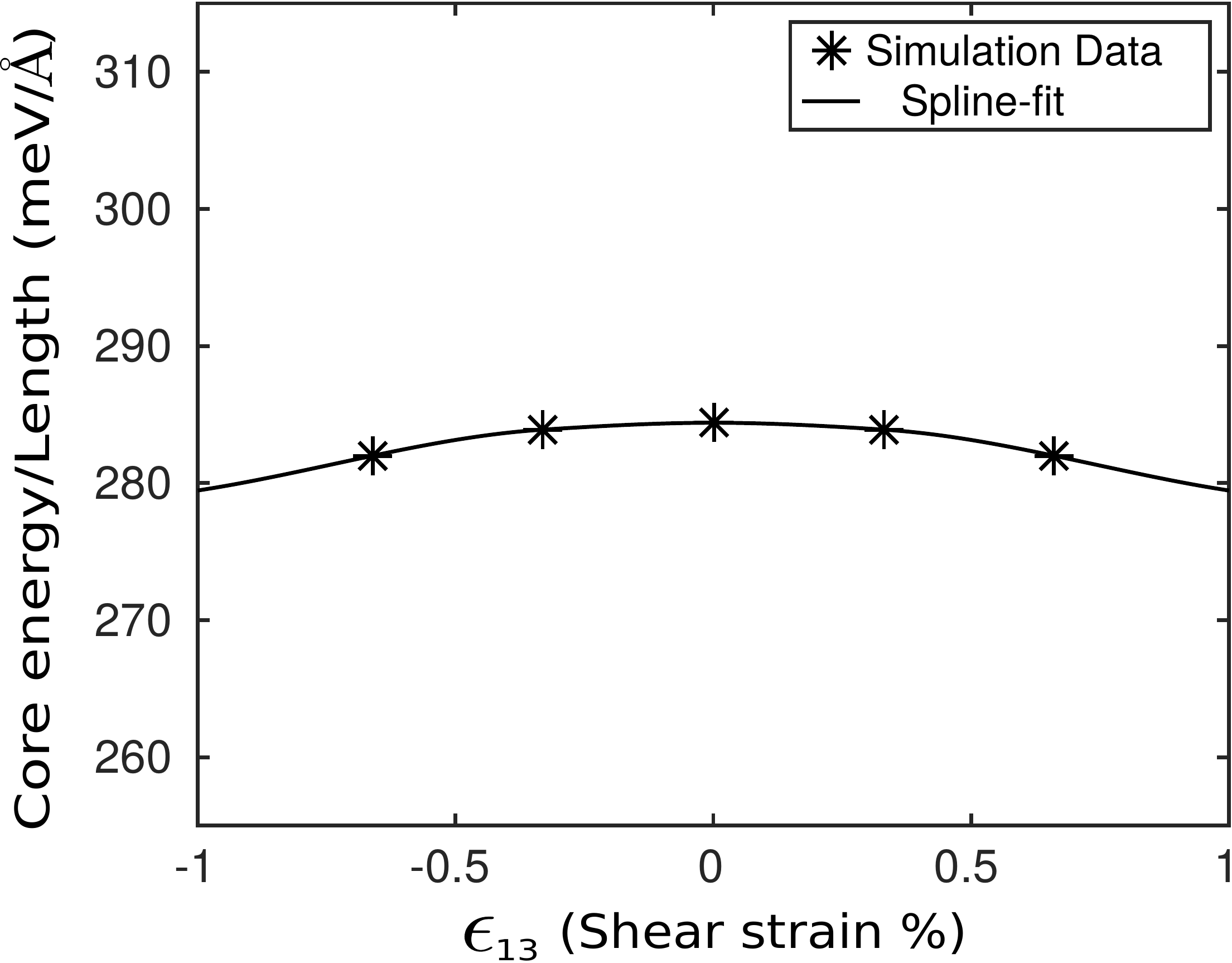}\label{fig:coreEnergy13}}
\caption{\label{fig:coreEnergyShear}\small{Core-energy per unit length of dislocation line of relaxed Shockley partials as a function of shear strains: (a) $\epsilon_{12}$; (b) $\epsilon_{13}$.}}
\end{figure}

\begin{figure}[htbp]
\centering
\subfigure[]{\includegraphics[width=0.41\textwidth]{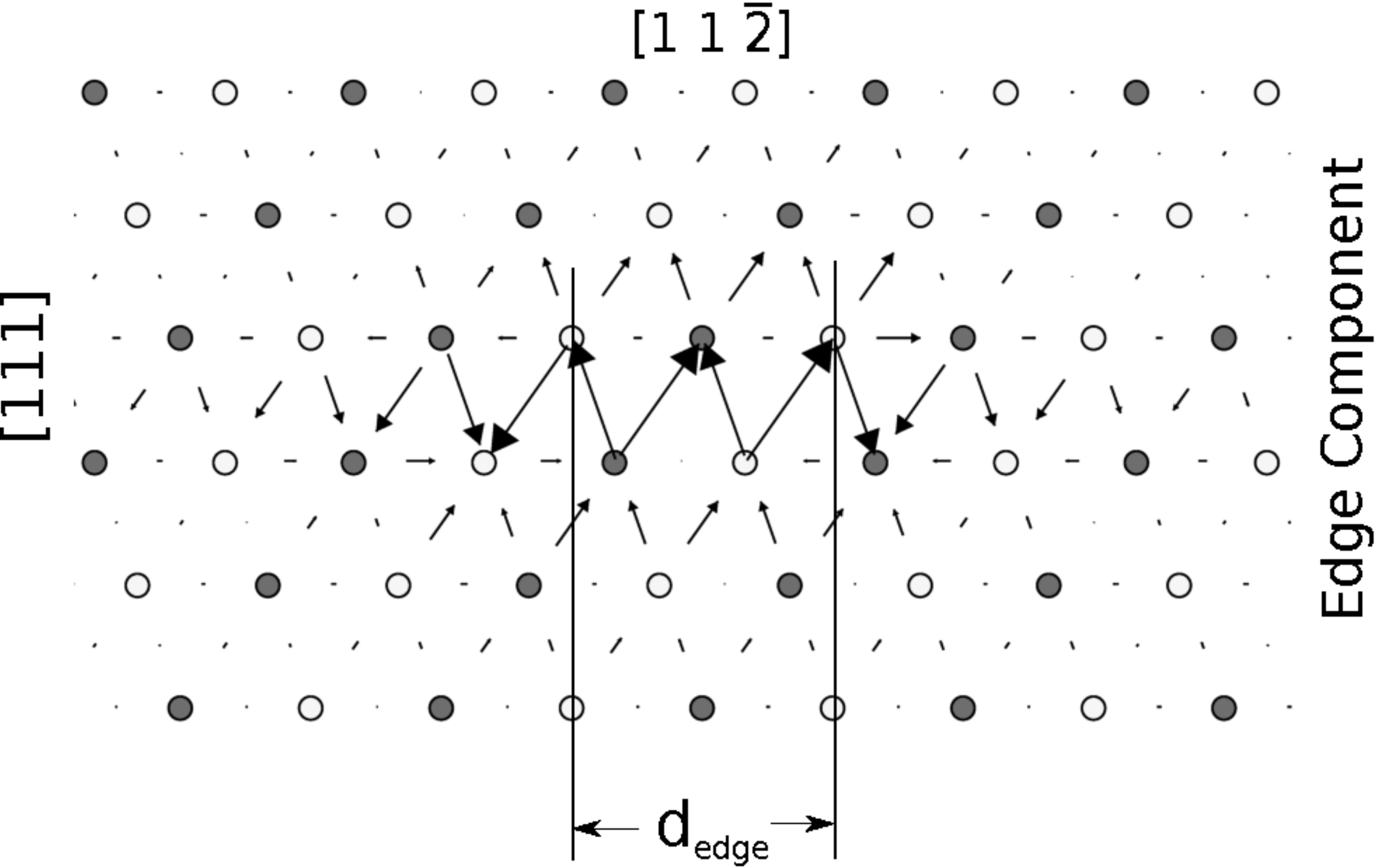}\label{fig:dd12EdgeP}}\\
\subfigure[]{\includegraphics[width=0.45\textwidth]{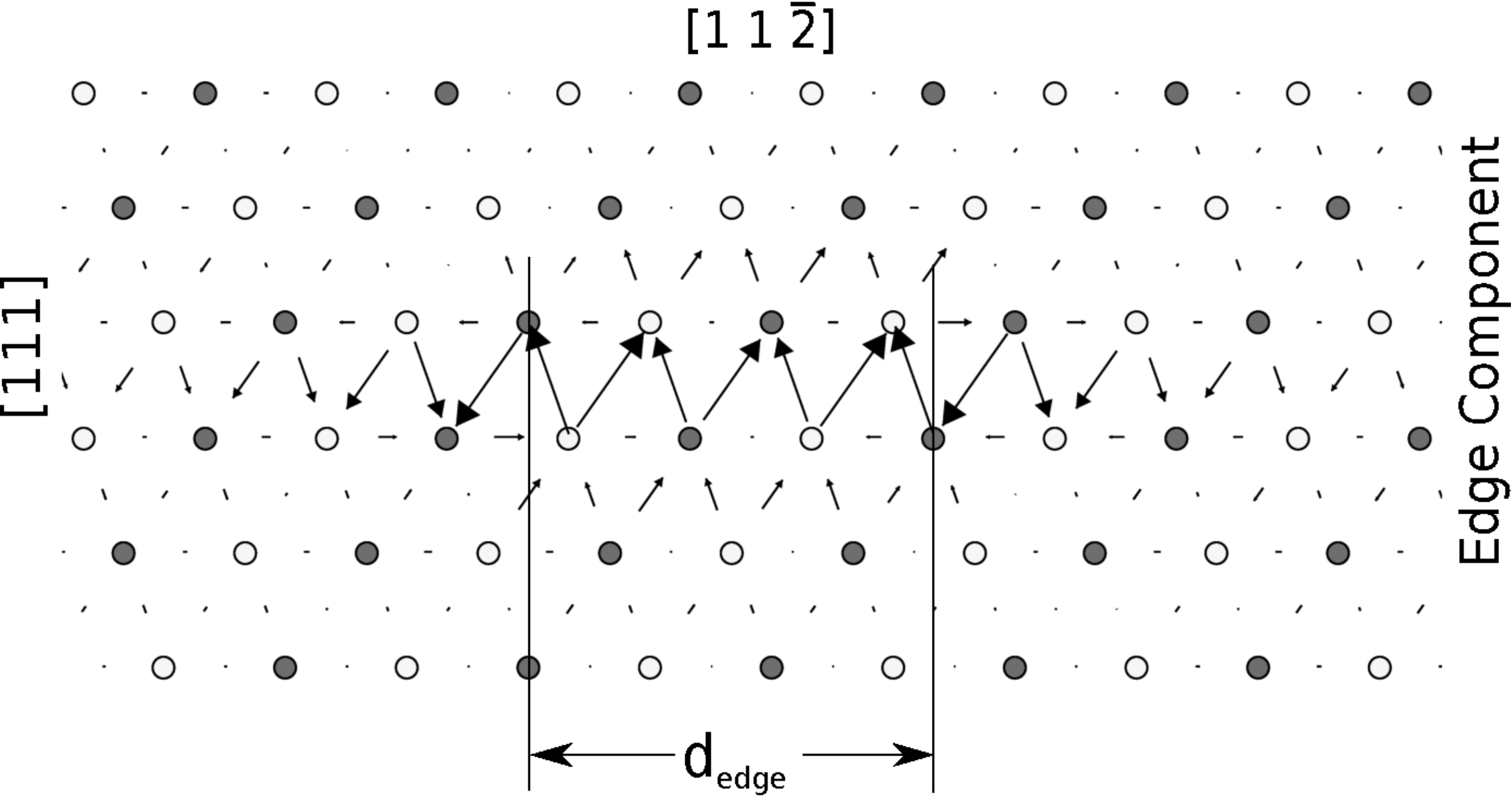}\label{fig:dd12EdgeM}}
\caption{\label{fig:dd2}\small{Differential displacement plots of the edge components of Shockley partials: (a) $\epsilon_{12}$ = 0.66\% ; (b) $\epsilon_{12}$ = -0.66\%.}}
\end{figure}

Finally, we consider the influence of macroscopic shear strains $\epsilon_{12}$ and $\epsilon_{13}$. We have not considered the $\epsilon_{23}$ strain, as this results in a net glide force on the screw dislocation and can result in dislocation glide upon overcoming the small Peierls barrier ($\approx$ 11 MPa, cf. ~\cite{Shin2013}). We considered shear strains of -0.66\%,-0.33\%, 0.33\% and 0.66\% in this study. Figure~\ref{fig:coreEnergyShear} shows the computed core energy dependence on $\epsilon_{12}$ and $\epsilon_{13}$ shear strains. We observe that the core energy dependence on $\epsilon_{13}$ is weak and symmetric, whereas the core energy dependence on $\epsilon_{12}$ strain is significant and is non-symmetric. This significant difference can be rationalized by taking note of the linear elastic forces on the Shockley partials due to these macroscopic shear strains. The $\epsilon_{13}$ shear strain causes climb forces to act on the screw components of Shockley partials. When the sign of $\epsilon_{13}$ is changed, the force direction is reversed but it has a symmetric influence on the core structure, which leads to the observed symmetry in the core energy dependence on $\epsilon_{13}$. On the other hand, $\epsilon_{12}$ shear strain results in equal and opposite glide forces on the edge component of Shockley partials, which can either increase or decrease the partial separation distance depending on the sign of $\epsilon_{12}$ shear strain. This leads to the asymmetry in the core energy dependence on $\epsilon_{12}$. This rationalization is also supported by investigating the change in the core-structure. For the $\epsilon_{12}$ strain, the partial separation in the edge-component of the DD plot changes considerably from 1.7 $|{\bf b}|$ at $\epsilon_{12}$ = 0.66\% to 2.9 $|{\bf b}|$ at $\epsilon_{12}$ = -0.66\% (cf. figure~\ref{fig:dd2}). On the other hand for $\epsilon_{13}$ strain, the partial separation distances are found to be unchanged for equal and opposite strains. 


\subsection{Core-force on an infinite straight dislocation}\label{sec:coreForce}
The study in section~\ref{sec:macroDeform} of the present work and \citet{Iyer2015} demonstrates that the core energy of a dislocation is significantly dependent on macroscopic deformations. This core energy dependence on the macroscopic deformation results in an additional configurational force on the  dislocation, beyond the Peach-Koehler force, which we refer to as the \emph{core-force}. The core-force on a unit line segment of an infinite straight dislocation due to an external strain field ${\boldsymbol \epsilon}^{\rm ext}$ is given by 
\begin{equation}
{f_{{\rm c},i}}({\boldsymbol \epsilon}^{\rm ext})= -\frac{\partial E_{\rm c}({\boldsymbol \epsilon}^{\rm ext})}{\partial{\epsilon}^{\rm ext}_{kl}} \frac{\partial {\epsilon}^{\rm ext}_{kl}}{\partial x_i}  , \label{eq:coreForceSimple}
\end{equation}
where $E_{c}({\boldsymbol \epsilon}^{\rm ext})$ is the core energy of the dislocation per unit-length. In the above expression, the core size is assumed to be smaller than the length scale on which ${\boldsymbol \epsilon}^{\rm ext}$ varies. 

We note that the core-force on an infinite straight dislocation depends on the spatial gradient of the strain field. Thus, the core-force can play an important role in governing the dislocation behavior in regions of inhomogeneous deformations, such as the case of multiple interacting dislocations which has been widely studied using 3D discrete dislocation dynamics calculations (cf. ~\citet{Arsenlis2002,Arsenlis2007,Bulatov2004,Kubin1992,Schwarz1999,Zbib1998}) to predict macroscopic deformation response in crystalline materials. Apart from this, core forces may be significant in other physical scenarios involving dislocations in an external inhomogeneous strain field, such as interaction of dislocations with other defects like grain boundaries and precipitates (cf. ~\cite{Dewald2007,Dewald2011,Koning2003,Shin2003}). In the next section, we develop a general framework for computing the core-force on a dislocation line segment in an aggregate of dislocations.

\section{Core-energetics based forces for an aggregate of dislocations}\label{sec:coreForceModel}
In this section, we first develop an electronic-structure informed energetics model for an aggregate of dislocations in an isotropic infinite elastic continua, and subsequently, we obtain the forces in a discrete setting from variations of the total energy with respect to the degrees of freedom. In particular, we focus on the contribution to the forces arising from the core energy dependence on macroscopic deformations. In the present work, we have used RS-OFDFT calculations from Section~\ref{sec:screw} and \citet{Das2016,Iyer2015} to inform the core energetics model for Aluminum, however, the framework developed here is not materials system or data specific, and can be used with other materials systems and core energetics data from other studies. We note that recent DDD models have used core energetics information (cf.~\cite{Martinez2008,Shishvan2008}) from atomistic calculations resulting in additional force terms beyond the linear elastic Peach-Koehler force. However, the core energy is assumed to be a constant in these models, whereas it is evident from our study that the core energy can have a strong dependence on macroscopic strains. As noted in Section~\ref{sec:coreForce}, for a simple case of an infinite edge dislocation, this dependence on macroscopic deformation results in another additional force that is proportional to external strain gradients and the slope of the core energy dependence on macroscopic strains (cf. equation \eqref{eq:coreForceSimple}). In regions of inhomogeneous deformations with large strain gradients, this additional force can be significant as will be evident from the case studies presented in Section~\ref{sec:caseStudies}. Moreover, in ~\ref{sec:corePostProc}, we demonstrate that the slopes of the core energy dependence on macroscopic strains in Aluminum computed using atomistic calculations are sensitive to the choice of the interatomic potential, which underscores the need for quantum-mechanical calculations to compute the core energetics.
     
Broadly, this section comprises of two main parts, and is supported by ~\ref{sec:corePostProc}. Section~\ref{sec:generalModel} discusses the choice of the underlying linear elastic model, the partitioning of the total energy of the system into core and elastic energies, and the assumptions which go into developing the core energetics model for an arbitrary dislocation aggregate. ~\ref{sec:corePostProc} discusses the post-processing of the core energy data from RS-OFDFT calculations to remove the elastic effects consistent with the energy partition and tabulates the relevant information. In Section~\ref{sec:forceDerivStr}, we consider a dislocation network discretized into straight segments, and derive the forces on the nodes of the segments using linear shape functions.

\subsection{Energetics model}\label{sec:generalModel}
We start by considering the elastic energy of the dislocation aggregate. We model the entire aggregate of dislocations as a collection of dislocation loops\footnote{We are interested in dislocation loops as dislocation lines cannot terminate inside the crystal.}, and denote this by $C$. In an infinite isotropic elastic continua, using classical linear elastic theory of dislocations for small displacements, the stress field due to $C$ at a spatial point $\bf{x}$ is expressed in terms of the line integral~\citep{Cai2006,Hirth1982}, 
\begin{align}
\sigma_{\alpha\beta}^{C} ({\bf x}) = & \frac{\mu}{8\pi} \oint_{C} \partial_i \partial_p \partial_p R\left[b^{\prime}_m \varepsilon_{im\alpha}\,{\rm d}x_{\beta}^{\prime}+b^{\prime}_m \varepsilon_{im\beta}\,{\rm d}x_{\alpha}^{\prime}\right] \nonumber \\
+ & \frac{\mu}{4\pi(1-\nu)} \oint_{C}b^{\prime}_m \varepsilon_{imk}\left(\partial_i\partial_{\alpha}\partial_{\beta}R - \delta_{\alpha \beta} \partial_i \partial_p \partial_p R\right)\,{\rm d}x_k^{\prime} \,, \label{eq:classicalStress}
\end{align}
where $\varepsilon_{ijk}$ denotes the cyclic tensor, $\partial_i \equiv \frac{\partial}{\partial x_i}$, $R=\|{\bf x}-{\bf x^{\prime}}\|$, ${\bf b}^{\prime}={\bf b}({\bf x}^{\prime})$ is the Burgers vectors at ${\bf x}^{\prime}$, and $\mu$ and $\nu$ are the isotropic shear modulus and Poisson's ratio, respectively. The total elastic energy of $C$, denoted by $E^{C}_{\rm el}$, is expressed as a double line integral~\citep{Cai2006,Wit1967b,Hirth1982}, 
\begin{align}
E^{C}_{\rm el}= & -\frac{\mu}{8\pi} \oint_C \oint_C \partial_k \partial_k R \, b_i b_j^{\prime} \, {\rm d}x_i \, {\rm d} x_j^{\prime} -\frac{\mu}{4\pi(1-\nu)} \oint_C \oint_C \partial_i \partial_j R \, b_i b_j^{\prime}\,{\rm d}x_k \, {\rm d} x_k^{\prime} \nonumber \\
+& \frac{\mu}{4\pi(1-\nu)} \left[ \oint_C \oint_C \partial_k \partial_k R \, b_i b_i^{\prime}\,{\rm d}x_j \, {\rm d} x_j^{\prime}- \nu \oint_C \oint_C \partial_k \partial_k R \, b_i b_j^{\prime}\,{\rm d}x_j \, {\rm d} x_i^{\prime} \right]\,, \label{eq:classicalEnergy}
\end{align}
where ${\bf b}={\bf b}({\bf x})$ and ${\bf b^{\prime}}={\bf b}({\bf x}^{\prime})$ are the Burgers vectors at ${\bf x}$ and ${\bf x^{\prime}}$, respectively.  We note that the partial derivatives of $R$ appearing in the integrand of the above expressions for stress and energy become singular as $R \rightarrow 0$. In order to circumvent this issue, non-singular linear elastic theories~\citep{Brown1964,Cai2006,Gavazza1976,indenbom1992} have been proposed to remove the singularity in the elastic fields. Gradient elasticity theories~\citep{Gutkin1999,Lazar2013,Lazar2014} are also inherently non-singular, but we do not use them here as our focus is to investigate the forces from electronic structure effects at the dislocation core beyond what is captured by the classical linear elasticity theory. Among the non-singular formulations by ~\citet{Brown1964,Cai2006,Gavazza1976,indenbom1992}, the formulations of ~\citet{Gavazza1976} and ~\citet{Cai2006}, have the desirable variational property, i.e., the Peach-Koehler force on any point on the dislocation line is equal to the derivative of the total elastic energy with respect to its spatial position.  In the~\citet{Gavazza1976} formulation, a tubular region around the dislocation is excluded from the elastic energy calculation, while in the~\citet{Cai2006} formulation, the Burgers vector is smeared using an isotropic spreading function characterized by a parameter $a$ that quantifies the spread radius. In developing our model, we adopt the latter formulation as the former has difficulty dealing with dislocation lines with sharp corners which can be encountered at situations like junction nodes or cross-slip nodes. The total elastic energy of $C$ using the non-singular formulation of~\citet{Cai2006} is given by
\begin{align}
E^{C}_{\rm el}(a)= & -\frac{\mu}{8\pi} \oint_C \oint_C \partial_k \partial_k R_a b_i b_j^{\prime}\,{\rm d}x_i \, {\rm d} x_j^{\prime} -\frac{\mu}{4\pi(1-\nu)} \oint_C \oint_C \partial_i \partial_j R_a b_i b_j^{\prime}\, {\rm d}x_k \, {\rm d} x_k^{\prime} \nonumber \\
+& \frac{\mu}{4\pi(1-\nu)} \left[ \oint_C \oint_C \partial_k \partial_k R_a b_i b_i^{\prime}\, {\rm d}x_j \, {\rm d} x_j^{\prime}- \nu \oint_C \oint_C \partial_k \partial_k R_a b_i b_j^{\prime}\, {\rm d}x_j \, {\rm d} x_i^{\prime} \right], \label{eq:nsEnergy}
\end{align}
where $R_a=\sqrt{R^2+a^2}$. Choosing $a>0$, the spatial derivatives of $R_a$ are no longer singular, and we assume this going forward. We refer to~\citet{Cai2006} for the non-singular expression for the stress field, which, analogous to equation~\eqref{eq:classicalStress}, is expressed as a line integral. The non-singular stress field can be expressed in a condensed manner as  
\begin{equation}
\sigma_{\alpha\beta}^{C} ({\bf x};a)=   \oint_{C} \bar{\sigma}_{\alpha\beta} \left({\boldsymbol \xi}({\bf x}^{\prime}(s^{\prime})), {\bf b}({\bf x}^{\prime}(s^{\prime})) , {\bf x} -{\bf x}^{\prime}(s^{\prime});a \right) {\rm d} s^{\prime}, \label{eq:contourStress} 
\end{equation}
where $\bar{\sigma}_{\alpha\beta}$ denotes the per unit-length contribution to the stress field at the spatial point ${\bf x}$ from the dislocation line at ${\bf x}^{\prime}(s^{\prime})$, with  $s^{\prime}$ being the length parametric variable of $C$. The other dependencies of $\bar{\sigma}_{\alpha\beta}$ are ${\boldsymbol \xi}({\bf x}^{\prime})$, which is the tangent line direction at ${\bf x}^{\prime}$ such that ${\boldsymbol \xi}({\bf x}^{\prime}){\rm d} s^{\prime}= {\rm d} {\bf x}^{\prime}$ , and ${\bf b}({\bf x}^{\prime})$ is the Burgers vector at ${\bf x}^{\prime}$. The dependency on the isotropic elastic constants, $\mu$ and $\nu$, is implicit. Similarly, the non-singular strain field is also expressed as  
\begin{equation}
\epsilon_{\alpha\beta}^{C} ({\bf x};a)=   \oint_{C} \bar{\epsilon}_{\alpha\beta} \left({\boldsymbol \xi}({\bf x}^{\prime}(s^{\prime})), {\bf b}({\bf x}^{\prime}(s^{\prime})),{\bf x} -{\bf x}^{\prime}(s^{\prime});a\right) {\rm d} s^{\prime}\, ,\label{eq:contourStrain}
\end{equation}
where $\bar{\epsilon}_{\alpha\beta}$ is the per unit-length contribution to the strain field at spatial point ${\bf x}$ from the dislocation line at ${\bf x}^{\prime}(s^{\prime})$, with the same dependencies as the stress field. This representation for the strain field will be used in the subsequent development of the core energetics model. We note that this representation can also be extended to the anisotropic case, where the anisotropic stress and strain fields of a dislocation loop are expressed as a line integral~\citep{Mura1982}, as well as to gradient elasticity theories~\citep{Lazar2013,Lazar2014}.
   
\subsubsection{Non-elastic core energetics model}\label{sec:coreEnergyModel}

We now develop a model that accounts for the dislocation core energy in the total energy of $C$, and its dependence on macroscopic strain, which is informed by the core energetics data obtained from RS-OFDFT electronic structure calculations. To this end, we refer to the total energy of $C$ inside the tubular domain corresponding to the core-size (core-domain) as the core energy, and denote this by $E_{\rm c}^{C}$. A portion of $E_{\rm c}^{C}$ is the non-singular elastic energy of the core-domain, which we denote as $E_{\rm cel}^{C}(a)$ (core-domain elastic energy or elastic core-energy). We refer to the remaining part of the core energy as the non-elastic core-energy, denoted by $E_{\rm cnel}^{C}(a)$, which includes contributions from atomistic and quantum mechanical effects inside the dislocation core. Thus, we have the following partitioning for the the core energy:
\begin{equation}
E_{\rm c}^{C}= E_{\rm cel}^{C}(a)+E_{\rm cnel}^{C}(a). \label{eq:coreEnergyPartition}
\end{equation}
We note that as $E_{\rm cel}^{C}$ depends on the choice of the spread radius ($a$) in the non-singular approximation, $E_{\rm cnel}^{C}$ also depends on $a$. Changing $a$ changes the partitioning of total core energy $E_{\rm c}^{C}$, but this has no effect on the dislocation properties (energy and forces) as these are governed by $E_{\rm c}^{C}$, which is informed from electronic structure calculations and is independent of $a$. Furthermore, the energy outside the core-domain is solely elastic energy and is independent of the spread radius $a$---the non-singular elastic fields converge to the classical elastic fields at distances beyond $a$, and commonly used values of $a$, $a<$ 2$|{\bf b}|$~\citep{Martinez2008,Shishvan2008}, are much smaller than the RS-OFDFT calculated core-size of 7--10 $|{\bf b}|$. Thus, the total energy of $C$, which we denote as $E_{\rm tot}^{C}$, is independent of $a$, and is given by 
\begin{equation}
E_{\rm tot}^{C}= E_{\rm el}^{C}(a)+E_{\rm cnel}^{C}(a) \,. \label{eq:genEnergyPartition}
\end{equation}

We now present a model for $E_{\rm c}^{C}$ using reasonable approximations, where the model is directly informed by the core-energetics data from RS-OFDFT, and we subsequently extract the non-elastic core-energy $E_{\rm cnel}^{C}(a)$ using the partitioning in equation~\eqref{eq:coreEnergyPartition}. In our model, we ignore the non-elastic effects arising from the direct core-core interactions between any two points on $C$, which is a reasonable approximation as the average separation between dislocations in DDD simulations is larger than the RS-OFDFT estimated core-size of 7--10 $|{\bf b}|$. Under this approximation, $E_{\rm c}^{C}$ can be expressed as a line integral on $C$. Further, we approximate the core energy per unit length at any point ${\bf x}(s)$ with that of an infinite straight dislocation having the same local Burgers vector (${\bf b}(s)$), the same tangent line direction (${\boldsymbol \xi}(s)$), and embedded in the same macroscopic environment, i.e., the infinite straight dislocation subjected to the same external strain field as being felt by the dislocation core at ${\bf x}(s)$. In other words, we ignore the possible explicit dependence of $E_{\rm c}^{C}$ on the curvature of $C$. Thus, under these approximations, the core energy is given by  
\begin{equation}
E_{\rm c}^{C}= \oint_{C} E_{\rm c}^{\rm str}({\boldsymbol \xi}(s),{\bf b}(s),{\boldsymbol \epsilon}^{{\rm ext}_{\rm loc}}({\bf x}(s)+{\bf x}_{\rm c}))\, {\rm d} s, \qquad {\bf x}_{\rm c} \in \Omega_{\rm c}\,, \label{eq:coreEngNetwork1}
\end{equation}
where $E_{\rm c}^{\rm str}$ denotes core energy per unit length of an infinite straight dislocation. In the above expression, ${\boldsymbol \epsilon}^{{\rm ext}_{\rm loc}}({\bf x}(s)+{\bf x}_{\rm c})$ denotes the external strain field experienced by all points ${\bf x}_{\rm c}$ in the dislocation core domain $\Omega_{\rm c}$, where ${\bf x}_{\rm c}$ is the position vector with respect to the core center at ${\bf x}(s)$. The external strain field is due to the entire dislocation network $C$ excluding the dislocation lines at and near the dislocation core, which is computed using a cut-off procedure that will be discussed subsequently. The symbol ``${\rm loc}$" is used to denote that the strain tensor is expressed in a local coordinate frame at ${\bf x}(s)$, which is aligned such that the axis labelled `1' lies on the slip plane at ${\bf x}(s)$ and is perpendicular to ${\boldsymbol \xi}(s)$, axis labelled `2' is perpendicular to the slip plane, and the axis labelled `3' is along ${\boldsymbol \xi}(s)$. This choice of the local coordinate axes is the same as those employed in RS-OFDFT calculations and thus allows an immediate parametrization of the core energy dependence on macroscopic strains from the RS-OFDFT core energetics data. The transformation of the strain tensor field $\bar{\boldsymbol {\epsilon}}$ (the integrand in equation~\eqref{eq:contourStrain}) from the global frame to the local frame is given by 
\begin{equation}
\bar{{\epsilon}}^{{\rm loc}}_{\alpha\beta}\left({\boldsymbol \xi}(s^{\prime}), {\bf b}(s^{\prime}),{\bf x} -{\bf x^{\prime}}(s^{\prime})\right)=R_{k\alpha}({\boldsymbol \xi}(s),{\bf b}(s))R_{l\beta}({\boldsymbol \xi}(s),{\bf b}(s)) \bar{{\epsilon}}_{kl}\left({\boldsymbol \xi}(s^{\prime}), {\bf b}(s^{\prime}),{\bf x} -{\bf x^{\prime}}(s^{\prime})\right),\label{eq:tensorTransform} 
\end{equation} 
where the rotation matrix ${\bf R}({\boldsymbol \xi}(s),{\bf b}(s))$ maps the components of any vector in the local coordinate frame at ${\bf x}(s)$ to the components of the same vector in the global coordinate frame. We note that the dependency on the smearing parameter, $a$, is ignored in the above expression as the separation between the dislocations is assumed to be larger than $a$. Using equations~\eqref{eq:contourStrain} and \eqref{eq:tensorTransform}, as the rotation matrix is independent of $s^{\prime}$, the external strain field in the local coordinate frame, ${\boldsymbol \epsilon}^{{\rm ext}_{\rm loc}}({\bf x})$, is given by  
\begin{gather}
{\epsilon}^{{\rm ext}_{\rm loc}}_{\alpha\beta}({\bf x})=\int_{C-\rho_{\rm cut}} \bar{\epsilon}_{\alpha\beta}^{\rm loc} \left({\boldsymbol \xi}(s^{\prime}), {\bf b}(s^{\prime}),{\bf x} -{\bf x^{\prime}}(s^{\prime})\right) \, {\rm d} s^{\prime},  \label{eq:extStrainContour1}
\end{gather} 
where $\rho_{\rm cut}$ is a small cut-off radius characterizing a spherical region centered at ${\bf x}(s)$, and the dislocation lines inside this region are excluded in evaluating the above line integral. We note that the use of this cut-off approach is solely to restrict the sources of the external strain field to those outside of the dislocation core, and is not an attempt to regularize the external strain field at the core as the regularization is already achieved by virtue of the non-singular elastic model. We further note that the external strain field in the above expression is considered to be independent of $a$, as the average separation distance in DDD calculations is greater than the values used for $a$, beyond which the strain fields are independent of the choice of $a$. Thus, by extension, $E_{\rm c}^{C}$ is also independent of the choice of $a$. We now consider reasonable simplifications to equations~\eqref{eq:coreEngNetwork1} and ~\eqref{eq:extStrainContour1}, which enable us to use the available RS-OFDFT data to inform the core energy.\newline
{\it Simplification} (1): In equation~\eqref{eq:coreEngNetwork1}, the core energy per unit length of the straight dislocation, $E^{\rm str}_{\rm c}$, in all generality, is a function of the external strain field at all points inside the core. However, parametrizing $E^{\rm str}_{\rm c}$ in the strain field function space is intractable. A reasonable approximation will be to simplify this dependence to a homogeneous mean-field strain dependence. To this end, we simplify the dependence to the external strain field at the dislocation line (${\bf x}_{\rm c}={\bf 0}$), given by  
\begin{equation}
E_{\rm c}^{C}= \oint_{C} E_{\rm c}^{\rm str}({\boldsymbol \xi}(s),{\bf b}(s),{\boldsymbol \epsilon}^{{\rm ext}_{\rm loc}}({\bf x}(s)))\, {\rm d} s \,. \label{eq:coreEngNetwork2}
\end{equation}
We note that this approximation is reasonable when the distance between dislocations is much larger than the core-size. 
\newline
{\it Simplification} (2): We note that the dependence of $E^{\rm str}_{\rm c}$ on ${\boldsymbol \xi}(s)$ and ${\bf b}(s)$ arises from the changing character for the dislocation that depends on the angle between the line direction and the Burgers vector, given by 
\begin{equation}
\theta(s)= {\rm arccos}\left(\frac{{\boldsymbol \xi}(s) \cdot {\bf b}(s)}{ \|{\bf b}(s)\|} \right)\,. \label{eq:orientationAngle}
\end{equation} 
Parametrizing $E^{\rm str}_{\rm c}$ as a function of $\theta(s)$ can be very tedious and time consuming especially using electronic structure calculations. To this end, we adopt the commonly used approximation, where the core energy of a dislocation with mixed character is interpolated from the core energies of the edge and screw dislocations as 
\begin{equation}
E_{\rm c}^{\rm str}({\boldsymbol \xi}(s),{\bf b}(s),{\boldsymbol \epsilon}^{{\rm ext}_{\rm loc}}({\bf x}(s))) =  E_{\rm c}^{\rm edge}({\boldsymbol \epsilon}^{{\rm ext}_{\rm loc}}({\bf x}(s))) {\rm sin}^{2}(\theta(s)) + E_{\rm c}^{\rm screw}({\boldsymbol \epsilon}^{{\rm ext}_{\rm loc}}({\bf x}(s))) {\rm cos}^{2}(\theta(s)) \,. \label{eq:mixedCore}
\end{equation} 
\newline
{\it Simplification} (3): In general, the dependence of $E_{\rm c}^{\rm str}$ on ${\boldsymbol \epsilon}^{{\rm ext}_{\rm loc}}$ can be non-linear. However, when the distance between dislocations is sufficiently large, the external strain fields at the dislocation line are small. Thus, we can further simplify the dependence of $E_{\rm c}^{\rm str}$ on ${\boldsymbol \epsilon}^{{\rm ext}_{\rm loc}}$ by using a Taylor expansion to first order about ${\boldsymbol \epsilon}^{{\rm ext}_{\rm loc}}=0$ as
\begin{equation}
E_{\rm c}^{C}= \oint_{C} \left[ E_{\rm c}^{\rm str}({\boldsymbol \xi}(s),{\bf b}(s),{\boldsymbol \epsilon}^{{\rm ext}_{\rm loc}}({\bf x}(s))={\bf 0}) 
       + S_{\alpha\beta}({\boldsymbol \xi}(s),{\bf b}(s)){\epsilon}^{{\rm ext}_{\rm loc}}_{\alpha\beta}({\bf x}(s))\right] \,{\rm d} s \,, \label{eq:coreEngNetworkLinear}
\end{equation}
where $S_{\alpha\beta}({\boldsymbol \xi}(s),{\bf b}(s))$ is the slope of the core energy dependence on external strain evaluated at zero strain,
\begin{equation}
S_{\alpha\beta}({\boldsymbol \xi}(s),{\bf b}(s))= \left.\frac{\partial E_{\rm c}^{\rm str}\left({\boldsymbol \xi}(s),{\bf b}(s),{\boldsymbol \epsilon}^{{\rm ext}_{\rm loc}}\right)}{\partial {\epsilon}^{{\rm ext}_{\rm loc}}_{\alpha\beta}} \right|_{{\boldsymbol \epsilon}^{{\rm ext}_{\rm loc}}={\bf 0}}.\label{eq:coreEngSlope}
\end{equation}
The values of the slopes for dislocations with mixed character are evaluated, using equation~\eqref{eq:mixedCore}, from the slopes of the pure edge and screw dislocations, which, in turn, are available using RS-OFDFT calculations. 

Next, using the partition in equation~\eqref{eq:coreEnergyPartition}, we obtain $E_{\rm cnel}^{C}(a)$ from $E_{\rm c}^{C}$ by subtracting the non-singular linear elastic energy. The values of the slopes at zero strain in equation~\eqref{eq:coreEngSlope} are also accordingly obtained by removing the contribution from the non-singular elastic energy. We refer to~\ref{sec:corePostProc} for details of this post-processing. Finally, as a counterpart of equation~\eqref{eq:coreEngNetwork2}, we obtain 
\begin{equation}
E_{\rm cnel}^{C}(a)= \oint_{C} E_{\rm cnel}^{\rm str}({\boldsymbol \xi}(s),{\bf b}(s),{\boldsymbol \epsilon}^{{\rm ext}_{\rm loc}}({\bf x}(s));a)\, {\rm d} s \,,\label{eq:nelCoreEngNetwork2}
\end{equation} 
and as the counterpart of the linearized equations~\eqref{eq:coreEngNetworkLinear}, we obtain
\begin{equation}
E_{\rm cnel}^{C}(a)= \oint_{C} \left[ E_{\rm cnel}^{\rm str}({\boldsymbol \xi}(s),{\bf b}(s),{\boldsymbol \epsilon}^{{\rm ext}_{\rm loc}}({\bf x}(s))={\bf 0};a)
       + \hat{S}_{\alpha\beta}({\boldsymbol \xi}(s),{\bf b}(s);a){\epsilon}^{{\rm ext}_{\rm loc}}_{\alpha\beta}({\bf x}(s))\right] \, {\rm d} s\,, \label{eq:nelCoreEngNetworkLinear}
\end{equation}
where
\begin{equation}
\hat{S}_{\alpha\beta}({\boldsymbol \xi}(s),{\bf b}(s);a)= \left.\frac{\partial E_{\rm cnel}^{\rm str}\left({\boldsymbol \xi}(s),{\bf b}(s),{\boldsymbol \epsilon}^{{\rm ext}_{\rm loc}};a\right)}{\partial {\epsilon}^{{\rm ext}_{\rm loc}}_{\alpha\beta}} \right|_{{\boldsymbol \epsilon}^{{\rm ext}_{\rm loc}}={\bf 0}}.\label{eq:nelCoreEngSlope}
\end{equation}
  
We note that an important limitation of the aforementioned model stems from neglecting direct interactions between dislocation cores at different points on the dislocation aggregate. This is a reasonable assumption when distances between the dislocations are much larger compared to the core-size. However, there are practical situations where this assumption fails---e.g. when radius of curvature of the dislocation line is comparable to core-size, dislocations passing each other at distances less than core size, and dislocation core reactions like annihilation and junction formation. In such situations, the energetics have to be obtained from direct electronic structure calculations of the dislocation-dislocation interactions, which are still out of reach for such large systems. We note that, even for these situations, the energetics model developed here is an improvement over existing models that ignore the core energy dependence on external strains.  

\subsection{Derivation of nodal core forces in a discretized network of dislocation line segments}\label{sec:forceDerivStr}
In a 3D nodal discrete dislocation network, it is common to discretize the dislocation line into straight line segments forming discretized polygonal loops. The network is represented by a set of nodes with position vectors $\{{\bf r}^{i}\}$, which are connected by straight segments. The other set of degrees of freedom corresponds to ${\bf b}^{ij}$, which denotes the perfect dislocation Burgers vector of line-segment $ {\bf l}^{ij}={\bf r}^{j}-{\bf r}^{i}$ pointing from node $i$ to node $j$. Here  $ij$ denotes the index of the line segment, ${\bf l}^{ij}$. Overall, we denote the network as $C\equiv \{{\bf r}^{i},{\bf b}^{ij}\}$. Constraints are imposed on ${\bf b}^{ij}$ and the nodal connections such that the Burgers vector at each node is conserved, and dislocation lines cannot end in the crystal. Additionally, ${\bf b}^{ij}$ are assumed to be constant when the nodal positions $\{{\bf r}^{i}\}$ are updated without changing the node connectivity. We refer to ~\cite{Arsenlis2002,Arsenlis2007,Bulatov2004}  for a comprehensive description of the 3D DDD implementation including mobility laws, time integration, topological rearrangements, treatment of dislocation core reactions, and computational strategies. In this study, we are concerned with the nodal forces arising from non-elastic core energy contributions. 

Applying the core energetics model we developed in Section~\ref{sec:generalModel} to a discrete dislocation network $C\equiv \{{\bf r}^{i},{\bf b}^{ij}\}$, we obtain the non-elastic core energetics as (cf. equation~\eqref{eq:nelCoreEngNetwork2}),
\begin{equation}
 E_{\rm cnel}^{C}(a)=\sum_{ij \, \in \,U}\int^{L_{ij}}_0 E_{\rm cnel}^{\rm str}({\boldsymbol \xi}^{ij},{\bf b}^{ij},{\boldsymbol \epsilon}^{{\rm ext}_{\rm loc}}({\bf x}^{ij}(s_{ij}));a)\, {\rm d} s_{ij} \,. \label{eq:engPartStr}
\end{equation}
In the above, $L_{ij}=\|{\bf l}^{ij}\|$ is the length of ${\bf l}^{ij}$, ${\boldsymbol \xi}^{ij}=\frac{{\bf l}^{ij}}{\|{\bf l}^{ij}\|}$ is the unit vector corresponding to ${\bf l}^{ij}$, ${\bf x}^{ij}(s_{ij})$ is the position vector of a point on ${\bf l}^{ij}$, parametrized by $s_{ij}$, which is the parametric length coordinate of ${\bf l}^{ij}$ measured from node $i$ towards node $j$. The index $ij$ is summed over the set $U$, which represents the collection of all distinct line segments in the dislocation network. In particular, $ij$ and $ji$ are \emph{not} considered as distinct as they refer to the same segment but with reversed directions for the unit vector and Burgers vector. Further, for computing the external strain field, we follow a more convenient cut-off procedure suited to the segment discretization rather than the spherical cut-off procedure used in equation~\eqref{eq:extStrainContour1}. This is given by 
\begin{equation}
{\boldsymbol \epsilon}^{{\rm ext}_{\rm loc}}({\bf x}^{ij}(s_{ij}))= \sum_{kl \, \in \, U_{ij}^{\prime}} {\boldsymbol \epsilon}^{{kl}_{\rm loc}}({\boldsymbol \xi}^{kl},{\bf b}^{kl},{\bf x}^{ij}(s_{ij})) ,\label{eq:strExtStrain}
\end{equation}
where ${\boldsymbol \epsilon}^{{kl}_{\rm loc}}({\boldsymbol \xi}^{kl},{\bf b}^{kl},{\bf x}^{ij}(s_{ij}))$ is the strain field contribution of ${\bf l}^{kl}$ segment at the spatial point ${\bf x}^{ij}(s_{ij})$, expressed in the local frame attached to ${\bf l}^{ij}$ in which the electronic structure core-energetics data is available, and the set $U_{ij}^{\prime}$ comprises of all distinct line segments in the network excepting those that have $i$ or $j$ as one of their nodes. In other words, the set $U_{ij}^{\prime}$ excludes the segment ${\bf l}^{ij}$ and its immediate neighbours. We note that this approach and the spherical cut-off approach converge to the same external strain field as $\max_{{ij}\in U}L_{ij} \rightarrow 0$ and $\rho_{\rm cut} \rightarrow 0$, respectively. In the above expression, ${\boldsymbol \epsilon}^{{kl}_{\rm loc}}({\boldsymbol \xi}^{kl},{\bf b}^{kl},{\bf x}^{ij}(s_{ij}))$ is expressed using equation~\eqref{eq:tensorTransform} as
\begin{equation}
{\epsilon}^{{kl}_{\rm loc}}_{\alpha\beta}({\boldsymbol \xi}^{kl},{\bf b}^{kl},{\bf x}^{ij}(s_{ij})) = R_{\eta\alpha}({\boldsymbol \xi}^{ij},{\bf b}^{ij})R_{\omega\beta}({\boldsymbol \xi}^{ij},{\bf b}^{ij}) {\epsilon}^{kl}_{\eta\omega}({\boldsymbol \xi}^{kl},{\bf b}^{kl},{\bf x}^{ij}(s_{ij})) \,, \label{eq:transformSeg}
\end{equation}
which transforms the strain field contribution from ${\bf l}^{kl}$ segment at the spatial point ${\bf x}^{ij}(s_{ij})$ in the global frame to the local frame attached to ${\bf l}^{ij}$. Next, we substitute equation~\eqref{eq:strExtStrain} in equation~\eqref{eq:engPartStr}, and linearize the dependence of the non-elastic core energy on the external macroscopic strain using equations~\eqref{eq:nelCoreEngNetworkLinear} and ~\eqref{eq:nelCoreEngSlope}, to simplify $E_{\rm cnel}^{C}(a)$ as
\begin{align}
E_{\rm cnel}^{C}(a)=& \sum_{ij \, \in \,U} \int^{L_{ij}}_0 E_{\rm cnel}^{\rm str}({\boldsymbol \xi}^{ij},{\bf b}^{ij},{\boldsymbol \epsilon}^{{\rm ext}_{\rm loc}}({\bf x}^{ij}(s_{ij}));a)\, {\rm d} s_{ij}, \nonumber\\
\approx & \sum_{ij \, \in \,U}E_{\rm cnel}^{\rm str}({\boldsymbol \xi}^{ij},{\bf b}^{ij},{\bf 0};a) L_{ij} 
 + \sum_{ij \, \in \,U} \sum_{kl \, \in \, U_{ij}^{\prime}}   \hat{S}_{\alpha\beta}({\boldsymbol \xi}^{ij},{\bf b}^{ij};a)\int^{L_{ij}}_0{\epsilon}^{{kl}_{\rm loc}}_{\alpha\beta}({\boldsymbol \xi}^{kl},{\bf b}^{kl},{\bf x}^{ij}(s_{ij}))\, {\rm d} s_{ij}\,. \label{eq:strCoreEng}
\end{align}
For keeping the subsequent analysis concise, we rewrite the above expression in a condensed form as
\begin{equation}
E_{\rm cnel}^{C}(a)=\sum_{ij \, \in \,U} T_{\rm self}^{ij} + \sum_{ij \, \in \,U} \sum_{kl \, \in \, U^{\prime}_{ij}} T_{\rm ext}^{(ij,kl)} \, .\label{eq:strCoreEngCond}
\end{equation}
 
The force on node $i$, ${\bf F}^{i}$, is the negative derivative of the total energy with respect to the position ${\bf r}^{i}$, i.e.,
\begin{align}
{\bf F}^{i}=& -\frac{\partial E_{\rm el}^{C}(a)}{\partial {\bf r}^{i}} -\frac{\partial E_{\rm cnel}^{C}(a)}{\partial {\bf r}^{i}} \nonumber\\
 =& {\bf F}^{i}_{\rm el} + {\bf F}^{i}_{\rm c},\label{eq:totalNodeForce}
\end{align} 
where ${\bf F}^{i}_{\rm el}$ and ${\bf F}^{i}_{\rm c}$ are the nodal elastic force and core force contributions. ${\bf F}^{i}_{\rm el}$, has been analytically determined in previous literature~\citep{Arsenlis2007,Cai2006} for the non-singular elastic formulation, by applying the principle of virtual work and using the Peach-Koehler formula. The focus in this study is on obtaining the expressions for ${\bf F}^{i}_{\rm c}$. Using the representation in~\eqref{eq:strCoreEngCond}, ${\bf F}^{i}_{\rm c}$ is given by
\begin{subequations}
\begin{align}
\begin{split}
{\bf F}^{i}_{\rm c}=& -\frac{\partial E_{\rm cnel}^{C}(a)}{\partial {\bf r}^{i}}= \sum_j {\bf f}_{\rm cs}^{\{i\}j}+\sum_{j}\sum_{kl \, \in \, U^{\prime}_{ij}} {\bf f}_{\rm ce}^{(\{i\}j,kl)}+\sum_{j}\sum_{kl \, \in \, U^{\prime}_{ij}} \tilde{\bf f}_{\rm ce}^{(\{i\}j,kl)},\label{eq:strForceBreaka} 
\end{split}\\
\begin{split}
{\bf f}_{\rm cs}^{\{i\}j}=& -\frac{\partial T_{\rm self}^{ij}}{\partial {\bf r}^{i}} \,{\rm ;}\quad {\bf f}_{\rm ce}^{(\{i\}j,kl)}= -\frac{\partial T_{\rm ext}^{(ij,kl)}}{\partial {\bf r}^{i}} \,{\rm ;}\quad \tilde{\bf f}_{\rm ce}^{(\{i\}j,kl)}= -\frac{\partial T_{\rm ext}^{(kl,ij)}}{\partial {\bf r}^{i}} \,,\label{eq:strForceBreakb} 
\end{split}
\end{align}\label{eq:strForceBreak} 
\end{subequations}
where $j$ runs over all nodes which have a connection to node $i$, and the superscript notation `$\{i\}$' denotes that the force is with respect to perturbation of node $i$. In the above, the first term, ${\bf f}_{\rm cs}^{\{i\}j}$, is force resulting from change in $T_{\rm self}^{ij}$ due to perturbation of ${\bf r}^{i}$, but this does not account for the force resulting from the core energy dependence on external macroscopic strains. We note that this term is already incorporated into current DDD implementations \citep{Arsenlis2007,Martinez2008}, but, we still analyze it here for the sake of completeness. The next two terms in the nodal force manifest from the core energy dependence on external strain, and these are not considered in current DDD implementations. To elaborate, ${\bf f}_{\rm ce}^{(\{i\}j,kl)}$ is the contribution to ${\bf F}^{i}_{\rm c}$ arising from the core energy change of segment ${\bf l}^{ij}$ due its dependence on the external strain field of another segment ${\bf l}^{kl}$ ($kl\, \in \, U^{\prime}_{ij}$), and $\tilde{\bf f}_{\rm ce}^{(\{i\}j,kl)}$ is the contribution arising from the core energy change of the segment ${\bf l}^{kl}$ ($kl \, \in \, U^{\prime}_{ij}$) due to its dependence on the strain field of ${\bf l}^{ij}$.

We now derive the expressions for each of the terms contributing to ${\bf F}^{i}_{\rm c}$ from the first order perturbations in $T_{\rm self}^{ij}$, $T_{\rm ext}^{(ij,kl)}$ and $T_{\rm ext}^{(kl,ij)}$ resulting from a perturbation of the nodal position ${\bf r}^{i}$, while holding the other nodes fixed. Noting that the segments are always constrained to remain straight, we define a linear shape function attached to node $i$ having a compact support over all segments ${\bf l}^{ij}$ attached to it,
\begin{align}
N_i({\bf x}^{ij}(s_{ij}))= & \frac{\|{\bf x}^{ij}(s_{ij})-{\bf r}_j \|}{\| {\bf r}_i- {\bf r}_j    \|} \,,\label{eq:shapeFunction1}
\end{align}
which can also be expressed as
\begin{align}
N_i(s_{ij})       = & \frac{L_{ij}-s_{ij}}{L_{ij}} \,.\label{eq:shapeFunction2}
\end{align}
We first start with computing the first order perturbations in $T_{\rm self}^{ij}$ with respect to ${\bf r}^{i}$, and aim to write them in the form 
\begin{equation}
\delta T_{\rm self}^{ij} = - {\bf f}_{\rm cs}^{\{i\}j}\cdot \delta {\bf r}^{i}\,. \label{eq:selfPertAim}
\end{equation}
Using equations~\eqref{eq:strCoreEng} and ~\eqref{eq:strCoreEngCond},  
\begin{align}
\delta T_{\rm self}^{ij}= & \frac{\partial{E_{\rm cnel}^{\rm str}({\boldsymbol \xi}^{ij},{\bf b}^{ij},{\bf 0};a)}}{{\partial \xi_{\alpha}^{ij}}}
\frac{\partial \xi_{\alpha}^{ij}}{\partial r^{i}_{\beta}}\delta r^{i}_{\beta}L_{ij}- E_{\rm cnel}^{\rm str}({\boldsymbol \xi}^{ij},{\bf b}^{ij},{\bf 0};a)\xi_{\beta}^{ij}\delta r^{i}_{\beta}  \nonumber\\
= & \frac{\partial{E_{\rm cnel}^{\rm str}({\boldsymbol \xi}^{ij},{\bf b}^{ij},{\bf 0};a)}}{{\partial \xi_{\alpha}^{ij}}}\{ \xi_{\alpha}^{ij}\xi_{\beta}^{ij}-\delta_{\alpha\beta}\}
\delta r^{i}_{\beta} - E_{\rm cnel}^{\rm str}({\boldsymbol \xi}^{ij},{\bf b}^{ij},{\bf 0};a)\xi_{\beta}^{ij}\delta r^{i}_{\beta} \,,\label{eq:selfPert1}
\end{align}
where $\delta_{\alpha\beta}$ denotes the Kronecker delta function.
Comparing equations~\eqref{eq:selfPertAim} and ~\eqref{eq:selfPert1}, we can extract the force vector
\begin{equation}
{f_{{\rm cs},\beta}^{\{i\}j}}=\frac{\partial{E_{\rm cnel}^{\rm str}({\boldsymbol \xi}^{ij},{\bf b}^{ij},{\bf 0};a)}}{{\partial \xi_{\alpha}^{ij}}}\{ \delta_{\alpha\beta}-\xi_{\alpha}^{ij}\xi_{\beta}^{ij}\}+E_{\rm cnel}^{\rm str}({\boldsymbol \xi}^{ij},{\bf b}^{ij},{\bf 0};a)\xi_{\beta}^{ij} \,,\label{eq:strSelfForce}
\end{equation}
where the first term is the contribution to the force which tends to rotate the line segment orientation to lower its core energy, while the second term represents a line tension force to reduce the length of the segment. Next, we consider the perturbations in $T_{\rm ext}^{(ij,kl)}$ with respect to ${\bf r}^{i}$. From equations~\eqref{eq:strCoreEng},~\eqref{eq:strCoreEngCond} and ~\eqref{eq:transformSeg}, $T_{\rm ext}^{(ij,kl)}$ is given by 
\begin{equation}
T_{\rm ext}^{(ij,kl)} = \hat{S}_{\alpha\beta}({\boldsymbol \xi}^{ij},{\bf b}^{ij};a)R_{\eta\alpha}({\boldsymbol \xi}^{ij},{\bf b}^{ij})R_{\omega\beta}({\boldsymbol \xi}^{ij},{\bf b}^{ij})  \int^{L_{ij}}_0{\epsilon}^{kl}_{\eta\omega}({\boldsymbol \xi}^{kl},{\bf b}^{kl},{\bf x}^{ij}(s_{ij}))\, {\rm d} s_{ij}.\label{eq:strEngExt1} 
\end{equation} 
The first order perturbations in $T_{\rm ext}^{(ij,kl)}$ with respect to ${\bf r}^{i}$ results from the following perturbations, given by
\begin{subequations}
\begin{flalign}
\begin{split}
\delta \hat{S}_{\alpha\beta}({\boldsymbol \xi}^{ij},{\bf b}^{ij};a)=&\frac{\partial \hat{S}_{\alpha\beta}({\boldsymbol \xi}^{ij},{\bf b}^{ij};a)}{\partial \xi_{p}^{ij}}
\frac{\partial \xi_{p}^{ij}}{\partial r^{i}_{q}}\delta r^{i}_{q} \\
=& \frac{\partial \hat{S}_{\alpha\beta}({\boldsymbol \xi}^{ij},{\bf b}^{ij};a)}{\partial \xi_{p}^{ij}}
\frac{\{ \xi_{p}^{ij}\xi_{q}^{ij}-\delta_{pq}\}}{L_{ij}}\delta r^{i}_{q}\\
=& -t^{\hat{S}}_{\alpha \beta q}\delta r^{i}_{q} \,, \label{eq:strSlopePert1}
\end{split}\\ 
\begin{split}
 \delta  \left( R_{\eta\alpha}({\boldsymbol \xi}^{ij},{\bf b}^{ij})R_{\omega\beta}({\boldsymbol \xi}^{ij},{\bf b}^{ij})\right)  
= & \left[ \frac{\partial R_{\eta\alpha}({\boldsymbol \xi}^{ij},{\bf b}^{ij})}{\partial \xi_{p}^{ij}} R_{\omega\beta}({\boldsymbol \xi}^{ij},{\bf b}^{ij}) +  R_{\eta\alpha}({\boldsymbol \xi}^{ij},{\bf b}^{ij})  \frac{\partial R_{\omega\beta}({\boldsymbol \xi}^{ij},{\bf b}^{ij})}{\partial \xi_{p}^{ij}} \right]\frac{\partial \xi_{p}^{ij}}{\partial r^{i}_{q}}\delta r^{i}_{q} \\
= & \left[  \frac{\partial R_{\eta\alpha}({\boldsymbol \xi}^{ij},{\bf b}^{ij})}{\partial \xi_{p}^{ij}} R_{\omega\beta}({\boldsymbol \xi}^{ij},{\bf b}^{ij}) +  R_{\eta\alpha}({\boldsymbol \xi}^{ij},{\bf b}^{ij})  \frac{\partial R_{\omega\beta}({\boldsymbol \xi}^{ij},{\bf b}^{ij})}{\partial \xi_{p}^{ij}} \right] \frac{\{ \xi_{p}^{ij}\xi_{q}^{ij}-\delta_{pq}\}}{L_{ij}}\delta r^{i}_{q}\\
=& -t^{R}_{\alpha \beta \omega \eta q} \delta r^{i}_{q} \,,  \label{eq:strRotPert1}
\end{split}\\ 
\begin{split}
\delta\left( \int^{L_{ij}}_0{\epsilon}^{kl}_{\eta\omega}({\boldsymbol \xi}^{kl},{\bf b}^{kl},{\bf x}^{ij}(s_{ij}))\, {\rm d} s_{ij} \right)= &
\left[ \int^{L_{ij}}_0 \frac{\partial {\epsilon}^{kl}_{\eta\omega}({\boldsymbol \xi}^{kl},{\bf b}^{kl},{\bf x}^{ij}(s_{ij}))}{\partial x^{ij}_q}  N_{i}(s_{ij}) \, {\rm d} s_{ij} \right.\\  & \left. - \frac{\xi^{ij}_q}{L_{ij}}\int^{L_{ij}}_0  {\epsilon}^{kl}_{\eta\omega}({\boldsymbol \xi}^{kl},{\bf b}^{kl},{\bf x}^{ij}(s_{ij})) \, {\rm d} s_{ij} \right]\delta r_q^{i} \\
=& -\left(t^{\epsilon_1}_{\eta \omega q} +t^{\epsilon_2}_{\eta \omega q}\right)\delta r^{i}_{q} \,. \label{eq:strStrainPert1}
\end{split}
\end{flalign}\label{eq:strExtPert1}
\end{subequations}
In the above, the first contributions results from the orientation dependence of the slopes of the core energy vs external macroscopic strain, originating from the difference in the slopes of the edge and screw dislocations. The second term results from the orientation dependence of the rotation matrix, which is solely geometric in nature. The third contribution is composed of two terms. The first term results from the perturbations to the spatial positions of the points on segment ${\bf l}^{ij}$, which then perturbs the external strain field experienced by these points. The second term captures the contribution arising from the change in the length of ${\bf l}^{ij}$. In the above perturbations, we used the fact that the position of segment ${\bf l}^{kl}$ remains unchanged with respect to perturbations of ${\bf r}^{i}$ owing to the cut-off procedure. Using equations~\eqref{eq:strExtPert1} and~\eqref{eq:strForceBreak}, we arrive at the force 
\begin{multline}
{f_{{\rm ce},q}^{(\{i\}j,kl)}} =t^{\hat{S}}_{\alpha \beta q} R_{\eta\alpha}({\boldsymbol \xi}^{ij},{\bf b}^{ij})R_{\omega\beta}({\boldsymbol \xi}^{ij},{\bf b}^{ij}) \int^{L_{ij}}_0{\epsilon}^{kl}_{\eta\omega}({\boldsymbol \xi}^{kl},{\bf b}^{kl},{\bf x}^{ij}(s_{ij}))\, {\rm d} s_{ij} \\
+ t^{R}_{\alpha \beta \omega \eta q} \hat{S}_{\alpha\beta}({\boldsymbol \xi}^{ij},{\bf b}^{ij};a)  \int^{L_{ij}}_0{\epsilon}^{kl}_{\eta\omega}({\boldsymbol \xi}^{kl},{\bf b}^{kl},{\bf x}^{ij}(s_{ij})) \, {\rm d} s_{ij}\\ 
+ t^{\epsilon_1}_{\eta \omega q} \hat{S}_{\alpha\beta}({\boldsymbol \xi}^{ij},{\bf b}^{ij};a)R_{\eta\alpha}({\boldsymbol \xi}^{ij},{\bf b}^{ij})R_{\omega\beta}({\boldsymbol \xi}^{ij},{\bf b}^{ij}) +t^{\epsilon_2}_{\eta \omega q} \hat{S}_{\alpha\beta}({\boldsymbol \xi}^{ij},{\bf b}^{ij};a)R_{\eta\alpha}({\boldsymbol \xi}^{ij},{\bf b}^{ij})R_{\omega\beta}({\boldsymbol \xi}^{ij},{\bf b}^{ij}).\label{eq:strExtForce1}
\end{multline}
 Next, we consider first order perturbations of $T_{\rm ext}^{(kl,ij)}$ with respect to ${\bf r}^{i}$. Using equations~\eqref{eq:strEngExt1} and ~\eqref{eq:contourStrain}, we have
\begin{align}
T_{\rm ext}^{(kl,ij)} =&  \hat{S}_{\alpha\beta}({\boldsymbol \xi}^{kl},{\bf b}^{kl};a)R_{\eta\alpha}({\boldsymbol \xi}^{kl},{\bf b}^{kl})R_{\omega\beta}({\boldsymbol \xi}^{kl},{\bf b}^{kl})  \int^{L_{kl}}_0{\epsilon}^{ij}_{\eta\omega}({\boldsymbol \xi}^{ij},{\bf b}^{ij},{\bf x}^{kl}(s_{kl})) \, {\rm d} s_{kl}\nonumber\\
=& \hat{S}_{\alpha\beta}({\boldsymbol \xi}^{kl},{\bf b}^{kl};a)R_{\eta\alpha}({\boldsymbol \xi}^{kl},{\bf b}^{kl})R_{\omega\beta}({\boldsymbol \xi}^{kl},{\bf b}^{kl})  \int^{L_{kl}}_0   \int^{L_{ij}}_0 {\bar{\epsilon}}_{\eta\omega}^{ij}({\boldsymbol \xi}^{ij},{\bf b}^{ij},{\bf x}^{kl}(s_{kl})-{\bf x}^{ij}(s_{ij})) \, {\rm d} s_{ij} {\rm d} s_{kl} \nonumber\\
= & D_{\eta \omega} \int^{L_{kl}}_0  \int^{L_{ij}}_0 {\bar{\epsilon}}_{\eta\omega}^{ij}({\boldsymbol \xi}^{ij},{\bf b}^{ij},{\bf x}^{kl}(s_{kl})-{\bf x}^{ij}(s_{ij})) \, {\rm d} s_{ij} {\rm d} s_{kl} \,,\label{eq:strEngExt2} 
\end{align} 
and
\begin{align}
\delta T_{\rm ext}^{(kl,ij)}=& D_{\eta \omega} \frac{\{ \xi_{p}^{ij}\xi_{q}^{ij}-\delta_{pq}\}}{L_{ij}} \int^{L_{kl}}_0  \int^{L_{ij}}_0 \frac{\partial{\bar{\epsilon}}_{\eta\omega}^{ij}({\boldsymbol \xi}^{ij},{\bf b}^{ij},{\bf x}^{kl}(s_{kl})-{\bf x}^{ij}(s_{ij}))}{\partial \xi^{ij}_p} \, {\rm d} s_{ij} {\rm d} s_{kl} \,\delta r^{i}_{q}\nonumber\\
& + D_{\eta \omega} \int^{L_{kl}}_0  \int^{L_{ij}}_0 \frac{\partial{\bar{\epsilon}}_{\eta\omega}^{ij}({\boldsymbol \xi}^{ij},{\bf b}^{ij},{\bf x}^{kl}(s_{kl})-{\bf x}^{ij}(s_{ij}))}{\partial x^{ij}_q} N_{i}(s_{ij}) \, {\rm d} s_{ij} {\rm d} s_{kl} \, \delta r^{i}_{q} \nonumber\\
&   - D_{\eta \omega} \frac{\xi_q^{ij}}{L_{ij}} \int^{L_{kl}}_0  \int^{L_{ij}}_0 {\bar{\epsilon}}_{\eta\omega}^{ij}({\boldsymbol \xi}^{ij},{\bf b}^{ij},{\bf x}^{kl}(s_{kl})-{\bf x}^{ij}(s_{ij}))\, {\rm d} s_{ij} {\rm d} s_{kl} \, \delta r^{i}_{q}\nonumber\\
=& - {\tilde{f}_{{\rm ce}, q}^{(\{i\}j,kl)}}  \delta r^{i}_{q} \,.\label{eq:strExtForce2} 
\end{align}
In the above, the first term results from the perturbation in the core energy of ${\bf l}^{kl}$ due to the perturbation in the strain field of segment ${\bf l}^{ij}$'s resulting from a change in its orientation. The second and the third term manifest from the perturbations in the strain field of segment ${\bf l}^{ij}$ associated with its position and length, respectively. In summary, equations~\eqref{eq:strSelfForce},~\eqref{eq:strExtForce1}, and ~\eqref{eq:strExtForce2} provide the various contributions to the nodal core force ${\bf F}^{i}_{\rm c}$ in equation~\eqref{eq:strForceBreak}. In ~\ref{sec:coreForcePartials}, we extend the above analysis to the case where the perfect dislocations are dissociated into partials.

The various terms in the nodal core force can broadly be divided into three types based on their dependence on the strain field. These three types are: (\rom{1}) terms which are proportional to the spatial gradients of the strain field, (\rom{2}) which are proportional to the strain field or its gradients with respect to the dislocation orientation, and (\rom{3}) which are independent of the strain field. Starting with type-(\rom{1}) contributions, we note that they arise from the core energy dependence on external strain and perturbations in the external strain due to perturbations in the displacement vector connecting the dislocation source (producing the external strain) to the core location (cf. third term in equation ~\eqref{eq:strExtForce1} and second term in equation ~\eqref{eq:strExtForce2}). 
Next, the type-(\rom{2}) contributions also arise from the core energy dependence on external strain, but, due to perturbations in the length and orientation of the dislocation segments (cf. first, second and fourth terms in equation~\eqref{eq:strExtForce1}, and first and third terms in equation~\eqref{eq:strExtForce2}). 
Finally, type-(\rom{3}) contributions (cf. equation~\eqref{eq:strSelfForce}) manifest from the perturbations of the core energy solely due to perturbation in the length and orientation of dislocation segments without accounting for any dependence of the core energy on the external strain. We note that the current DDD frameworks (cf. ~\cite{Arsenlis2007,Martinez2008}) only account for type-(\rom{3}) contributions to the core force. Further, these three different types of force contributions have different decay behavior in dislocation interactions. Type-(\rom{1}) contributions are proportional to the spatial gradient of the strain field, thus are short-ranged decaying as $\mathcal{O}(\frac{1}{d^{2}})$, where $d$ is the distance between the two interacting dislocation segments $ij \in U$ and $kl \in U^{\prime}_{ij}$. On the other hand, type-(\rom{2}) contributions are long-ranged decaying as $\mathcal{O}(\frac{1}{d})$ (the same decay of the elastic force), while type-(\rom{3}) contributions do not depend on $d$. Thus, we expect type-(\rom{1}) contributions to be significant in comparison with the elastic force at smaller distances. In Section~\ref{sec:caseStudies}, we use case studies to numerically compare the type-(\rom{1}) core force contributions with the elastic force for a wide range of dislocation interactions. Studies that also include non-trivial type-(\rom{2}) contributions will be considered in a future work.

\section{Case studies using core-energetics based forces}\label{sec:caseStudies}
In this section, we consider case studies involving pairs of simple dislocation structures in fcc Al to find the spatial extent to which the core force contribution (cf.~Section~\ref{sec:forceDerivStr}) is significant in comparison to the longer ranged Peach-Koehler force. In our case studies, we consider the systems as an aggregate of dislocation loops \footnote{Infinite straight dislocations can be treated as loops with radius approaching infinity.} only allowing for rigid body translations of the individual loops. In other words, we do not allow any changes in the shape, size and orientation of the dislocation loops. A full-fledged treatment without any restrictions on the degrees of freedom requires the efficient implementation of the core forces in a 3D DDD framework, which we will pursue in a future work. Further, in these case studies, we treat the dislocations as perfect dislocations for computing the non-singular elastic fields. This is a reasonable assumption here, as the combined elastic fields of the partials converges to that of the perfect dislocation elastic fields beyond the core size. 

We denote the $i$th loop in a system of $n$ loops as $C_{i}$.  Following Section~\ref{sec:forceDerivStr}, the core force on the dislocation loop $C_{i}$ (denoted as  ${\bf F}^{i}_{\rm c}$) corresponding to its rigid body translations, while keeping other loops fixed, is obtained as
\begin{align}
{\bf F}^{i}_{\rm c}= \oint_{C_i} {\bf f}_{\rm c}^i({\bf x}^{i}(s_{i}))\, {\rm d} s_{i}= \oint_{C_i} \left\{ \sum_{j \neq i} {\bf f}_{\rm ce}^{(i,j)}({\bf x}^{i}(s_{i})) +\sum_{j \neq i} \tilde{\bf f}_{\rm ce}^{(i,j)}({\bf x}^{i}(s_{i})) \right\} \,{\rm d} s_i \, ,\label{eq:caseCore}
\end{align}
where
\begin{gather}
f_{{\rm ce},q}^{(i,j)}({\bf x}^{i}(s_{i}))= -\hat{S}_{\alpha\beta}\left({\boldsymbol \xi}^{i}(s_{i}),{\bf b}^{i};a\right)R_{\eta\alpha}\left({\boldsymbol \xi}^{i}(s_{i}),{\bf b}^{i}\right)R_{\omega\beta}\left({\boldsymbol \xi}^{i}(s_{i}),{\bf b}^{i}\right) 
\oint_{C_j} \frac{\partial{\bar{\epsilon}}_{\eta\omega}^{j}({\boldsymbol \xi}^{j} (s_{j}),{\bf b}^{j},{\bf x}^{i}(s_{i})-{\bf x}^{j}(s_{j}))}{\partial x^{i}_q}
\,{\rm d}s_{j},\label{eq:caseCore1}
\end{gather}
and
\begin{gather}
\tilde{f}_{{\rm ce},q}^{(i,j)}({\bf x}^{i}(s_{i}))= - \oint_{C_j} \hat{S}_{\alpha\beta}\left({\boldsymbol \xi}^{j}(s_{j}),{\bf b}^{j};a\right)R_{\eta\alpha}\left({\boldsymbol \xi}^{j}(s_{j}),{\bf b}^{j}\right)R_{\omega\beta}\left({\boldsymbol \xi}^{j}(s_{j}),{\bf b}^{j}\right) 
\frac{\partial{\bar{\epsilon}}_{\eta\omega}^{i}({\boldsymbol \xi}^{i}(s_{i}),{\bf b}^{i},{\bf x}^{j}(s_{j})-{\bf x}^{i}(s_{i}))}{\partial x^{i}_q}\,{\rm d}s_{j} \,.\label{eq:caseCore2}
\end{gather}
In the above, ${\boldsymbol \xi}^{i}(s_{i})$ is the tangent line direction at ${\bf x}^{i}(s_{i})$ on $C_i$, ${\bf b}^{i}$ is the Burgers vector of $C_i$,  and $\bar{\boldsymbol \epsilon}^{i}({\boldsymbol \xi}^{i}(s_{i}),{\bf b}^{i},{\bf x}-{\bf x}^{i}(s_{i}))$ is the per unit-length strain field contribution at ${\bf x}$ due to the dislocation line at a point ${\bf x}^{i}(s_{i})$ on loop  $C_i$. Based on the classification of core force contributions in Section~\ref{sec:forceDerivStr}, the above core force corresponds to the type-(\rom{1}) contribution arising from pairwise interactions of loops $C_i$ with $C_j$. 
 
\begin{figure}[htbp]
\centering
\includegraphics[width=0.8\textwidth]{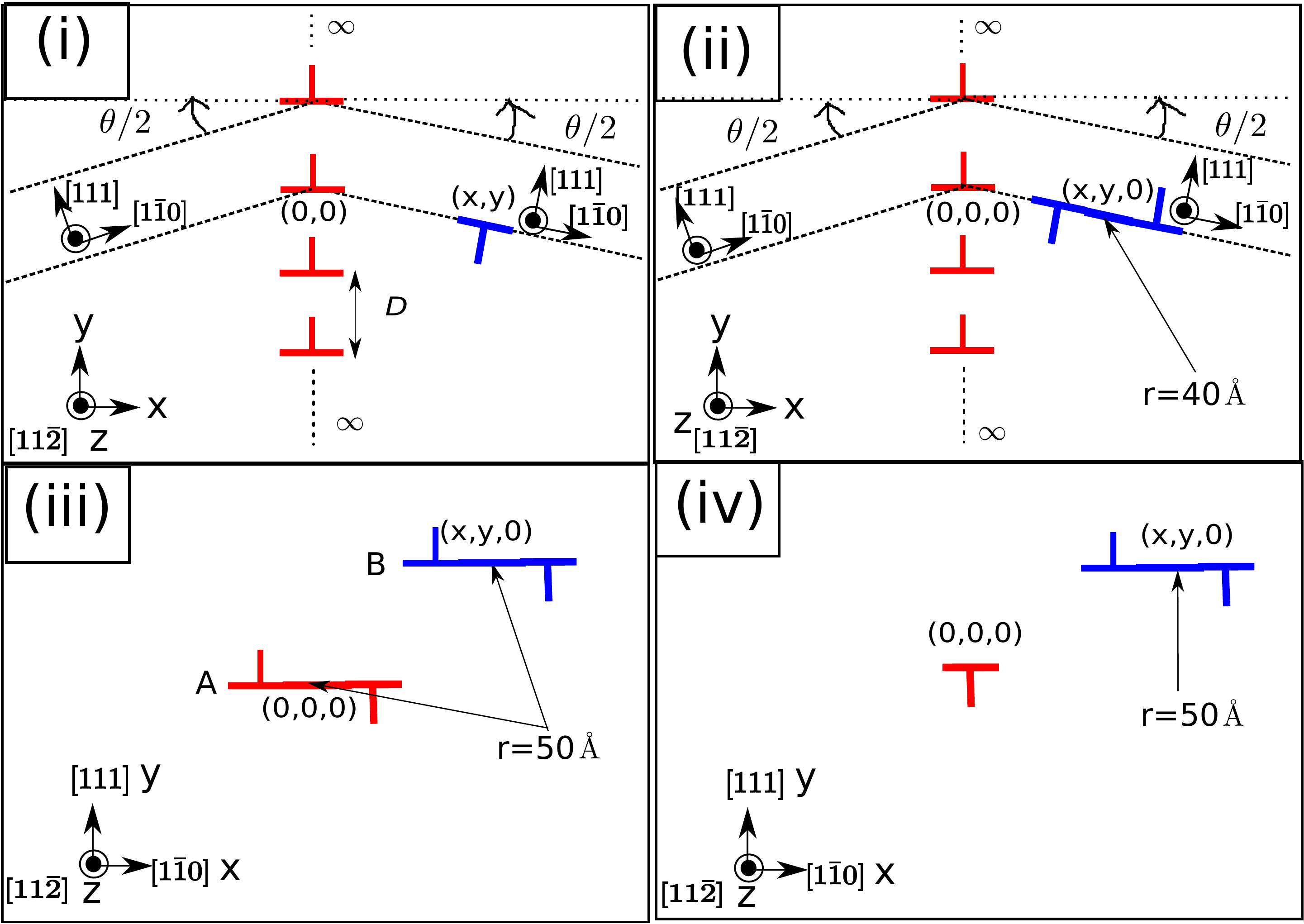}
\caption{\label{fig:caseStudySchematic} Schematic of the case studies.}
\end{figure}

We now undertake a numerical study on assessing the significance of the core force in comparison to the Peach-Koehler force. To this end, we consider the following case studies: (i) interaction of a straight edge dislocation with a low-angle tilt grain boundary, (ii) interaction of a circular glide loop with a low-angle tilt grain boundary (iii) interaction of two circular glide loops, and (iv) interaction of a circular glide dislocation loop with a straight edge dislocation. The schematics of the geometry are shown in figure~\ref{fig:caseStudySchematic}. The Burgers vectors of the dislocations are defined with respect to line directions taken as the following: along [$1$$1$$\bar{2}$] for straight edge dislocations, and for the glide loops, the line direction curls around the [111] direction in the right hand sense. In each of these case studies, the forces are evaluated with respect to translational perturbations of the blue-colored dislocation system, while keeping the red-colored dislocation system fixed. The values of the fcc Al material constants used in the calculations are obtained from RS-OFDFT, which are provided in Table~\ref{tab:matParam}. The value of the smearing parameter, $a$, in the non-singular elastic formulation is chosen to be $1|{\bf b}|$ for all the case studies. Corresponding to this choice of $a$, Table ~\ref{tab:coreDataPerf} provides the values of the non-elastic core energy and its slopes with respect to external macroscopic strains. 

\begin{figure}[htbp]
\centering
\subfigure[]{
\includegraphics[width=0.8\textwidth]{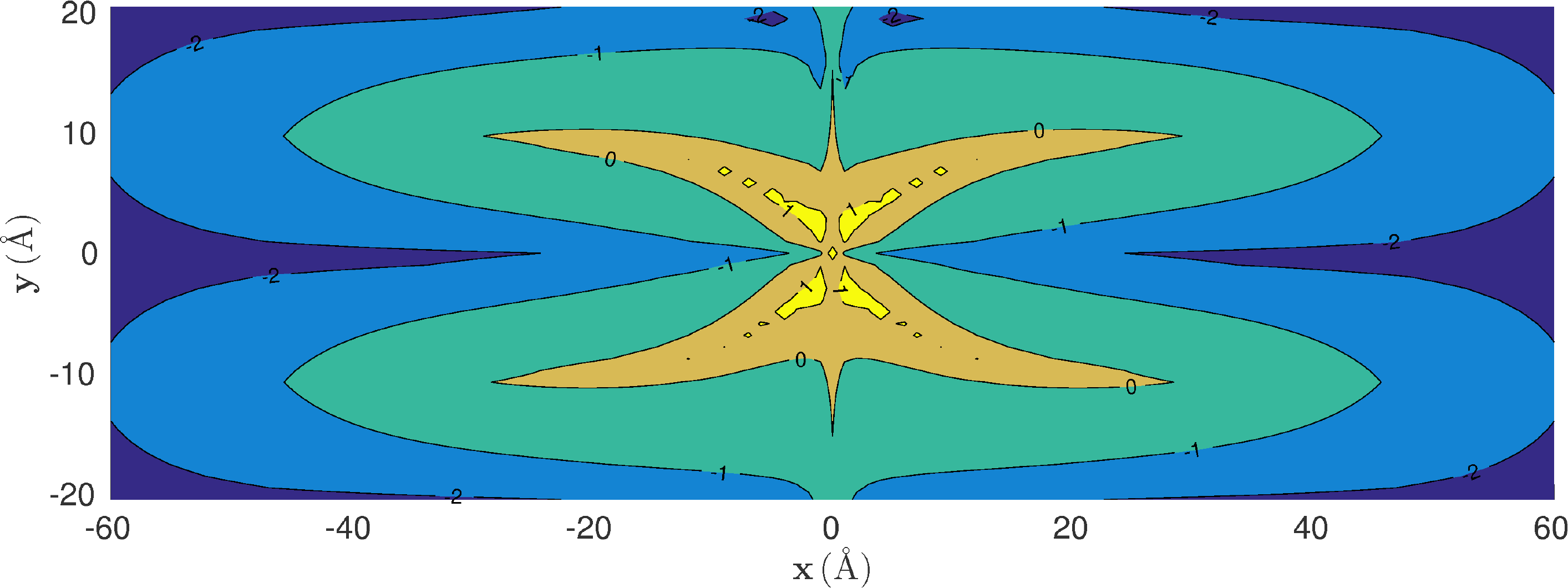}
\label{fig:GBa}}
\subfigure[]{
\includegraphics[width=0.8\textwidth]{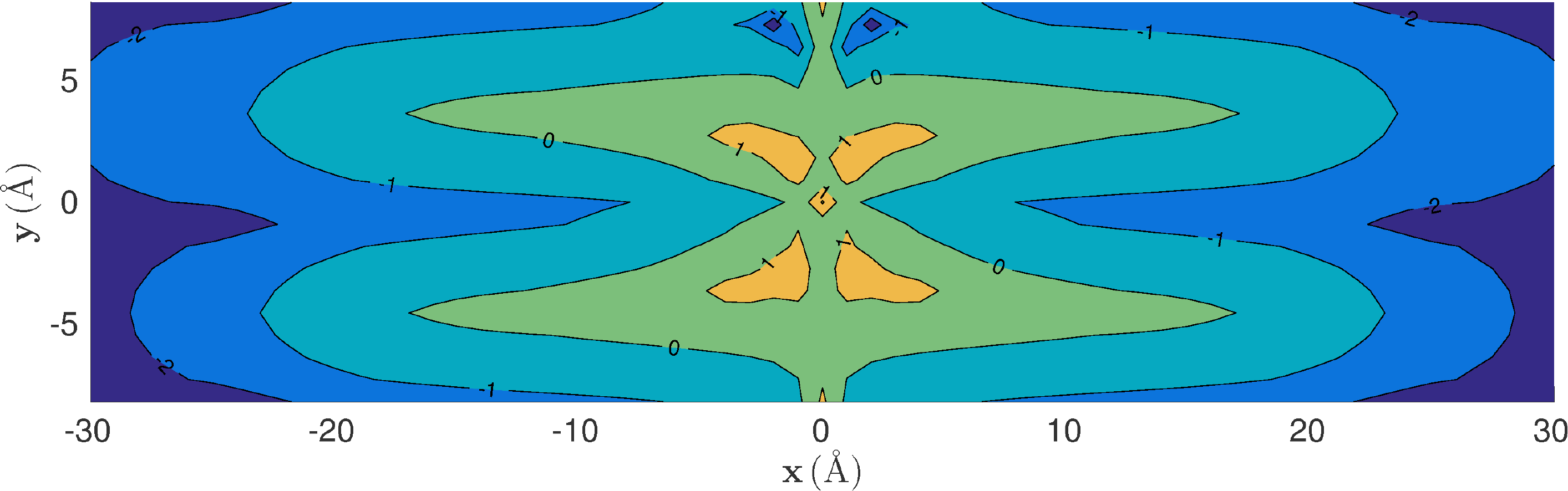}
\label{fig:GBb}}
\caption{Case study (i): Contour plot of ${\rm log}_{\rm 10}\left(R_{\rm str}(\{x,y\})\right)$ for the interaction between a straight edge dislocation and a low-angle title grain boundary for tilt angles a) $\theta$ = 4$^{\circ}$ b) $\theta$ = 10$^{\circ}$. The range of the $\rm{y}$-axis in these plots is $[-\frac{D}{2}, \frac{D}{2}]$.}
\end{figure}

{\it Case study} (i): 
We consider the interaction between a $\frac{a_{0}}{2}$[$1$$\bar{1}$$0$] low-angle tilt grain boundary (LATGB) with tilt axis [$1$$1$$\bar{2}$], and a straight negative edge dislocation with Burgers vector $\frac{-a_{0}}{2}$[$1$$\bar{1}$$0$]. LATGB's are conventionally modeled as an uniformly spaced array of edge dislocations up to tilt angles of $\approx$ 15$^{\circ}$~\citep{Read1950}). We denote the tilt angle as $\theta$,  as shown in figure~\ref{fig:caseStudySchematic}. The relationship between $\theta$ and the dislocation array spacing, denoted as $D$, is given by
\begin{equation}
D=\frac{|{\bf b}^{arr}|}{2\,{\rm sin}(\theta/2)}\, ,
\end{equation}
where ${\bf b}^{arr}=$ $\frac{a_{0}}{2}$[$1$$\bar{1}$$0$]  is the Burgers vector of each edge dislocation in the array. We consider two tilt angles of  $\theta= {\rm 4}^{\circ}$ and  $\theta={\rm 10}^{\circ}$. In both these systems, we evaluate the Peach-Koehler force and the core force per unit length on the straight edge dislocation. The quantity of interest to us is the ratio of glide component (along [$1$$\bar{1}$$0$] on the rotated frame) of the core force, ${F}_{\rm c}^{\rm str, gl}$, to the glide component of the Peach-Koehler force, ${F}_{\rm PK}^{\rm str, gl}$, with a regularization ($c_0$), given by 
\begin{equation}
R_{\rm str}(\{x,y\})=\left(\frac{ |{F}_{\rm c}^{\rm str, gl}(\{x,y\})|}{ |{F}_{\rm PK}^{\rm str, gl}(\{x,y\})| +c_0}\right)\,, \label{eq:ratioStr}
\end{equation}
where $\{x,y\}$ is the position of the straight edge dislocation in the un-rotated frame attached to the GB. The regularization is used to avoid singularities in the ratio at the points where ${F}_{\rm PK}^{\rm str, gl}$ vanishes. The value of $c_0$, a positive regularization constant, is chosen to be 
\begin{equation}
c_0=10 \times |{\rm PN}_{\rm f}|\,, 
\end{equation}
where ${\rm PN}_{\rm f}$ is the Peierls-Nabarro force per unit length of the straight edge dislocation computed using the Peierls stress to be 1.6 MPa~\citep{Shin2013}. Figures~\ref{fig:GBa} and ~\ref{fig:GBb} show the contour plots of ${\rm log}_{\rm 10}\left(R_{\rm str}(\{x,y\})\right)$ for tilt angles of  $\theta= {\rm 4}^{\circ}$ and  $\theta={\rm 10}^{\circ}$, respectively. For better presentability of the contour, we truncate the range of ${\rm log}_{\rm 10}\left(R_{\rm str}(\{x,y\})\right)$ to $[-3,3]$, the values to the left and right of this range being fixed at $-3$ and $3$, respectively. We also adopt this truncation procedure for the subsequent case studies. In the contour plots, the regions of interest are those with ${\rm log}_{\rm 10}\left(R_{\rm str}(\{x,y\})\right) > -1$, which from equation~\eqref{eq:ratioStr} correspond to
\begin{equation}
|{F}_{\rm c}^{\rm str, gl}(\{x,y\})| >  \frac{|{F}_{\rm PK}^{\rm str, gl}(\{x,y\})|}{10} \quad \& \quad  |{F}_{\rm c}^{\rm str, gl}(\{x,y\})| > |{\rm PN}_{\rm f}|.
\end{equation}
In the $x$ direction, these regions extend upto $x=4$ nm from the grain boundary in figure~\ref{fig:GBa}, and $x=2$ nm from the grain boundary in figure~\ref{fig:GBb}. In the $y$ direction, these regions almost completely fills up the separation between the dislocations in the GB. There are also areas inside these regions with separation distances of $<$ 2 nm, where ${\rm log}_{\rm 10}\left(R_{\rm str}(\{x,y\})\right) > 0$.  Here, the core force is greater in magnitude compared to the Peach-Koehler force, and is also greater than ten times the Peierls-Nabarro force. Thus, we can expect the core force to influence the physical processes involved in dislocation--GB interactions like dislocation pile-up, dislocation absorption and dislocation transmission that play an important role in governing mechanical properties of polycrystalline materials. 

\begin{figure}[htbp]
\centering
\subfigure[]{
\includegraphics[width=0.8\textwidth]{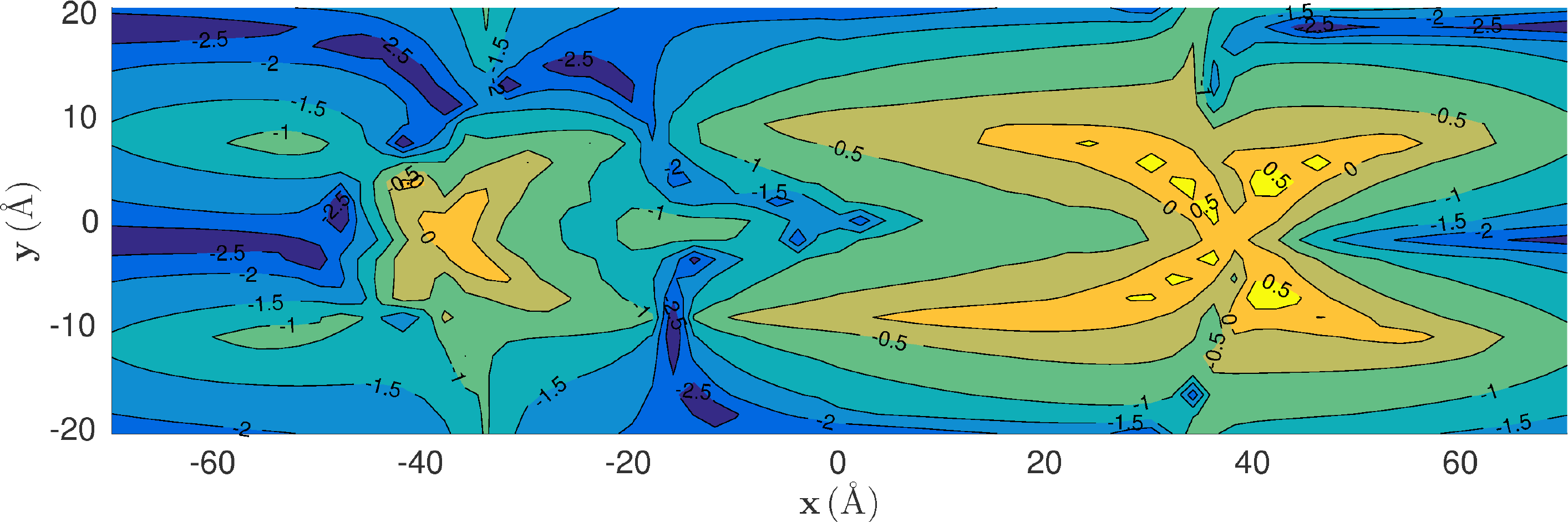}
\label{fig:GB2a}}
\subfigure[]{
\includegraphics[width=0.8\textwidth]{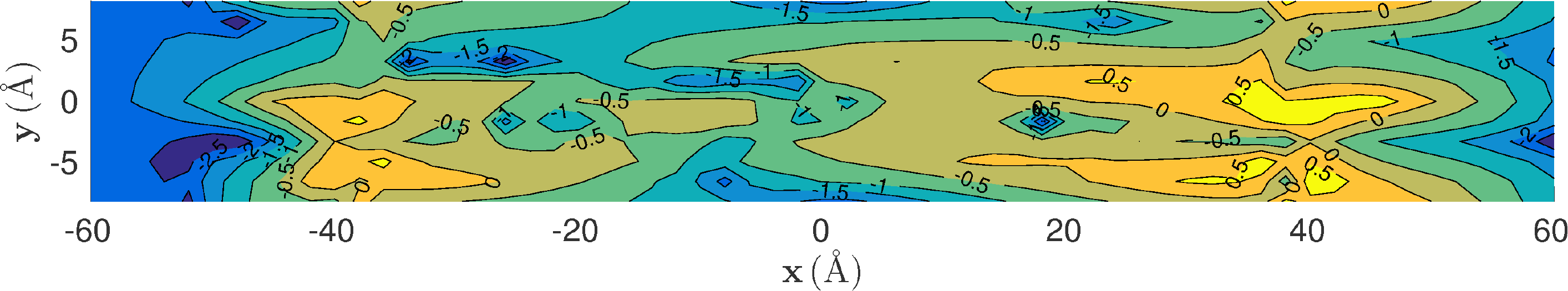}
\label{fig:GB2b}}
\caption{Case study (ii): Contour plot of ${\rm log}_{\rm 10}\left(R_{\rm loop}(\{x,y\})\right)$ for the interaction between a glide loop and a low-angle tilt grain boundary for tilt angles a) $\theta$ = 4$^{\circ}$ b) $\theta$ = 10$^{\circ}$.  The range of the $\rm{y}$-axis in these plots is $[-\frac{D}{2}, \frac{D}{2}]$.}
\end{figure}

{\it Case study} (ii): In this case study, we consider the interaction  between a $\frac{a_{0}}{2}$[$1$$\bar{1}$$0$] LATGB with tilt axis [$1$$1$$\bar{2}$], and a circular (111)  glide loop of radius 40 $\text{\AA}$  with Burgers vector $\frac{-a_{0}}{2}$[$1$$\bar{1}$$0$]. We evaluate the total Peach-Koehler force and core force on the dislocation loop for two different values of LATGB tilt angles of $\theta= {\rm 4}^{\circ}$ and  $\theta={\rm 10}^{\circ}$. As in the previous case study, we consider the regularized ratio of glide component of the core force on the loop, ${F}_{\rm c}^{\rm loop, gl}$, to the glide component of the Peach-Koehler force on the loop, ${F}_{\rm PK}^{\rm loop, gl}$, given by
\begin{equation}
R_{\rm loop}(\{x,y\})=\left(\frac{ |{ F}_{\rm c}^{\rm loop, gl}(\{x,y\})|}{ |{F}_{\rm PK}^{\rm loop, gl}(\{x,y\})| +c_0}\right), \label{eq:ratioLoop}
\end{equation}
where
\begin{equation}
c_0=10 \times {\rm Per}_{\rm loop} \times |{\rm PN}_{\rm f}|\,, \label{eq:loopReg}
\end{equation} 
with ${\rm Per}_{\rm loop}$ denoting the perimeter of the circular glide loop.  Figures~\ref{fig:GB2a} and~\ref{fig:GB2b} show the contour plots for tilt angles of $\theta= {\rm 4}^{\circ}$ and  $\theta={\rm 10}^{\circ}$, respectively. The regions with ${\rm log}_{\rm 10}\left(R_{\rm loop}(\{x,y\})\right) > -1$ extend up to distances of 4--6 nm between the loop center and the LATGB.

\begin{figure}[htbp]
\centering
\includegraphics[width=0.5\textwidth]{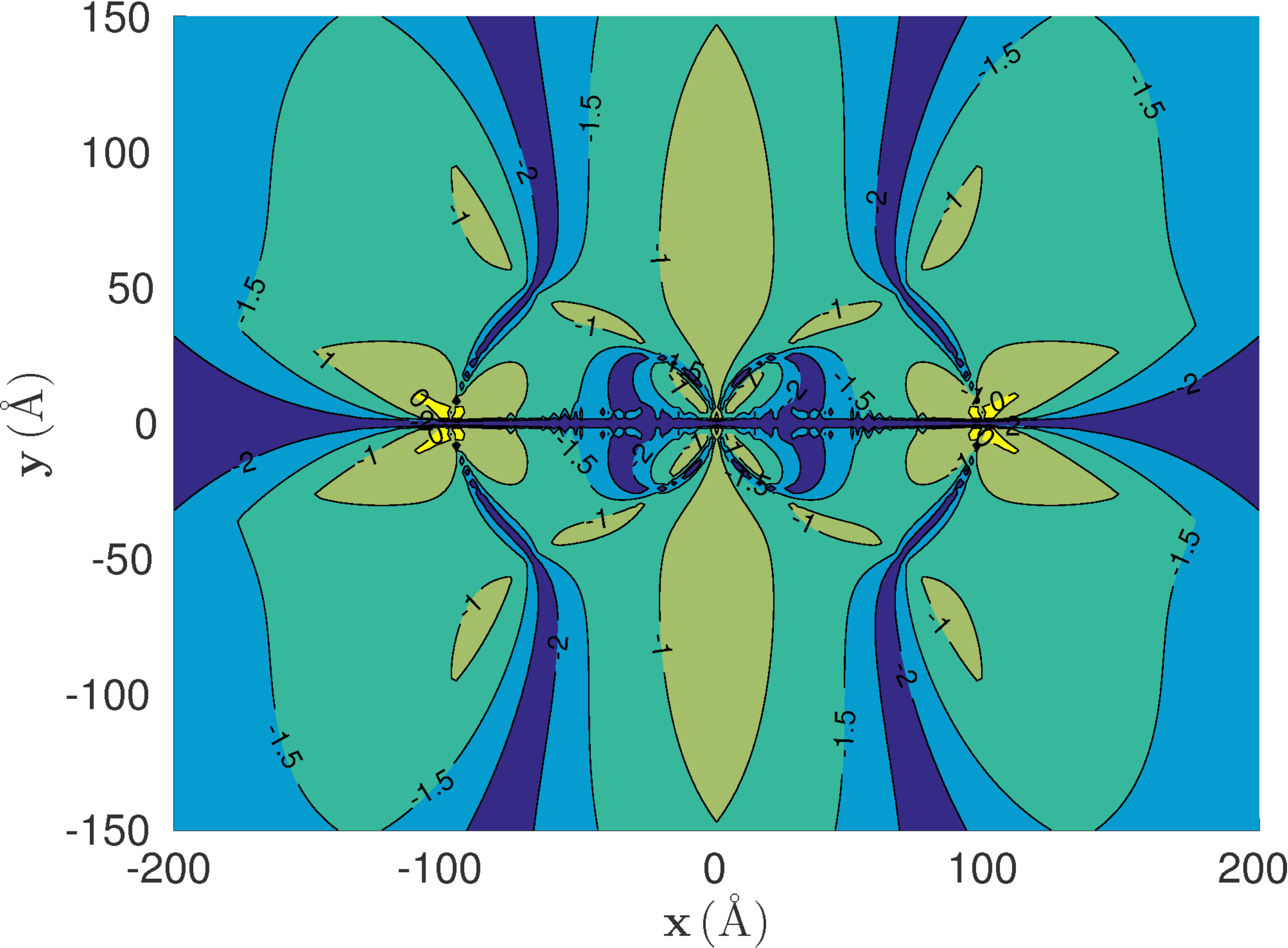}
\caption{\label{fig:loopLoop}Case study (iii): Contour plot of ${\rm log}_{\rm 10}\left(R_{\rm loop}(\{x,y\})\right)$ for the interaction between two glide loops.}
\end{figure}

{\it Case study} (iii): 
Here we consider the interaction between two circular (111) glide loops of radius 50 $\text{\AA}$ with equal Burgers vector $\frac{a_{0}}{2}$[$1$$\bar{1}$$0$]. The center of the loop A is fixed at the origin $\{0,0,0\}$, while loop B's center is located at a variable position $\{x,y,0\}$. We compute ${F}_{\rm c}^{\rm loop, gl}$ and ${F}_{\rm PK}^{\rm loop, gl}$ on glide loop B, and obtain the contour plot of ${\rm log}_{\rm 10}\left(R_{\rm loop}(\{x,y\})\right)$ using equations~\eqref{eq:ratioLoop} and ~\eqref{eq:loopReg}, which is shown in figure ~\ref{fig:loopLoop}. We observe that the regions with ${\rm log}_{\rm 10}\left(R_{\rm loop}(\{x,y\})\right) > -1$ extend up to very significant distances of $\approx$ 15 nm between the loop centers.

\begin{figure}[htbp]
\centering
{\includegraphics[width=0.5\textwidth]{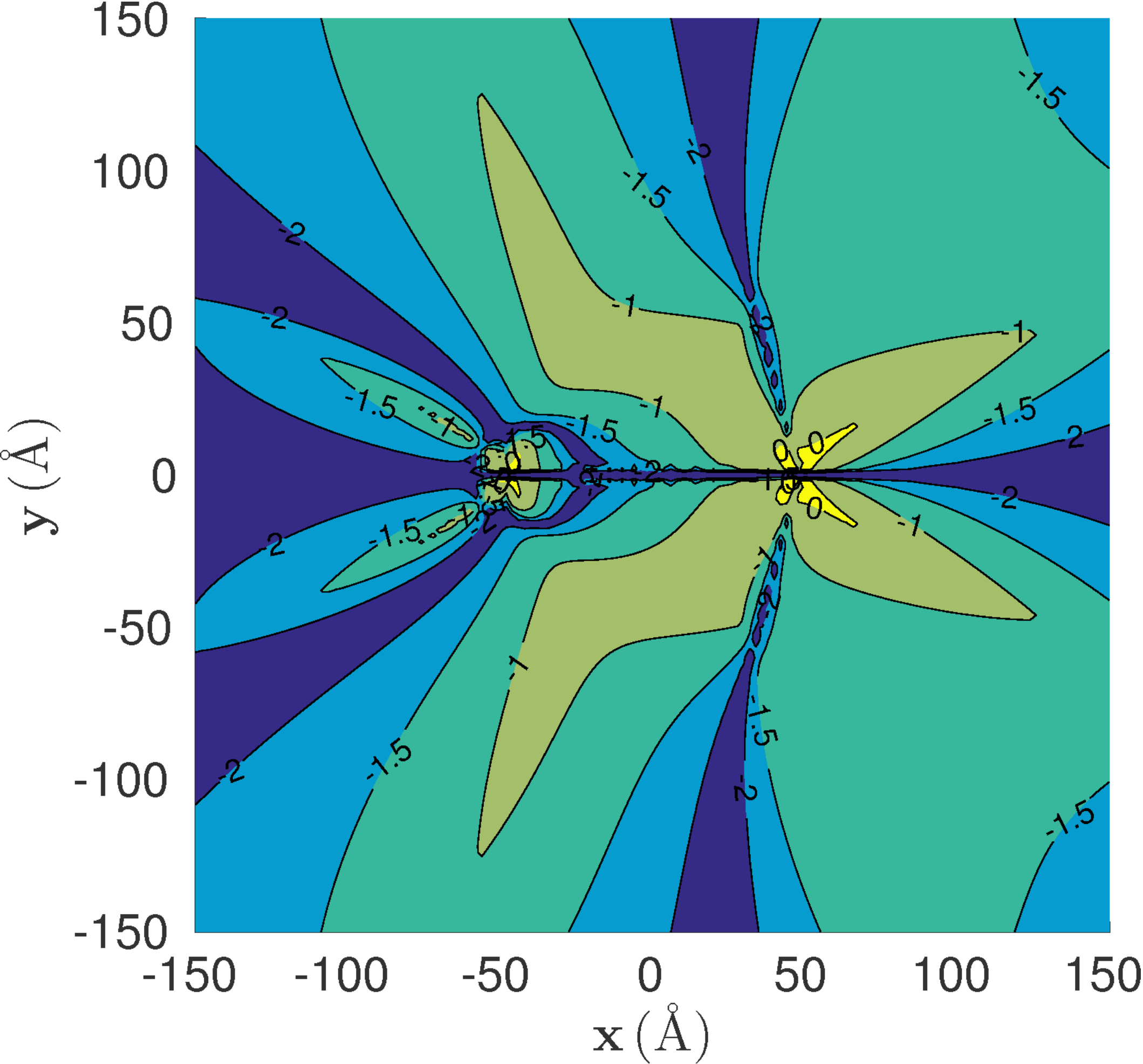}}
\caption{\label{fig:loopStr}Case study (iv): Contour plot of ${\rm log}_{\rm 10}\left(R_{\rm loop}(\{x,y\})\right)$ for the interaction between a glide loop and a straight edge dislocation.}
\end{figure}

{\it Case study} (iv): 
In the final case study, we consider the interaction between a circular (111) glide loop of radius 50 $\text{\AA}$ with Burgers vector $\frac{a_{0}}{2}$[$1$$\bar{1}$$0$], and a straight negative edge dislocation with Burgers vector $\frac{-a_{0}}{2}$[$1$$\bar{1}$$0$]. The straight edge dislocation is fixed at the origin while the glide loop's center has a variable position, $\{x,y,0\}$. We compute ${F}_{\rm c}^{\rm loop, gl}$ and ${F}_{\rm PK}^{\rm loop, gl}$ on the glide loop. Figure ~\ref{fig:loopStr} shows the contour plot of ${\rm log}_{\rm 10}\left(R_{\rm loop}(\{x,y\})\right)$. We observe that core forces are considerable up to distances of $\approx$ 10 nm between the glide loop center and the straight edge dislocation.

\section{Conclusions}\label{sec:Conclusions}
In summary, there are two parts to this study. In the first part of this study, we studied the core structure and core energetics of an isolated screw dislocation in Aluminum using a real-space formulation of orbital-free DFT with finite-element discretization (RS-OFDFT). This study is complementary to an earlier RS-OFDFT study by ~\cite{Das2016,Iyer2015} on an isolated edge dislocation in Aluminum. In order to directly compute the core energetics, we employed mixed boundary conditions corresponding to an isolated dislocation embedded in the bulk---i.e., periodic boundary conditions along the dislocation line and bulk Dirichlet boundary conditions on the electronic fields obtained from the Cauchy-Born hypothesis along the boundary on the other two directions for fixed atomic positions on the boundary that are determined based on the elastic field of the dislocation. The local real-space formulation and the finite-element discretization are essential to realizing these boundary conditions, which are not accessible using the widely employed plane-wave discretization in electronic structure calculations. We computed the dislocation energies of a perfect screw dislocation for a series of increasing domain sizes, and identified the region up to which the perturbations in the electronic structure are significant to the dislocation energetics. This allowed us to unambiguously characterize the core size, where the physics cannot be described using a homogenized continuum theory. We estimate, from an energetic viewpoint, the core size of the perfect screw dislocation in Aluminum to be $\approx$ 7 $|{\bf b}|$. This core size corresponds well with the estimated core size of $\approx$ 10 $|{\bf b}|$ for the perfect edge dislocation in Aluminum~\citep{Iyer2015}. However, these core sizes are much larger than the conventional estimates of  1--3 $|{\bf b}|$ based on displacement fields. Upon ionic relaxation, the perfect screw dislocation dissociated into two Shockley partials with partial separation distances of 8.2 \AA~and 6.6 \AA~ measured from the screw and edge component differential displacement plots, respectively. These partial separation distances compare closely with other estimates from prior electronic structure studies. We also computed the core energy under externally applied affine volumetric, uniaxial and shear deformations, and found that, in general, the core energy was strongly dependent on the macroscopic deformations with non-zero slopes at zero deformation. Similar observations were reported for the core energetics of edge dislocation in~\cite{Das2016,Iyer2015}. This suggests that the dependence of the core energy on macroscopic deformations is a fundamental characteristic of dislocation energetics. In addition, we also computed the dependence of the core energy on macroscopic deformations using atomistic calculations, and found that the values of the slopes at zero deformation not only differ considerably from the RS-OFDFT results, but that they are also widely sensitive to the choice of the interatomic potential. This underscores the need for accurate large-scale electronic structure calculations to compute these properties. 

In the second part of this study, we developed an energetics model for a dislocation aggregate, denoted by $C$, in an isotropic infinite elastic continua, which incorporates the dependence of the core energetics on macroscopic deformations. The underlying elastic model is chosen to be the non-singular elastic formulation by~\cite{Cai2006}. Under the physically reasonable assumption that the core size is smaller than the structural length-scales of the dislocation aggregate, the core energetics of isolated edge and screw dislocations are used to describe the total core energy, which is expressed as a line integral on $C$. The core energy per unit length at each point on $C$ depends on the external strain field at the point resulting from all parts of $C$ excepting a chosen cut-off region around the point. Further, we subtract the non-singular elastic effects from the total core energy to obtain the non-elastic core energy that is dependent on external strain. Next, we extended the developed energetics model to a nodal dislocation network, and derived the nodal force associated with the non-elastic core energy contribution, which we referred to as the nodal core force. Finally, we considered case studies to compare the magnitude of the Peach-Koehler force with the magnitude of the core force contribution arising from the macroscopic deformation dependence of the core energy. These case studies involved interactions of grain boundary-straight dislocation, grain boundary-glide loop, glide loop-glide loop and glide loop-straight dislocation in fcc Aluminum. Numerically computing the Peach-Koehler and core force along the slip direction in these case studies, we found that even up to distances of 10-15 nm between dislocations, the magnitude of the core force is significant with respect to the Peach-Koehler force (being at least 10\% of the Peach-Koehler force), while also being larger than the Peierls-Nabarro force. Furthermore, for some configurations with distances of $< 2$ nm, the magnitude of the core force was found to be comparable or more than the elastic force.

Based on the analysis of the core force expressions and the results of the case studies, we anticipate that the core force may influence key elemental mechanisms of dislocation enabled hardening, such as structure and strength of various dislocation junctions, critical passing stress around obstacles etc. For such studies of elemental mechanisms, the nodal core force expressions developed here can be readily incorporated into DDD implementations. However, extending this to large-scale bulk plasticity simulations using DDD codes like ParaDis requires a computationally efficient implementation of the nodal core force terms in these frameworks. In this regard, the asymptotic behavior of various terms contributing to the core force can be exploited, along with obtaining analytical expressions, which will be pursued in a subsequent work.  Obtaining the dislocation core energetics for material systems where orbital free DFT kinetic energy functionals are not well developed or are not sufficiently accurate, such as transition metals and covalently bonded systems, requires the use of Kohn-Sham DFT for studying such systems. While much progress has been made in developing large-scale real-space methods for Kohn-Sham DFT with reduced order-scaling~\citep{Motamarri2014, Ponga2016,SuryanarayanaJMPS2013} to tackle mechanics related problems, further development is necessary to realize the bulk Dirichlet boundary conditions that are needed to obtain the core energetics of an isolated dislocation in the bulk, and this forms an important future direction. Further, extending the Kohn-Sham studies, often conducted under the pseudopotential approximation, to all-electron Kohn-Sham DFT calculations will provide useful data to assess the accuracy of pseudopotentials in describing the dislocation cores, and this presents another worthwhile direction for future studies. We anticipate that the recent methodological developments towards large-scale real-space all-electron Kohn-Sham DFT calculations~\citep{Bikash2016,MGBO2016} will be useful in these efforts.

\section*{Acknowledgments}
We gratefully acknowledge the support from the U.S. Department of Energy, Office of Basic Energy Sciences, Division of Materials Science and Engineering under Award No. DE-SC0008637 that funds the Predictive Integrated Structural Materials Science (PRISMS) center at University of Michigan, under the auspices of which this work was performed. This work used the Extreme Science and Engineering Discovery Environment (XSEDE), which is supported by National Science Foundation grant number ACI-1053575. This research used resources of the National Energy Research Scientific Computing Center, a DOE Office of Science User Facility  supported by the Office of Science of the U.S. Department of Energy under Contract No. DE-AC02-05CH11231. We also acknowledge Advanced Research Computing at University of Michigan for providing additional computing resources through the Flux computing platform. 

\clearpage

\appendix
\section{Computation of non-elastic core energy from RS-OFDFT dislocation core energetics data}\label{sec:corePostProc}
Here, we discuss the post-processing of the core-energetics data from RS-OFDFT calculations to estimate the non-elastic core energy of the dislocation aggregate, $E_{\rm cnel}^{C}$. The non-elastic core energy is obtained by subtracting from the core energy of a dislocation aggregate, $E_{\rm c}^{C}$, the elastic energy of the dislocation (cf.~equation~\eqref{eq:coreEnergyPartition}). In the model presented in section~\ref{sec:generalModel}, as the core energy of a dislocation aggregate is computed in terms of the core energies of straight edge and screw dislocations, $E_{\rm c}^{\rm edge/screw}({\boldsymbol \epsilon}^{\rm ext})$ (cf. equation~\eqref{eq:mixedCore}), it is sufficient to compute the non-elastic core energies of straight edge and screw dislocations---$E_{\rm cnel}^{\rm edge/screw}({\boldsymbol \epsilon}^{\rm ext})$.  
 
To this end, we consider the domain corresponding to the dislocation core, denoted by $\Omega_{\rm c}$, as determined from RS-OFDFT calculations---10 $|{\bf b}|$ for the edge dislocation~\citep{Iyer2015} and 8.7 $|{\bf b}|$ for the screw dislocation (cf. section~\ref{sec:screw}) in Aluminum. The coordinate axes 1---2---3 are aligned such that the axis labelled `1' lies on the slip plane and is perpendicular to the line direction, axis labelled `2' is perpendicular to the slip plane, and the axis labelled `3' is along the line direction. In the present work, as we restrict our model and study to a dislocation aggregate of perfect dislocations, though an extension to consider dissociated partials is possible following the lines of~\citet{Martinez2008}, we subtract the non-singular elastic contribution of perfect dislocation from the RS-OFDFT core energy data that corresponds to relaxed Shockley partials. Thus, the energetics associated with core structure relaxation are present in the non-elastic core energy. Further, in the context of DDD calculations, the average distance between dislocations is much larger compared to the core size, beyond which the combined elastic fields of the Shockley partials converges to the elastic field of the perfect dislocation. We now consider the non-singular elastic energy of isolated straight edge and screw dislocations under an external homogeneous strain, ${\boldsymbol \epsilon}^{\rm ext}$, using a choice for  $a$ and the isotropic elastic constants for Al computed from RS-OFDFT (cf. Table~\ref{tab:matParam}). The non-singular elastic energy per unit length of dislocation line in the domain $\Omega_{\rm c}$ is given by 
\begin{align}
E_{\rm cel}^{\rm edge/screw}({\boldsymbol \epsilon}^{\rm ext};a) = & \frac{1}{2} \int_{\Omega_{\rm c}} \big( {\boldsymbol \sigma^{\rm d}}({\bf x}_{\rm c},a) + {\boldsymbol \sigma^{\rm ext}} \big)  : \big( {\boldsymbol \epsilon^{\rm d}}({\bf x}_{\rm c},a) + {\boldsymbol \epsilon^{\rm ext}} \big) \, {\rm d}A \nonumber \\
= & \frac{1}{2}\int_{\Omega_{\rm c}}  {\boldsymbol \sigma^{\rm d}}({\bf x}_{\rm c},a) : {\boldsymbol \epsilon^{\rm d}}({\bf x}_{\rm c},a)\, {\rm d}A 
+ \frac{1}{2}\int_{\Omega_{\rm c}}  {\boldsymbol \sigma^{\rm ext}} : {\boldsymbol \epsilon^{\rm ext}} \, {\rm d}A 
+ \int_{\Omega_{\rm c}}  {\boldsymbol \sigma^{\rm d}}({\bf x}_{\rm c},a) : {\boldsymbol \epsilon^{\rm ext}}\, {\rm d}A \,,\label{eq:cel1}
\end{align}
where ${\boldsymbol \sigma}^{\rm d}({\bf x}_{\rm c},a)$ and ${\boldsymbol \epsilon}^{\rm d}({\bf x}_{\rm c},a)$  are the non-singular stress and strain fields, respectively, of the straight edge or screw dislocation at any point ${\bf x}_{\rm c} \in \Omega_{\rm c}$, and ${\bf x}_{\rm c}$ is measured with respect to the dislocation line. In the above, as we are interested in the regime of small external homogenous strains, we assume that ${\boldsymbol \sigma}^{\rm d}$ and ${\boldsymbol \epsilon}^{\rm d}$ are independent of ${\boldsymbol \epsilon^{\rm ext}}$.   

Next, we turn to the core energies of the isolated straight edge and screw dislocations, and the estimation of the corresponding non-elastic core energies. While the core energy, $E_{\rm c}^{\rm edge/screw}({\boldsymbol \epsilon}^{\rm ext})$, is defined as the total energy inside $\Omega_{\rm c}$ measured with respect to the unstrained perfect crystal, the core-energetics data from RS-OFDFT calculations, presented in section~\ref{sec:screw}, measure the excess energy with respect to a perfect crystal undergoing the same external homogenous strain (cf. equation~\eqref{eq:formationEngAffine}). Thus, $E_{\rm c}^{\rm edge/screw}({\boldsymbol \epsilon}^{\rm ext})$ is related to the RS-OFDFT core energetics data, which we denote by $E_{\rm c, \, data}^{\rm edge/screw}({\boldsymbol \epsilon}^{\rm ext})$, as
\begin{equation}
 E_{\rm c}^{\rm edge/screw}({\boldsymbol \epsilon}^{\rm ext})=E_{\rm c, data}^{\rm edge/screw}({\boldsymbol \epsilon}^{\rm ext}) + \frac{1}{2}\int_{\Omega_{\rm c}}  {\boldsymbol \sigma^{\rm ext}} : {\boldsymbol \epsilon^{\rm ext}} \, {\rm d}A  \,. \label{eq:data1}
\end{equation}      
The non-elastic core energy, following the partitioning in equation~\eqref{eq:coreEnergyPartition}, is given by 
\begin{equation}
 E_{\rm cnel}^{\rm edge/screw}({\boldsymbol \epsilon}^{\rm ext}) = E_{\rm c}^{\rm edge/screw}({\boldsymbol \epsilon}^{\rm ext}) - E_{\rm cel}^{\rm edge/screw}({\boldsymbol \epsilon}^{\rm ext})\,. \label{eq:APartition}
\end{equation}
In the regime of small external homogeneous strains, linearizing the above equation about ${\boldsymbol \epsilon}^{\rm ext}=0$  (cf. equations~\eqref{eq:coreEngNetworkLinear} - ~\eqref{eq:coreEngSlope}), we obtain
\begin{align}
E_{\rm cnel}^{\rm edge/screw}({\boldsymbol \epsilon}^{\rm ext};a) \, \approx \,\,& E_{\rm cnel}^{\rm edge/screw}({\boldsymbol \epsilon}^{\rm ext}={\bf 0};a) + \hat{S}^{\rm edge/screw}_{\alpha\beta}(a){\epsilon}^{\rm ext}_{\alpha\beta} \nonumber \\
= & \left( E_{\rm c}^{\rm edge/screw}({\boldsymbol \epsilon}^{\rm ext}={\bf 0})  - \frac{1}{2}\int_{\Omega_{\rm c}}  {\boldsymbol \sigma^{\rm d}}({\bf x}_{\rm c},a) : {\boldsymbol \epsilon^{\rm d}}({\bf x}_{\rm c},a)\, {\rm d}A \right) \nonumber \\
& + \left(S^{\rm edge/screw}_{\alpha\beta}-\int_{\Omega_{\rm c}}   \sigma^{\rm d}_{\alpha\beta}({\bf x}_{\rm c},a) \,{\rm d}A\right){\epsilon}^{\rm ext}_{\alpha\beta} \,. \label{eq:Aextract}
\end{align}  
In the above, $ \hat{S}^{\rm edge/screw}_{\alpha\beta} = \frac{\partial E_{\rm cnel}^{\rm edge/screw}}{\partial {\epsilon}^{\rm ext}_{\alpha\beta}}\Big|_{{\boldsymbol \epsilon}^{\rm ext}=\bf{0}}$ and $S^{\rm edge/screw}_{\alpha\beta}= \frac{\partial E_{\rm c}^{\rm edge/screw}}{\partial {\epsilon}^{\rm ext}_{\alpha\beta}}\Big|_{{\boldsymbol \epsilon}^{\rm ext}=\bf{0}}$. We note that, from equation~\eqref{eq:data1}, $E_{\rm c}^{\rm edge/screw}({\boldsymbol \epsilon}^{\rm ext}={\bf 0}) = E_{\rm c, data}^{\rm edge/screw}({\boldsymbol \epsilon}^{\rm ext}={\bf 0})$, and, further, $S^{\rm edge/screw}_{\alpha\beta}=\frac{\partial E_{\rm c, data}^{\rm edge/screw}}{\partial {\epsilon}^{\rm ext}_{\alpha\beta}}\Big|_{{\boldsymbol \epsilon}^{\rm ext}=\bf{0}}$. Thus, the values of ${\bf S}^{\rm edge/screw}$ are directly obtained from the RS-OFDFT calculations in this work for a screw dislocation in Section~\ref{sec:macroDeform}, and for an edge dislocation from ~\citet{Das2016, Iyer2015}. Although we lack the core energetics data for the shear strains causing the perfect edge or screw dislocations to glide, from symmetry, the slopes corresponding to these glide shear strains, $S^{\rm edge}_{12}$ and $S^{\rm screw}_{23}$, should be zero. Table~\ref{tab:coreDataTotal} provides the data for $E_{\rm c}^{\rm edge/screw}({\boldsymbol \epsilon}^{\rm ext}={\bf 0})$ and ${\bf S}^{\rm edge/screw}$. Table~\ref{tab:coreDataPerf} provides the corresponding non-elastic core energetics data, $E_{\rm cnel}^{\rm edge/screw}({\boldsymbol \epsilon}^{\rm ext}={\bf 0})$ and $\hat{\bf S}^{\rm edge/screw}$, computed based on equation~\eqref{eq:Aextract} using $a=1|{\bf b}|$. We note that though the non-elastic core energy at zero strain has a small value, the values of the slopes are considerable. Furthermore, we note that the values of $\hat{\bf S}^{\rm edge/screw}$ and ${\bf S}^{\rm edge/screw}$ are equal, which follows from equation~\eqref{eq:Aextract} as 
\begin{equation}
\int_{\Omega_{\rm c}}   {\boldsymbol \sigma}^{\rm d}({\bf x}_{\rm c},a) \,{\rm d}A ={\bf 0} \,, \label{eq:nelslope}
\end{equation}
which in turn is readily observed from the anti-symmetries in the analytical expressions for the stress fields of the perfect edge and screw dislocations~\citep{Cai2006}. Finally, in Table~\ref{tab:coreDataAtomistics}, we have also provided values of ${\bf S}^{\rm edge/screw}$ computed using atomistic calculations. Comparing with Table~\ref{tab:coreDataTotal}, we observe that they vary widely based on the choice of the EAM potential, which underscores the need for electronic structure calculations to compute these quantities. 

\begin{table}
\begin{center}
\caption{\label{tab:matParam}\small{Material parameters of fcc Al computed using RS-OFDFT with Wang-Govind-Carter (WGC) kinetic energy functional~\citep{Wang1999}, local density approximation (LDA) for the exchange-correlation energy~\citep{Perdew1981}, and Goodwin-Needs-Heine pseudopotential~\citep{Goodwin1990}. The isotropic elastics constants, $\mu$ and $\nu$ are computed from the fcc cubic elastic constants using Voigt average \citep{Hirth1982}.}}
\begin{tabular}{ | m{15em} | m{1cm}|} 
\hline
Cell relaxed lattice constant (Bohr) & 7.63  \\ 
Isotropic shear modulus, $\mu$  (GPa) & 22  \\ 
Isotropic Poisson's ratio, $\nu$ & 0.35  \\ 
\hline 
\end{tabular}
\end{center}
\end{table}

\begin{table}
\begin{center}
\caption{\label{tab:coreDataTotal}\small{Core energy of edge and screw dislocations in Al, and their slopes with respect to external strains at zero strain, directly obtained from RS-OFDFT calculations. All values are in eV/\AA.}}
\begin{tabular}{ |c|c|c|c|}
\hline
\multicolumn{2}{ |c| }{Edge} &  \multicolumn{2}{ c| }{Screw} \\
\hline
$E_{\rm c}^{\rm edge}({\boldsymbol \epsilon}^{\rm ext}={\bf 0})$ & 0.401 & $E_{\rm c}^{\rm screw}({\boldsymbol \epsilon}^{\rm ext}={\bf 0})$ & 0.284  \\ 
\hline
$S^{\rm edge}_{11}$  & -2.3 & $S^{\rm screw}_{11}$  & -1.4\\
$S^{\rm edge}_{22}$  & -2.2 & $S^{\rm screw}_{22}$  & -1.2\\
$S^{\rm edge}_{33}$  & -1.0 & $S^{\rm screw}_{33}$  & -0.3\\
$S^{\rm edge}_{13}$  & 0.0 & $S^{\rm screw}_{13}$  & 0.0\\
$S^{\rm edge}_{23}$  & -2.9\tablefootnote{\label{first} The values of $\hat{S}^{\rm screw}_{12}$ and $\hat{S}^{\rm edge}_{23}$ are obtained from the RS-OFDFT calculated core energy dependence on Escaig strains of screw dislocation (cf. Section~\ref{sec:macroDeform}) and edge dislocation~\citep{Das2016,Iyer2015}, respectively. The non-zero values of the slopes are due to the change in the Shockley partial separation distance in response to the Escaig strains. However, as the details of the partials are absent in the mesoscale model used for the case studies in Section~\ref{sec:caseStudies}, we do not include these contributions there.}  & $S^{\rm screw}_{12}$  & 1.3\footref{first}\\
$S^{\rm edge}_{12}$  & - & $S^{\rm screw}_{23}$  & -\\
\hline
\end{tabular}
\end{center}
\end{table} 

\begin{table}
\begin{center}
\caption{\label{tab:coreDataPerf}\small{Non-elastic core energy of edge and screw dislocations in Al, and their slopes with respect to external strains at zero strain, obtained from the RS-OFDFT data by subtracting the non-singular elastic contribution with smearing parameter $a=1|{\bf b}|$. All values are in eV/\AA.}}
\begin{tabular}{ |c|c|c|c|}
\hline
\multicolumn{2}{ |c| }{Edge} &  \multicolumn{2}{ c| }{Screw} \\
\hline
$E_{\rm cnel}^{\rm edge}({\boldsymbol \epsilon}^{\rm ext}={\bf 0})$ & 0.046 & $E_{\rm cnel}^{\rm screw}({\boldsymbol \epsilon}^{\rm ext}={\bf 0})$ & 0.024  \\ 
\hline
$\hat{S}^{\rm edge}_{11}$  & -2.3 & $\hat{S}^{\rm screw}_{11}$  & -1.4\\
$\hat{S}^{\rm edge}_{22}$  & -2.2 & $\hat{S}^{\rm screw}_{22}$  & -1.2\\
$\hat{S}^{\rm edge}_{33}$  & -1.0 & $\hat{S}^{\rm screw}_{33}$  & -0.3\\
$\hat{S}^{\rm edge}_{13}$  & 0.0 & $\hat{S}^{\rm screw}_{13}$  & 0.0\\
$\hat{S}^{\rm edge}_{23}$  & -2.9 & $\hat{S}^{\rm screw}_{12}$  & 1.3\\
$\hat{S}^{\rm edge}_{12}$  & - & $\hat{S}^{\rm screw}_{23}$  & -\\
\hline
\end{tabular}
\end{center}
\end{table}

\makeatletter
\setlength{\@fptop}{0pt}
\makeatother

\begin{table}
\begin{center}
\caption{\label{tab:coreDataAtomistics}\small{Slopes of core energy of edge and screw dislocations in Al with respect to external strains at zero strain, directly obtained from atomistic calculations using two different EAM potentials for Al. All values are in eV/\AA.}}
\begin{tabular}{ |c|c|c|c|c|c|c|c|}
\hline
\multicolumn{4}{ |c| }{Al99.eam.alloy~\citep{Mishin1999}} &  \multicolumn{4}{ c| }{Al-LEA.eam.alloy~\citep{Liu2004}}\\
\hline
\multicolumn{2}{ |c| }{Edge} &  \multicolumn{2}{ c }{Screw} & \multicolumn{2}{ |c| }{Edge} &  \multicolumn{2}{ c| }{Screw}\\
\hline
$S^{\rm edge}_{11}$  & 0.0 & $S^{\rm screw}_{11}$  & -0.9 &  $S^{\rm edge}_{11}$  & -4.5 & $S^{\rm screw}_{11}$  & -7.7\\
$S^{\rm edge}_{22}$  & -3.0  &$S^{\rm screw}_{22}$  & -1.6 &  $S^{\rm edge}_{22}$  & -4.4 & $S^{\rm screw}_{22}$  & -3.9 \\
$S^{\rm edge}_{33}$  & -1.8 &$S^{\rm screw}_{33}$  & -2.0 &  $S^{\rm edge}_{33}$  & -2.1 & $S^{\rm screw}_{33}$  & -0.6\\
$S^{\rm edge}_{13}$  & 0.0  &$S^{\rm screw}_{13}$  & 0.0 &  $S^{\rm edge}_{13}$  & 0.0 & $S^{\rm screw}_{13}$  & 0.0\\
$S^{\rm edge}_{23}$  & -1.5  &$S^{\rm screw}_{12}$  & 0.3 &  $S^{\rm edge}_{23}$  & -3.5 & $S^{\rm screw}_{12}$  & 6.2 \\
\hline
\end{tabular}
\end{center}
\end{table}

\section{Derivation of nodal core forces for dislocations dissociated into partials}\label{sec:coreForcePartials}
Here, we briefly discuss on extending the derivation of the nodal core forces for a network of perfect dislocations presented in Section~\ref{sec:forceDerivStr} to the case where the perfect dislocation lines are dissociated into partials. Consideration of partials is crucial to correctly study various elementary mechanisms which govern plasticity and hardening in fcc materials, for e.g.~, dislocation junction formation and cross-slip mechanisms (cf.~\cite{Martinez2008,Shenoy2000}). We consider the DDD framework developed by ~\cite{Martinez2008} which accounts for partials and stacking faults in fcc crystals, and briefly discuss incorporating the core-energetics information into such a framework. Their methodology considers the dissociated partials as independent segments, and introduce a new degree of freedom, $\gamma {\bf n}^{ij}$ to each partial segment ${\bf l}^{ij}$, where $\gamma$ is the stacking fault energy per unit area, and ${\bf n}^{ij}$ is the unit normal to the slip plane on which the partials are dissociated. The force at any point on the segment corresponding to perturbations in the stacking fault energy is expressed as
\begin{equation}
f^{ij}_{{\rm sf},p}= \gamma \varepsilon_{pqk} \xi_{q} n_k^{ij} \,, \label{sfForcePoint}
\end{equation}
where the direction of ${\bf n}^{ij}$ is chosen such that the above force always points towards the other partial, or in other words it is attractive. The attractive force is balanced by the elastic repulsion between the partials and other external elastic forces, such as those resulting from Escaig stresses, producing the equilibrium stacking fault width. We refer to ~\cite{Martinez2008} for more details on topological considerations and conservation rules related to the consideration of partials. Focussing on the core energetics, similar to equation~\eqref{eq:Aextract}, we remove the non-singular elastic energy due to the Shockley partials and additionally the already accounted stacking fault energy effects from $E_{\rm c}^{\rm edge/screw}({\boldsymbol \epsilon}^{\rm ext}={\bf 0})$ and $ S^{\rm edge/screw}_{\alpha\beta}$, to obtain the non-elastic core energy and its slopes, denoted as $\check{E}_{\rm cnel}^{\rm edge/screw}({\boldsymbol \epsilon}^{\rm ext}={\bf 0};a)$ and $ \check{S}^{\rm edge/screw}_{\alpha\beta}(a)$, respectively. Further, as the partials are treated as independent segments, it is convenient to assume the non-elastic core energy of each partial to be half of the full dislocation non-elastic core energy,
\begin{equation}
\check{E}_{\rm cnel}^{\rm {edge}_{p}/{screw}_p}({\boldsymbol \epsilon}^{\rm ext};a)=\frac{1}{2}\left(\check{E}_{\rm cnel}^{\rm edge/screw}({\boldsymbol \epsilon}^{\rm ext}={\bf 0};a) + \check{S}^{\rm edge/screw}_{\alpha\beta}(a){\epsilon}^{\rm ext}_{\alpha\beta}\right) \,,
\end{equation}
where $\check{E}_{\rm cnel}^{\rm {edge}_{p}/{screw}_p}({\boldsymbol \epsilon}^{\rm ext};a)$ is the non-elastic core energy of the edge or screw partial per unit length of dislocation line, and ${\boldsymbol \epsilon}^{\rm ext}$ is the external strain field experienced by the partial excluding the strain field contribution from the other partial in the pair. To exclude the other partial, the external strain field cutoff procedure developed for the perfect dislocation network in equation~\eqref{eq:strExtStrain} can be modified as follows, 
\begin{equation}
{\boldsymbol \epsilon}^{{\rm ext}_{\rm loc}}({\bf x}^{ij}(s_{ij}))= \sum_{kl \, \in \, U_{ij}^{\prime\prime}} {\boldsymbol \epsilon}^{{kl}_{\rm loc}}({\boldsymbol \xi}^{kl},{\bf b}^{kl},{\bf x}^{ij}(s_{ij})) \, ,\label{eq:strDisExtStrain}
\end{equation}
where the set $ U_{ij}^{\prime\prime}$ includes all distinct segments in the network excepting the ones which have one of their nodes as $i$ or $j$ and their respective Shockley partials. Rest of the analysis for derivation of the core forces follows along similar lines as the case of the perfect dislocation presented in Section~\ref{sec:forceDerivStr}.

\clearpage
\section*{References}
\bibliography{mybibfile}

\end{document}